\pdfoutput=1

\documentclass{ws-jai}
\usepackage[flushleft]{threeparttable}

\usepackage{amssymb}
\usepackage{xspace}

\usepackage{afterpage}

\newcommand*{\SPRing}{{\it single-pixel ring}\xspace}
\newcommand*{\INFOCUS}{{\it InFOC$\mu$S}\xspace}
\newcommand*{\Integral}{{\it INTEGRAL}\xspace}

\begin{document}

\catchline{}{}{}{}{} 

\markboth{M.~Beilicke et al.}{Design and Performance of the X-ray polarimeter X-Calibur}

\title{DESIGN AND PERFORMANCE OF THE X-RAY POLARIMETER X-CALIBUR}


\author{M.~Beilicke$^1$,
F.~Kislat$^1$,
A.~Zajczyk$^1$,
Q.~Guo$^1$,
R.~Endsley$^1$,
M.~Stork$^1$,
R.~Cowsik$^1$,
P.~Dowkontt$^1$,
S.~Barthelmy$^2$,
T.~Hams$^2$,
T.~Okajima$^2$,
M.~Sasaki$^2$,
B.~Zeiger$^2$,
G.~De Geronimo$^3$,
M.G.~Baring$^4$,
and H.~Krawczynski$^1$
}


\address{$^1$Department of Physics and McDonnell Center for the Space
Sciences, Washington University in St.Louis, St. Louis, MO 63130, USA\\
$^2$Goddard Space Flight Center, NASA's Goddard Space Flight
Center, 8800 Greenbelt Rd., Greenbelt, Md., 20771\\
$^3$Microelectronics Group, Instrum. Div., Brookhaven Nat. Lab.,
Upton, NY, USA\\
$^4$Department of Physics and Astronomy, Rice University, 6100
Main MS-550, Houston, TX 77005-1827
}
\maketitle

\footnotetext{Corresponding author: M.~Beilicke (beilicke@physics.wustl.edu)}

\begin{history}
\received{(to be inserted by publisher)};
\revised{(to be inserted by publisher)};
\accepted{(to be inserted by publisher)};
\end{history}

\begin{abstract}

X-ray polarimetry promises to give qualitatively new information about 
high-energy astrophysical sources, such as binary black hole systems,
micro-quasars, active galactic nuclei, neutron stars, and gamma-ray 
bursts. We designed, built and tested a X-ray polarimeter, {\it 
X-Calibur}, to be used in the focal plane of the balloon-borne \INFOCUS 
grazing incidence X-ray telescope. X-Calibur combines a low-Z scatterer 
with a CZT detector assembly to measure the polarization of $20 - 80 \, 
\rm{keV}$ X-rays making use of the fact that polarized photons scatter 
preferentially perpendicular to the electric field orientation. 
X-Calibur achieves a high detection efficiency of $\simeq 80\%$. 
The X-Calibur detector assembly is completed, tested, and fully 
calibrated. The response to a polarized X-ray beam was measured 
successfully at the Cornell High Energy Synchrotron Source. This paper 
describes the design, calibration and performance of the X-Calibur 
polarimeter. In principle, a similar space-borne scattering polarimeter 
could operate over the broader $2-100 \, \rm{keV}$ energy band.

\end{abstract}


\section{Introduction} \label{sec:Introduction}

Only the most violent objects in the universe are capable of producing 
high-energy particles in non-thermal acceleration processes and emit 
photons with energies in the X-ray band and above. Spectral and 
morphological studies in the X-ray and gamma-ray bands have become 
established tools to study the non-thermal emission processes of various 
astrophysical sources \cite{Frederick2010}. However, many of the regions 
of interest (black hole vicinities, formation zones of relativistic 
jets, etc.) are too small to be spatially resolved with current and 
future instruments. Spectro-polarimetric X-ray observations are capable 
of providing additional information~-- namely (i) the energy-resolved 
fraction of linear polarization (e.g. what fraction of emission is 
polarized), and (ii) the projected orientation of the polarization plane 
(defined by the electric field vector of the photon) with respect to the 
emitting source. Various emission mechanisms of compact objects lead to 
comparable spectral signatures, but would differ in the polarization 
characteristics. The measurements of polarization properties would 
therefore help to constrain the geometry of the inner regions of 
relativistic plasma jets, mass-accreting black holes (BHs) and neutron 
stars \cite{Lei1997, Krawcz2011}.

So far, only a few space-borne missions have successfully measured 
polarization in the X-ray regime. The Crab nebula is the only source for 
which X-ray polarization has been established with a high level of 
confidence. In measurements with the OSO-8 satellite, the Crab exhibits 
a polarization fraction of $20 \%$ at energies of $2.6 - 5.2 \, 
\rm{keV}$ and a direction angle of $30^{\circ}$ with respect to the 
X-ray jet observed in the nebula \cite{Weisskopf1978}. At energies above 
$100 \, \rm{keV}$, measurements resulted in a polarization fraction of 
$46 \% \pm 10 \%$ with the direction aligned with the jet 
\cite{Dean2008, Forot2008}. A second astrophysical source emitting 
polarized X-rays was identified recently. \Integral observations of the 
X-ray binary Cygnus\,X-1 indicate a high fraction of polarization of 
$(67 \pm 30)\%$ in the $400 \, \rm{keV} - 2 \, \rm{MeV}$ band, whereas a 
$20\%$ upper limit was derived for the $250-400 \, \rm{keV}$ band 
\cite{Laurent2011}. Various authors have reported tentative evidence for 
polarized hard X-ray/soft gamma-ray emission from different gamma-ray 
bursts \cite{Coburn2003, Kalemci2007, Yonetoku2011, Kostelecky2013}. 
However, all of these detections have somewhat marginal significance, 
possibly being impacted by unknown systematic effects in their 
respective instrumentation.

Model predictions of polarized emission for various source types lie 
slightly below the sensitivity of the past OSO-8 mission, making future, 
more sensitive polarimetry missions particularly interesting. However, 
there are currently no dedicated missions in orbit that are capable of 
measuring X-ray polarization fractions in the $<10\%$ regime, a 
requirement to study the corresponding emission mechanisms for a variety 
of astrophysical source classes (see Sec.~\ref{sec:Science}). More 
recently, various experiments have been proposed that could change the 
situation as they combine broadband sensitivity with a high detection 
efficiency. The proposed polarimeters use focusing mirrors to collect 
photons from the source. Photoelectric effect polarimeters, like the 
{\it Gravity and Extreme Magnetism SMEX} (GEMS) mission \cite{GEMS} and 
XIPE \cite{XIPE}, track the direction of photo electrons ejected in 
photoelectric effect interactions of the X-rays. The {\it ASTRO-H} 
mission (to be launched in 2015) will carry the {\it Soft Gamma-Ray 
Imager}. The detector combines an X-ray and gamma-ray collimator with a 
Si scatterer and CdTe absorber. The mission will be able to do 
scattering polarimetry at $E > 50 \, \rm{keV}$ energies 
\cite{Tajima2010}. Our group is working on a proposal of the space borne 
scattering polarimeter PolSTAR which uses a lower-Z LiH stick as a 
scatterer to enable the detection of $3 - 80 \, \rm{keV}$ X-rays. The 
PolSTAR concept will be described in a forthcoming paper.  Missions like 
GEMS, PolSTAR and XIPE aim at obtaining $1 \%$ minimum detectable 
polarization fractions, see Eq.~(\ref{eqn:MDP}), for mCrab sources. 
These missions would allow very high-signal-to-noise detections of 
bright galactic sources (e.g. Cyg\,X-1, GRS\,1915+105, Her\,X-1) and 
could detect $\sim1 \%$ polarization fractions for extra-galactic 
sources (e.g. NGC\,4151).

Scattering polarimeters detect the direction into which photons scatter 
when interacting in the detector. Although different scattering 
processes dominate at different energies (coherent scattering below a 
few keV, Thomson scattering at intermediate energies, and Compton 
scattering at energies $> 20 \, \rm{keV}$), all scattering processes 
share the property that the photons scatter preferentially perpendicular 
to the polarization plane (electric field vector) of the incoming 
photon. For example, the angular dependence of Compton scattering 
processes is given by the Klein-Nishina cross section \cite{Evans1955}:

\begin{equation}
\label{eq:KleinNishinaCrossSection} 
\frac{\rm{d}\sigma}{\rm{d}\Omega} = \frac{r_{0}^{2}}{2} 
\frac{k_{1}^{2}}{k_{0}^{2}} \left[ \frac{k_{0}}{k_{1}} + 
\frac{k_{1}}{k_{0}} -2 \sin^{2} \theta \cos^{2} \eta \right], 
\end{equation}
where $\eta$ is the angle between the electric vector of the incident 
photon and the scattering plane, $r_{0}$ is the classical electron 
radius, $\bf{k}_{0}$ and $\bf{k}_{1}$ are the wave-vectors before and 
after scattering, and $\theta$ is the scattering angle. The azimuthal 
distribution of scattered events shows a sinusoidal modulation with a 
$180^{\circ}$ periodicity and a maximum at $\pm 90^{\circ}$ to the 
preferred electric field direction of a polarized X-ray signal.

In this paper we describe the design and performance of a scattering 
polarimeter, X-Calibur. The polarimeter utilizes a plastic scintillator 
as scatterer which can scatter $E > 20 \, \rm{keV}$ X-rays efficiently. 
This is ideal for the operation on a balloon where the energy threshold 
is well matched to the low energy cutoff of the transmissivity caused by 
the residual atmosphere above the balloon altitude of 125,000 feet. An 
assembly of Cadmium Zinc Telluride (CZT) detectors surrounds the 
scintillator in order to record the azimuthal distribution of the 
scattered photons, allowing one to reconstruct the polarization 
properties of the incoming X-ray beam. The background of charged 
particles and high-energy photons is suppressed by an active CsI shield.

The paper is structured as follows. Section~\ref{sec:Science} gives a 
brief overview over the scientific potential of hard X-ray polarimetry. 
The design of the X-Calibur polarimeter is described in 
Sec.~\ref{sec:XCLB}. The data analysis methods are described in 
Sec.~\ref{sec:Analysis}, followed by an outline of the simulation 
procedure in Sec.~\ref{sec:Simulations}. 
Section~\ref{sec:DetectorCharacterization} describes the calibration and 
characterization studies of the CZT detectors. Measurements with the 
assembled X-Calibur polarimeter to characterize the scattering 
scintillator and the shield are described in 
Sec.~\ref{sec:XCLB_Characterization}. Polarization measurements with 
X-Calibur are described in Sec.~\ref{sec:XCLB_PolarimetryMeasurements}. 
The paper ends with a summary and outlook in Sec.~\ref{sec:Conclusion}. 
The X-Calibur data presented in this paper were taken (i) in a 
laboratory environment at Washington University, (ii) at the Cornell 
High Energy Synchrotron Source (CHESS), and (iii) in Ft.~Sumner, NM, 
during a preparation campaign for an upcoming balloon flight.

\section{Scientific Potential} \label{sec:Science}

This section discusses the scientific potential for a scattering 
polarimeter such as X-Calibur from a balloon platform. X-rays from 
cosmic sources can be polarized owing to the anisotropy in the source 
geometry and/or the emission characteristics of various processes 
\cite{Lei1997}. Non-thermal emission, like synchrotron radiation, 
results in a large polarization fraction $r$. Synchrotron emission will 
result in linearly polarized photons with their electric fields oriented 
perpendicular to the magnetic field lines (projected); the observed 
polarization map can therefore be used to trace the magnetic field 
structure of the source, a common practice in radio and optical 
polarimetry. An electron population with a spectral energy distribution 
of $\rm{d}N/\rm{d}E \propto E^{-p}$ emitting in a uniform magnetic field 
will lead to an observable fraction of polarization of 
\cite{Korchakov1962}:

\begin{equation}
r_{\rm{sync}} = \frac{p+1}{p+7/3}, \quad \rm{with} \, (p + 3) = 
\frac{\alpha + 1} {\alpha + 5/3}.
\end{equation}
Here, $\alpha$ is the index of the X-ray power law spectrum. An observed 
polarization fraction close to this limit can therefore be interpreted 
as an indication of a highly ordered magnetic field since 
non-uniformities in the magnetic field will reduce $r_{\rm{sync}}$. The 
polarized synchrotron photons can in turn be inverse-Compton scattered 
by relativistic electrons~-- weakening the fraction of polarization (but 
not erasing it) and imprinting a scattering angle dependence to the 
observed fraction of polarization \cite{Krawczynski2012b}. Such 
inverse-Compton signals will usually (but not always) appear in hard 
gamma-rays, where polarimetry is difficult, due to multiple scattering 
in pair production detectors. Another important mechanism for polarizing 
photons is Thomson scattering which creates a polarization perpendicular 
to the scattering plane \cite{Rybicki1991}. Curvature radiation is 
polarized, as well. The scientific potentials of spectro-polarimetric 
observations over the broadest possible energy range are summarized 
below; more detailed discussions can be found in \citet{Krawcz2011} and 
\citet{Lei1997}.

\begin{figure}[t]
\begin{center}
\includegraphics[width=0.49\textwidth]{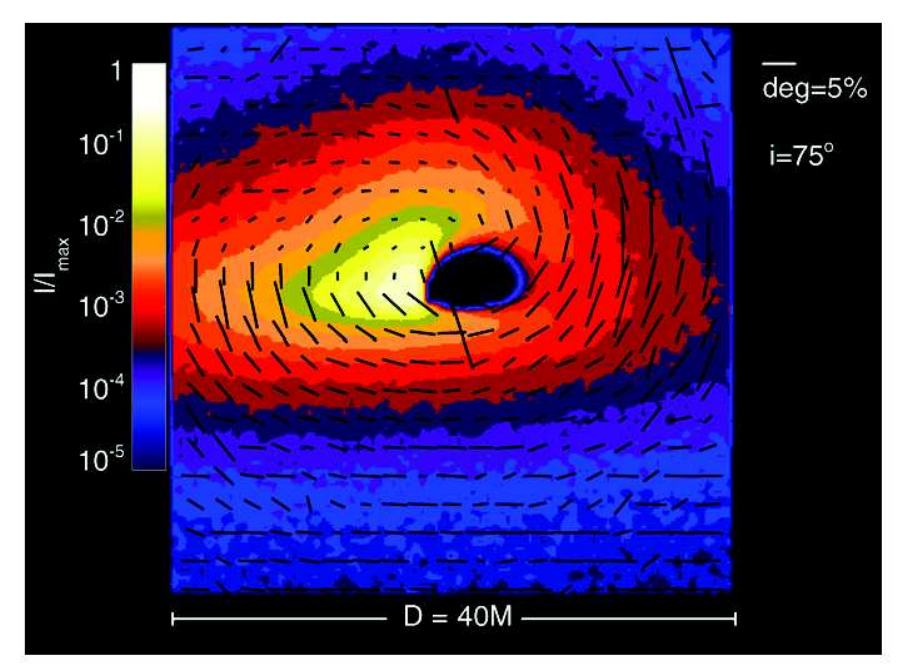}
\end{center}

\caption{Ray-traced image of direct radiation from a thermal disk around 
a BH including returning radiation (observer located at an inclination 
angle of $75^{\circ}$, gas on the left side of the disk moving toward 
the observer causing the characteristic increase in intensity due to 
relativistic beaming). The image is adopted from \citet{Schnittman2009}. 
The observed intensity is color-coded on a logarithmic scale and the 
energy-integrated polarization vectors are projected onto the image 
plane with lengths proportional to the fraction of polarization.}

\label{fig:BH_Polarization}
\end{figure}

{\it Binary black hole systems.} Particle scattering in a Newtonian 
accretion disk surrounding a BH will lead to the emission of polarized 
X-rays. Relativistic aberration and beaming, gravitational lensing, and 
gravito-magnetic frame-dragging will result in an energy-dependent 
fraction of polarization since photons with higher energies originate 
closer to the BH than the lower-energy photons \cite{Connors1977}. 
\citet{Schnittman2009} calculate the expected polarization signature 
including (i) the effects of deflection of photons emitted in the disk 
by the strong gravitational forces in the regions surrounding the BH and 
(ii) re-scattering these photons by the accretion disk 
\cite{Schnittman2009, Schnittman2010}. The resulting effect is a swing 
in the polarization direction from being horizontal at low energies to 
vertical at high energies, i.e., parallel to the spin axis of the BH. 
Spectro-polarimetric observations can therefore be used to constrain the 
mass and spin of the BH \cite{Schnittman2009}, as well as the 
inclination of the inner accretion disk and the shape of the corona 
\cite{Schnittman2010}, see Fig.~\ref{fig:BH_Polarization}. In principle, 
X-ray polarization can also be used to test General Relativity in the 
strong gravity regime \cite{Krawczynski2012c}.

{\it Pulsars.} High-energy particles in pulsar magneto-spheres are 
expected to emit synchrotron and/or curvature radiation which are 
difficult to distinguish from one another, solely based on the observed 
photon energy spectrum. However, since the orbital planes for 
accelerating charges that govern these two radiation processes are 
orthogonal to each other, their polarized emission will exhibit 
different behavior in position angle and polarization fraction as 
functions of energy and the rotation phase of the pulsar 
\cite{Dean2008}. An illustration of the models of phase dependence of 
the X-ray/gamma-ray polarization signatures in pulsars can be found in 
\citet{Dyks2004}. In magnetars, the highly-magnetized cousins of 
pulsars, polarization-dependent resonant Compton up-scattering is a 
leading candidate for generating the observed hard X-ray tails 
\cite{BaringHarding2007}. In both these classes, phase-dependent 
spectro-polarimetry can probe the emission mechanism, and provide 
insights into the magnetospheric locale of the emission region.

Cyclotron lines arising from transitions between Landau levels in 
intense magnetic fields that occur in the polar regions of neutron stars 
and magnetars are polarized. Also, the absorption scattering 
cross-sections from the Landau levels are dependent on the polarization 
state of the X-rays, so that the radiative transfer through such plasma 
will lead to a polarization of the emergent radiation. The first such 
cyclotron feature was observed by \citet{Truemper1978} in Her\,X-1 and 
is interpreted as due to an absorption line around $40 \, \rm{keV}$ 
\cite{Staubert2007}. Such features allow an estimate of the magnetic 
field strength \cite{Coburn2002}. Observations of polarized X-rays will 
strongly confirm that these features are indeed due to cyclotron lines 
in magnetic fields of $3-5 \times 10^{12} \, \rm{G}$. A detailed 
discussion of the theoretical aspects of cyclotron radiation is given in 
\citet{Leahy2010}.

{\it Pulsar wind nebulae.} The compact object (e.g. pulsar) resulting 
from a previous supernova explosion can be surrounded by a synchrotron 
emitting nebula. The nebula extends far beyond the magneto-sphere of the 
central pulsar, but its emission is believed to be driven by the pulsar 
which injects relativistic electrons/positrons that are further shock 
accelerated in the nebula. Spectro-polarimetric observations can be used 
to constrain the magnetic field and particle populations in such pulsar 
wind nebulae~-- such as the Crab nebula, the leading driver for this 
field of X-ray polarimetry. Given the more compact emission regions at 
high energies, these objects potentially show a higher polarization 
fraction at hard X-rays as compared to soft X-rays, reflecting the 
contrast between jet and more diffuse nebular contributions 
\cite{Forot2008}.

{\it Supernova remnants.} Supernova remnants (SNRs) present an 
opportunity to perform X-ray polarimetry, as well. The remnants possess 
tangled magnetic fields on large scales in their interiors, as is 
evidenced in the classic radio polarization map of the Crab nebula 
\cite{Velusamy1985}.  Both, radio and X-ray signals, are believed to be 
due to synchrotron emission, and so it is reasonably presumed that X-ray 
emission from SNRs should be significantly polarized. The X-ray spectra 
of such remnants are typically steeper than spectra in the radio, which 
leads to the expectation of higher polarization fractions in the X-ray 
band. Yet, since the electrons generating X-rays will diffuse on larger 
spatial scales than their radio-emitting counterparts do, the X-ray 
signals should capture the field morphology on larger scales. It may or 
may not be more coherent than the field structure on smaller (radio) 
scales. Due to instrumental limitations in angular resolution at X-ray 
energies, however, it will not be possible to resolve individual 
regions~-- depending on the angular size of the remnant. The 
observational challenge will therefore be to overcome the competition 
between compact regions causing highly polarized emission on one hand, 
and on the other hand an averaging effect of different emission regions 
with different orientations of their magnetic fields. An example of 
expectations for the SNR synchrotron polarization properties can be 
found in \citet{Bykov2009}.

{\it Relativistic jets in active galactic nuclei.} Relativistic 
electrons in jets of active galactic nuclei (AGN) emit polarized 
synchrotron radiation at radio/optical wavelengths. The same electron 
population is believed to produce hard X-rays by inverse-Compton 
scattering off a photon field. Simultaneous measurements of the 
polarization angle and the fraction of polarization in the radio to hard 
X-ray band could help to disentangle the following scenarios: (i) If the 
electrons mainly up-scatter the co-spatial synchrotron photon field 
(synchrotron self Compton), the polarization of the hard X-rays is 
expected to track the polarization at radio/optical wavelengths 
\cite{Poutanen1994}. The fraction of X-ray polarization could be close 
to the fraction of polarization of the synchrotron emission measured in 
the radio/optical bands and the polarization directions between radio, 
optical and X-rays should be identical. (ii) If the electrons dominantly 
up-scatter an external photon field (external Compton, e.g. photons of 
the cosmic microwave background or from the accretion disk surrounding 
the super-massive black hole) the hard X-rays will have a relatively 
small ($<$10\%) fraction of polarization \cite{McNamara2009}. Hadronic 
jet emission models for low-synchrotron-peaked AGN, on the other hand, 
predict an even higher fraction of polarization at high energies, 
compared to the leptonic SSC models \cite{Zhang2013}.

Polarization also allows one to test the structure of the magnetic field 
of the jet. Particles accelerated in a helical field which are moving 
through a standing shock can cause an X-ray synchrotron flare with a 
continuous (in time) swing in polarization direction. Such an event was 
observed from BL\,Lacertae at optical wavelengths \cite{Marscher2008}.

{\it Gamma-ray bursts.} Gamma-ray bursts are believed to be connected to 
hyper-nova explosions and the formation/launch of relativistic jets 
\cite{Woosley1993}. As in the case of the jets in AGN, the structure of 
their jets and the particle distribution responsible for gamma-ray 
bursts can be revealed by X-ray polarization measurements 
\cite{Kostelecky2013}. The X-ray emission of a gamma-ray burst, however, 
usually lasts for only a few minutes at most, so that rapid follow-up 
observations in the X-ray band below $30 \, \rm{keV}$ would be the main 
challenge for studying their polarization properties.


On a one-day balloon flight, we would achieve $5-15 \%$ MDPs, see 
Eq.~(\ref{eqn:MDP}), for between 1 and 4 sources. For the Crab, the 
phase resolved polarimetry would allow us to decide between emission 
models. For Cyg\,X-1 and GRS\,1915+105 we could test corona models, and 
for Her\,X-1, we could get a first estimate of the polarization fraction 
of the X-ray emission.

\section{Design of X-Calibur} \label{sec:XCLB}

\begin{figure*}[t]
\begin{center}
\includegraphics[width=0.99\textwidth]{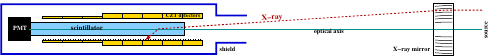}
\end{center}

\caption{Schematic view of the functionality of X-Calibur (not to 
scale). X-rays from an astrophysical source are focused with a grazing 
incidence mirror onto the scattering rod of the polarimeter. The 
scattered X-ray is recorded in one of the surrounding CZT detectors. The 
polarimeter is embedded by a shield to suppress background of particles 
not originating from the mirror.}

\label{fig:XCLB_Functionality}

\end{figure*}

\begin{figure*}[t]
\begin{center}
\includegraphics[height=0.38\textheight]{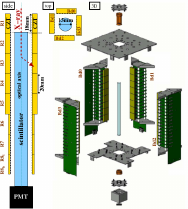}
\hfill
\includegraphics[height=0.38\textheight]{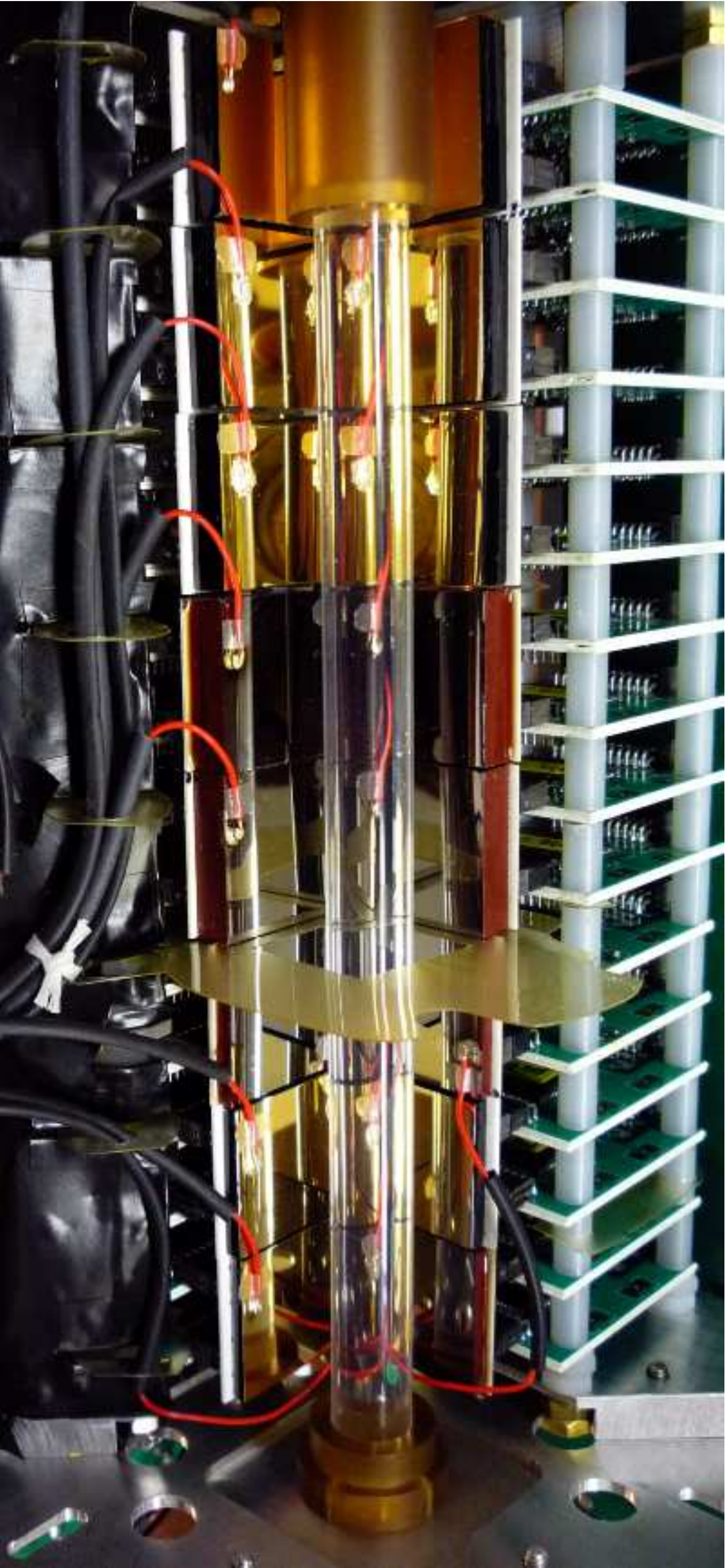}
\hfill
\includegraphics[height=0.38\textheight]{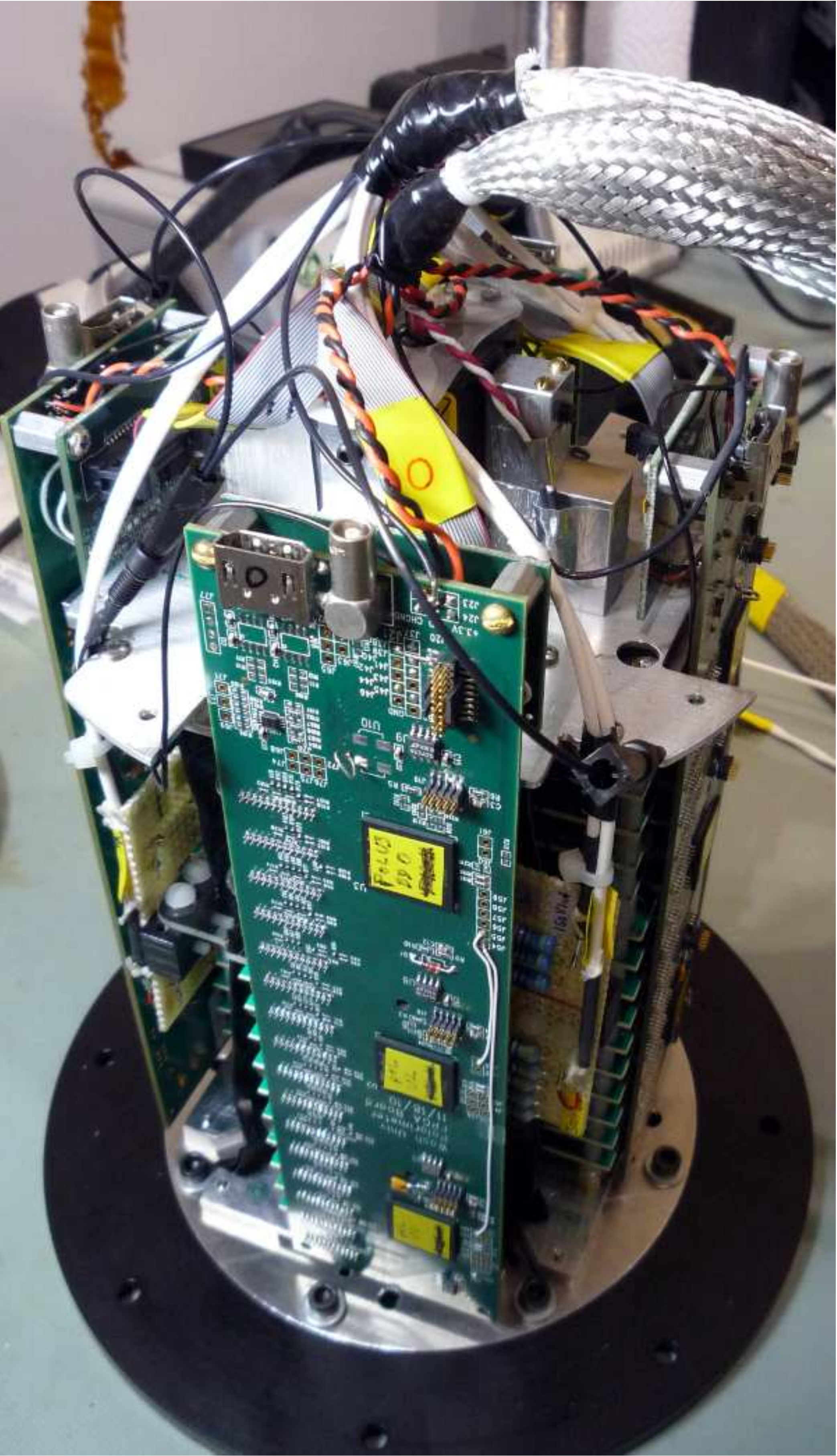}
\end{center}

\caption{Design of the X-Calibur polarimeter. {\bf Left:} Schematic 
illustration. Side view: incoming X-rays Compton-scatter in the 
scintillator rod (aligned with the optical axis, read out by a PMT) and 
are subsequently photo-absorbed in one of the surrounding CZT detectors 
($8\times8$ pixels each, see Fig.~\ref{fig:DetectorRing}). A group of 
four detectors surrounding the scintillator at a given depth are 
referred to as ring ({\it R1} to {\it R8}). Each ring can be further 
divided into top and bottom sub rings, as illustrated for ring {\it R8}. 
Top view: Detector ring viewed along the optical axis, looking into the 
X-ray beam. The polarization signature is imprinted in the azimuthal 
scattering distribution (see Fig.~\ref{fig:X-Calibur_Configurations}, 
top, for more details). The four sides of detectors are referred to as 
boards ({\it Bd0} to {\it Bd3}), each board comprising all eight 
detectors per side. 3D view (`exploded'): four sides of detector columns 
({\it Bd0} to {\it Bd3}) surround the central scintillator rod~-- 
covering the whole range of azimuthal scattering angles. {\bf Middle:} 
Partly assembled polarimeter with two detector sides removed for better 
visibility. The red wires provide the high voltage to the detectors. The 
read-out electronics is stacked at the backside of the detectors. {\bf 
Right:} Fully assembled polarimeter (flipped upside-down: X-ray entering 
from the bottom).}

\label{fig:Design}

\end{figure*}

\begin{table}[t!]

\begin{tabular}{rrrrr}

ID   &  Serial-No. & Date & $C_{\rm{ch}}$ & $C_{\rm{wu/ft}}$ \\
\hline \hline

\noalign{\smallskip}
\multicolumn{5}{l}{{\bf Endicott 5\,mm}} \\
\hline

EN$_{5}$1 & 672992-01  & 01/11 & 0/5 & 0/5 \\
EN$_{5}$2 & 672992-02  & 01/11 & 1/5 & 1/5 \\
EN$_{5}$3 & 672992-03  & 01/11 & 2/5 & 2/5 \\
EN$_{5}$4 & 672992-04  & 01/11 & 3/5 & 3/5 \\
EN$_{5}$5 & 672994-04  & 02/11 & 0/4 & 0/4 \\
EN$_{5}$6 & 672994-03  & 02/11 & 1/4 & 1/4 \\
EN$_{5}$7 & 672994-02  & 02/11 & 2/4 & 2/4 \\
EN$_{5}$8 & 672994-01  & 02/11 & 3/4 & --- \\

\noalign{\smallskip}
\multicolumn{5}{l}{{\bf Creative Electron 5\,mm}} \\
\hline

CE$_{5}$1 & 721613     & 03/11 \\

\noalign{\smallskip}
\multicolumn{5}{l}{{\bf QuikPak 5\,mm}} \\
\hline

QP$_{5}$1 & 3627       & 03/11 & --- & 3/4 \\
QP$_{5}$3 & 3611       & 03/11 & --- & --- \\
QP$_{5}$4 & 3834       & 03/11 & 3/1 & 3/1 \\
QP$_{5}$6 & 6292       & 03/11 & 0/1 & 0/1 \\
QP$_{5}$7 & 6345       & 03/11 & 2/1 & 2/1 \\
QP$_{5}$8 & 721c602    & 03/11 & 1/1 & 1/1 \\


QP$_{5}$13 & 6886       & 04/12 & 0/3 & 0/3 \\
QP$_{5}$14 & 6977       & 04/12 & 2/3 & 2/3 \\
QP$_{5}$16 & 10814      & 04/12 & 2/2 & 2/2 \\
QP$_{5}$17 & 10819      & 04/12 & 0/2 & 0/2 \\
QP$_{5}$18 & 10829      & 04/12 & 1/2 & 1/2 \\
QP$_{5}$19 & 10847      & 04/12 & 3/3 & 3/3 \\
QP$_{5}$20 & 10848      & 04/12 & 3/2 & 3/2 \\
QP$_{5}$21 & 10860      & 04/12 & 1/3 & 1/3 \\


\noalign{\smallskip}
\multicolumn{5}{l}{{\bf Endicott 2\,mm}} \\
\hline

EN$_{2}$1 & 674326-01  & 02/11 & 0/6 & 0/8 \\
EN$_{2}$2 & 674327-01  & 02/11 & 2/8 & 2/8 \\
EN$_{2}$3 & 674328-01  & 02/11 & 2/6 & 2/6 \\
EN$_{2}$4 & 674328-02  & 02/11 & 0/8 & 0/6 \\
EN$_{2}$5 & 674329-01  & 02/11 & --- & --- \\
EN$_{2}$6 & 674330-01  & 02/11 & 1/6 & 1/6 \\
EN$_{2}$7 & 674331-01  & 02/11 & 3/6 & --- \\
EN$_{2}$8 & 674332-01  & 02/11 & --- & 3/8 \\

\noalign{\smallskip}
\multicolumn{5}{l}{{\bf Creative Electron 2\,mm}} \\
\hline

CE$_{2}$1 & 2180       & 03/11 & 0/7 & 0/7 \\
CE$_{2}$2 & 720612     & 03/12 & 3/8 & 3/6 \\
CE$_{2}$3 & 720511     & 03/12 & 1/8 & 1/8 \\
CE$_{2}$4 & 721541i    & 03/12 & 1/7 & 1/7 \\
CE$_{2}$5 & 06172A     & 03/12 & 2/7 & 2/7 \\
CE$_{2}$6 & 726712     & 03/12 & 3/7 & 3/7 \\

\end{tabular}

\caption{CZT detector sample. ID as in this paper (two letters for the 
company, index for the detector thickness $[\rm{mm}]$, and a number), 
serial number, date of delivery [MM/YY], and position in polarimeter 
[Bd/ring] for the configurations $C_{\rm{ch}}$ and $C_{\rm{wu/ft}}$ (see 
Fig.~\ref{fig:X-Calibur_Configurations}).}

\label{tab:Detectors}

\end{table}

\begin{figure*}[t]
\begin{center}
\includegraphics[height=0.255\textheight]{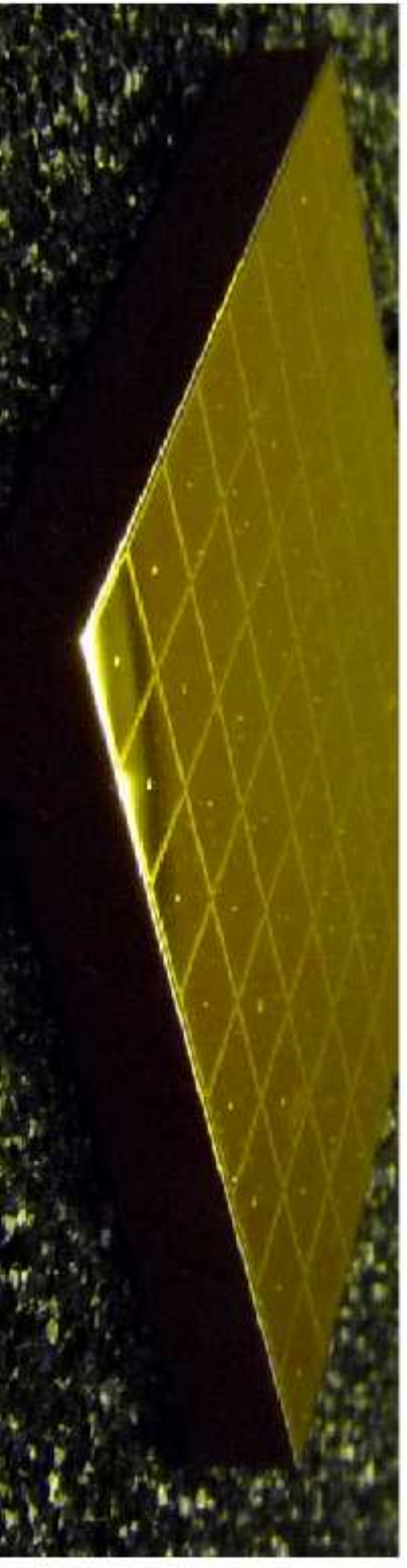}
\hfill
\includegraphics[height=0.255\textheight]{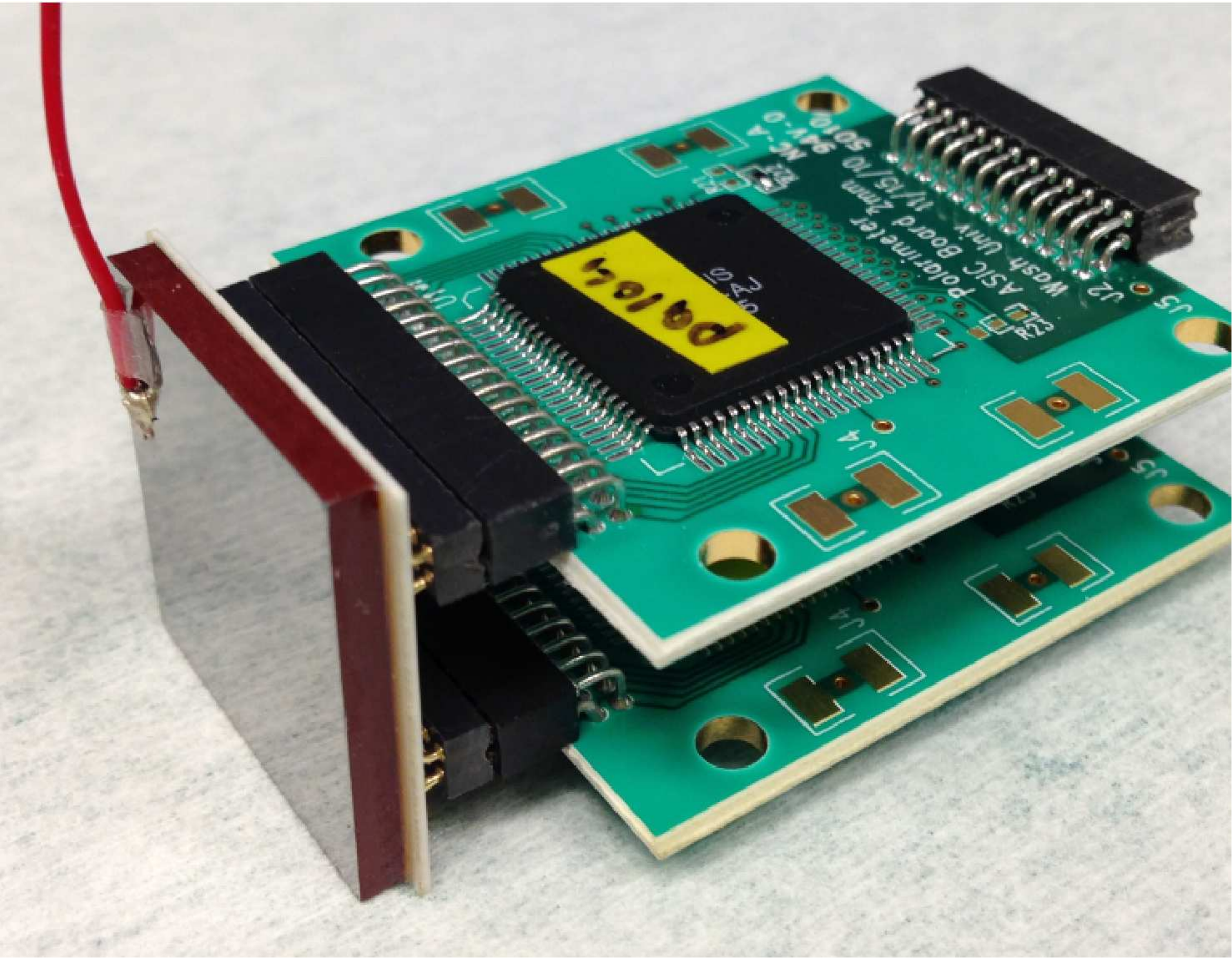}
\hfill
\includegraphics[height=0.255\textheight]{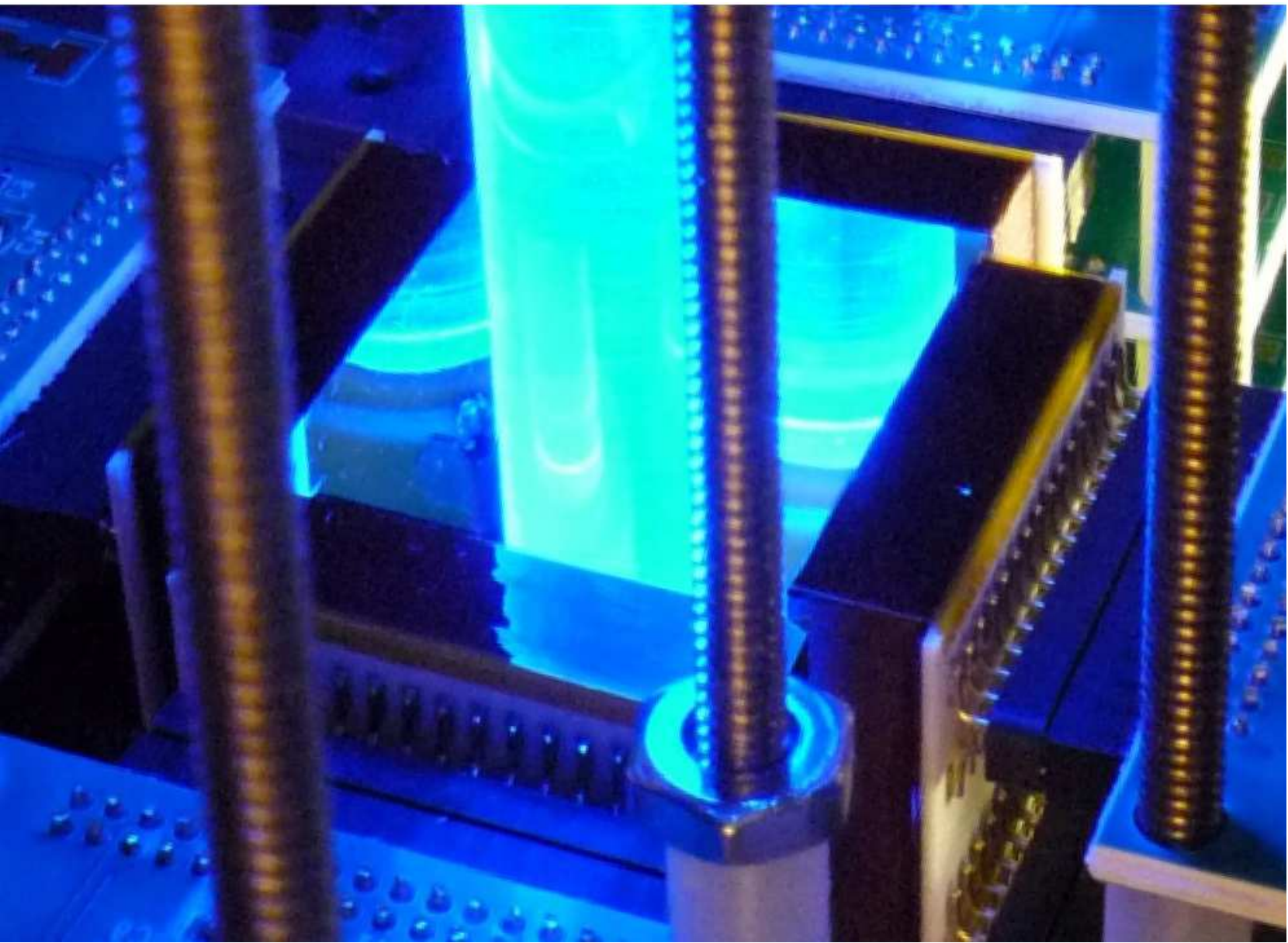}
\end{center}

\caption{Definition of a detector `ring'. {\bf Left:} $2 \times 2 \times 
0.2 \, \rm{cm}^{3}$ CZT detector with 64 pixels (anode side). {\bf 
Middle:} $2 \times 2 \times 0.2 \, \rm{cm}^{3}$ CZT detector bonded to a 
ceramic chip carrier which is plugged into 2 ASIC readout boards. The 
high-voltage cable is glued to the detector cathode (red wire). {\bf 
Right:} Four CZT detectors surrounding the scintillator (blueish glow) 
form a unit referred to as `ring'~-- covering the whole $360^{\circ}$ 
azimuthal scattering range with $4 \times 8 = 32$ pixels. The whole 
polarimeter is equipped with eight rings.}

\label{fig:DetectorRing}
\end{figure*}

The balloon-borne version of X-Calibur is a low-Z Compton scattering 
polarimeter that will allow one to measure polarization fractions in the 
$20 - 80 \, \rm{keV}$ band down to the percentage level. X-Calibur will 
be used in the focal plane of the X-ray mirror of the \INFOCUS telescope 
with a field-of-view of $10 \, \rm{arc min}$; X-Calibur does not provide 
imaging capabilities. Owing to the fact that a grazing incidence mirror 
reflects only under very shallow angles, it changes the polarization 
properties of X-rays by less than $1\%$ \cite{Katsuta2009}. The 
advantages of the X-Calibur design can be described as follows. (i) A 
high detection efficiency is achieved, using roughly $80 \%$ of photons 
impinging on the polarimeter. (ii) The use of a focusing optics instead 
of a large detector volume results in a compact instrument design that 
can be shielded efficiently~-- strongly reducing the background level. 
(iii) The continuous rotation of the polarimeter strongly reduces 
possible systematic effects that can hamper non-rotating polarimeters 
due to asymmetric azimuthal detector responses. These characteristics, 
as outlined in the following sections, make it a well-suited experiment 
to study several sources mentioned in Sec.~\ref{sec:Science} in a 
one-day balloon flight. The energy-dependent detection efficiency of the 
polarimeter depends on (i) the effective area and point-spread function 
of the X-ray mirror, (ii) the fraction of scatterings compared to 
competing interactions such as photo-absorption, and (iii) the 
geometrical detector coverage to record a high fraction of scattered 
X-rays (minimization of possible escape paths) \cite{Guo2010}. This 
section describes the overall design of the polarimeter, as well as the 
characteristics of its individual components.

\subsection{Design}

The conceptual design of the X-Calibur polarimeter is illustrated in 
Figures~\ref{fig:XCLB_Functionality} and \ref{fig:Design}. A low-Z 
scintillator rod aligned with the optical axis of the telescope is used 
as Compton-scatterer~-- leading to a polarization-dependent azimuthal 
scattering distribution that is recorded by the surrounding assembly of 
CZT detectors. The azimuthal distribution is resolved by $4\times8 = 32$ 
pixels for each of the $64$ depth bins along the optical axis. Detailed 
information on the simulations to optimize the X-Calibur design (as 
presented in this paper) can be found in \citet{Krawcz2011} and 
\citet{Guo2013}.

Throughout the paper, we refer to a detector ring as a set of four CZT 
detectors surrounding the scintillator rod on four sides at a given 
depth along the optical axis (see Fig.~\ref{fig:DetectorRing}, right). 
Ring {\it R1} is situated at the polarimeter entrance (top in 
Fig.~\ref{fig:Design}, left), and ring {\it R8} is situated at its rear 
end. Each ring covers the whole $360^{\circ}$ azimuthal scattering 
range. A ring can further be subdivided into sub rings, eg. {\it 
R8$_{\rm{t}}$} and {\it R8$_{\rm{b}}$} for the top and bottom half, 
respectively (see Fig.~\ref{fig:Design}, left). The smallest possible 
subdivision is a ring consisting of only one pixel row, referred to as 
\SPRing. All eight detectors situated on one of the four sides are 
referred to as detector board {\it Bd0} to {\it Bd3} (see 
Fig.~\ref{fig:Design}, left).

In the scintillator, photoelectric effect interactions dominate below 
$\sim$$15 \, \rm{keV}$. At $20 \, \rm{keV}$, however, the cross section 
of the photoelectric absorption already drops to $0.1 \, 
\rm{cm}^{2}/\rm{g}$ and can be neglected as compared to the cross 
section of Compton scattering for which X-Calibur is designed. This 
energy regime therefore defines the detection threshold of the 
polarimeter. The mean free path for Compton scattering in the 
scintillator material is $\approx$$4 \, \rm{cm}$, so that the length of 
the scattering rod of $14 \, \rm{cm}$ covers $\simeq 3.5$ path lengths. 
This translates into a $\simeq$$90 \%$ probability for 
Compton-scattering in the energy regime of $20 - 80 \, \rm{keV}$. For 
sufficiently energetic photons, the Compton interaction produces enough 
scintillation light to trigger a photo-multiplier tube (PMT) attached to 
the end of the scintillator. The trigger efficiency of the 
scintillator/PMT unit is discussed in 
Sec.~\ref{subsec:ScintillatorEfficiency}. The scattered X-rays are in 
turn photo-absorbed in the surrounding rings of high-Z CZT detectors. 
This combination of scatterer/absorber leads to a high fraction of 
unambiguously detected Compton events~-- in contrast to background 
events not entering the polarimeter along the optical axis for which the 
scintillator/CZT coincidences are strongly reduced (see 
Sec.~\ref{subsec:BG_Data}). Linearly polarized X-rays will preferably 
Compton-scatter perpendicular to their electric field vector, see 
Eq.~(\ref{eq:KleinNishinaCrossSection}). This will result in a 
modulation of the measured azimuthal scattering distribution that is 
used to determine the polarization properties of the X-rays (see 
Sec.~\ref{subsec:AnalysisPolarization}).

\subsection{The CZT Detectors}

X-ray photons that penetrate a semi-conductor X-ray detector deposit 
their energy in the detector volume, usually through photo-electric 
effect interactions. A charge cloud proportional to the X-ray energy is 
created and accelerated by a high-voltage electric field that is applied 
between the detector cathode and the pixels on the anode side. The 
moving charge is measured and digitized and the grid position of the 
corresponding pixel reflects the position of the interaction. CZT is the 
semi-conductor material of choice for X-ray detectors operating in the 
$E>5 \, \rm{keV}$ to MeV energy band with a high probability for 
photo-electric effect interactions, see for example \citet{VarPitch}.

The CZT detectors used in X-Calibur were ordered from different 
companies\footnote{Endicott Interconnect: http://www.evproducts.com, 
Quikpak/Redlen: http://redlen.ca, Creative Electron: 
http://creativeelectron.com}. The sample of detectors is listed in 
Tab.~\ref{tab:Detectors}. Each detector ($2\times2 \, \rm{cm}^{2}$) is 
contacted with a 64-pixel anode grid ($2.5 \, \rm{mm}$ pixel pitch) and 
a monolithic cathode facing the scintillator rod. Two different detector 
thicknesses ($2 \, \rm{mm}$ and $5 \, \rm{mm}$) are used in the 
polarimeter (Fig.~\ref{fig:Design}, left). Historically, our group had 
been working with detector thicknesses of $5 \, \rm{mm}$ and higher. 
Therefore, five of the polarimeter rings are equipped with $5 \, 
\rm{mm}$ detectors. The remaining three rings are equipped with $2 \, 
\rm{mm}$ detectors which are still sufficient to absorb more than $99\%$ 
of X-rays in the energy regime relevant for X-Calibur, and at the same 
time measure a lower level of background (which scales roughly with the 
detector volume). The cathodes of the detectors are biased at 
$V_{\rm{bi,5}} = -500 \, \rm{V}$ ($5 \, \rm{mm}$ detectors) and at 
$V_{\rm{bi,2}} = -150 \, \rm{V}$ ($2 \, \rm{mm}$ detectors), 
respectively.

Each CZT detector is permanently bonded (anode side) to a ceramic chip 
carrier which is plugged into the electronic readout board. 
Figure~\ref{fig:DetectorRing} (left) shows a single CZT detector unit 
with an $8 \times 8$ pixel matrix on the anode side as well as the 
readout electronics. Each CZT detector is read out by two digitizer 
boards, each consisting of a 32 channel ASIC and a 12-bit 
analog-to-digital converter. The ASIC was developed by G.~De~Geronimo 
(BNL) and E.~Wulf (NRL) \cite{Wulf2007}. The ASICs are operated at a 
medium amplification (gain) of $28.5 \, \rm{mV/fC}$ and a signal peaking 
time of $0.5 \, \mu \rm{s}$. These settings are a result of a previous 
optimization to achieve an optimal energy resolution and low noise. Each 
ASIC has a built-in capacitor that allows one to directly inject a 
programmable amount of charge into the individual readout channels for 
testing purposes. The readout noise of the ASIC is as low as $2.5 \, 
\rm{keV}$ FWHM (see Fig.~\ref{fig:TempStudiesNoise} in 
Sec.~\ref{subsec:TempStudies}). All 16 digitizer boards (reading eight 
CZT detectors) are read out by one harvester board ({\it Bd0}-{\it Bd3}, 
see Fig.~\ref{fig:Design}) transmitting the data to a PC-104 computer 
with a rate of 6.25~Mbits/s. X-Calibur comprises 2048 data channels. The 
time to read and process a triggered event is about $130 \, \mu\rm{s}$ 
(ASIC dead time). However, only the ASIC involved in the triggered event 
will be dead during the read-out. All other ASICs will still be 
sensitive and can store events that will be read out once the previous 
read-out cycle is completed.

\subsection{The Scintillator}

A plastic scintillator rod is used as Compton-scatterer. The advantage, 
compared to other scattering materials, is the scintillation light 
produced in the scattering interaction. The light is read by a PMT and 
can be used (optional) in the analysis. The EJ-200 scintillator 
(Hydrogen:Carbon ratio of $5.17:4.69$, $\left< Z \right> = 3.4$, $\rho 
\approx 1 \, \rm{g} / \rm{cm}^{3}$, decay time $2.1 \, \rm{nsec}$) is 
used, read by a Hamamatsu R7600U-200 PMT with a high quantum efficiency 
super-bi-alkali photo cathode. To increase the optical yield, the 
scintillator is wrapped in white {\it tyvek$^{\tiny \circledR}$} paper. 
The PMT signal is amplified and digitized. A discriminator tests whether 
the digitized PMT pulse exceeds a programmable trigger threshold and 
activates a corresponding flag ($f_{\rm{sci}}$). The flag is kept high 
for $6 \, \mu\rm{s}$ and is merged into the data stream of triggered CZT 
detector events. The trigger efficiency of the scintillator is studied 
in Sec.~\ref{subsec:ScintillatorEfficiency}. The $f_{\rm{sci}}$ flag 
allows one to select scintillator/CZT events from the data, which 
represent likely Compton-scattering candidates~-- strongly suppressing 
other backgrounds (see Sec.~\ref{subsec:BG_Data}). However, the 
polarimeter can be operated without the PMT trigger information, with 
the scintillator acting only as a passive scatterer.

\begin{figure*}[t]
\begin{center}
\includegraphics[height=0.31\textheight]{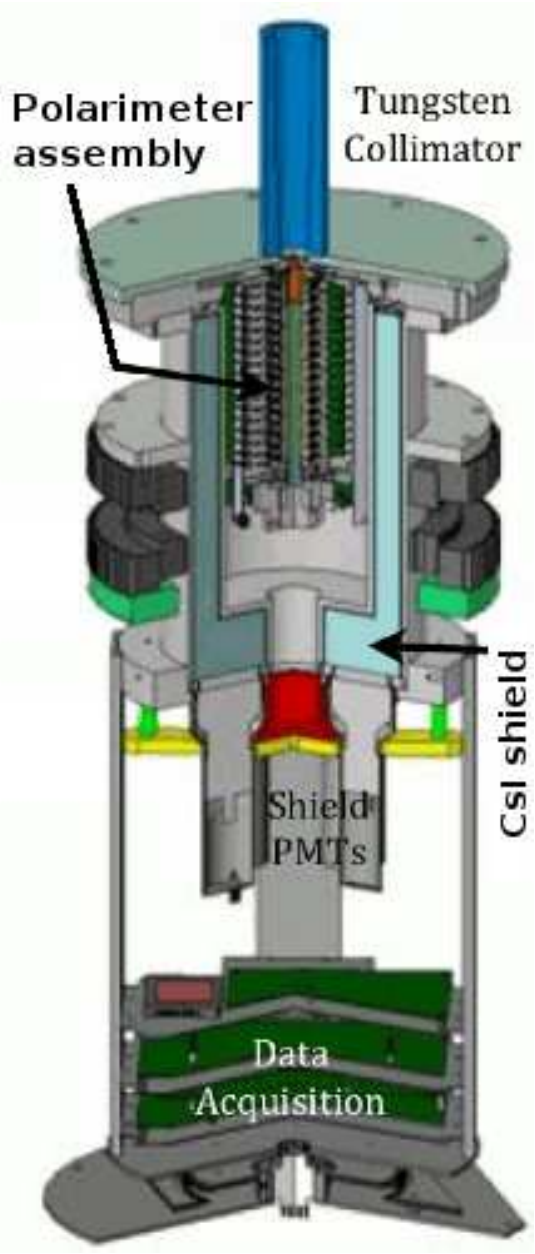}
\includegraphics[height=0.31\textheight]{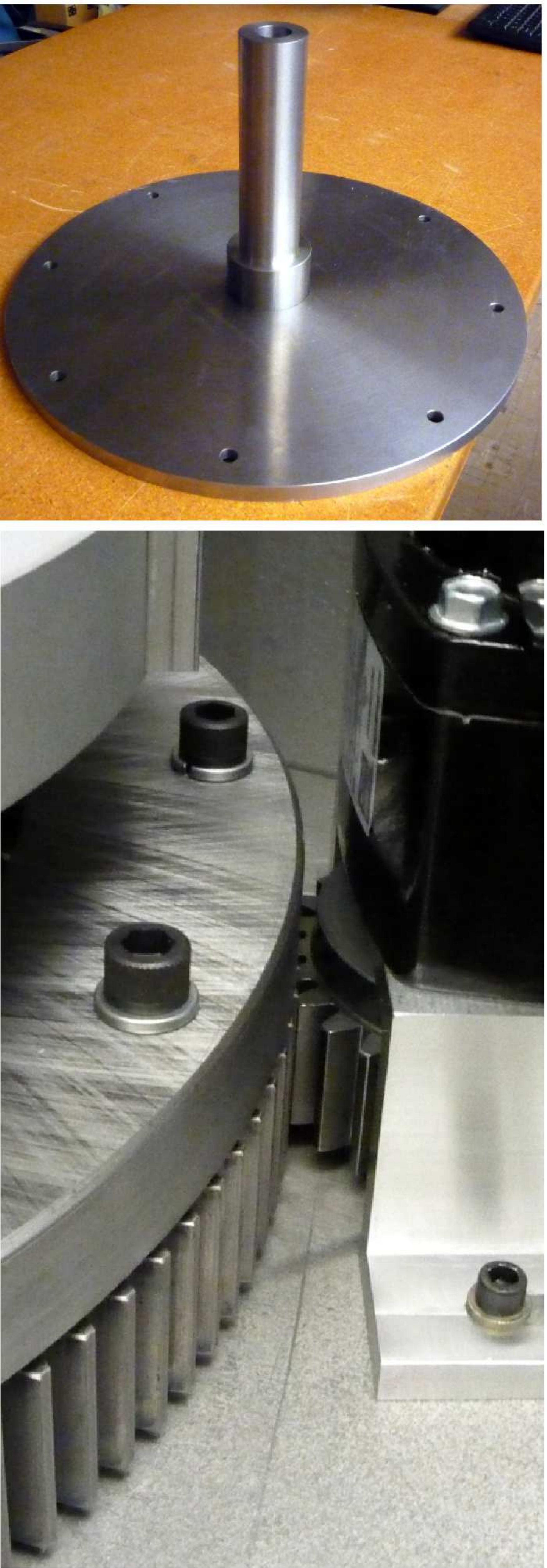}
\hfill
\includegraphics[height=0.31\textheight]{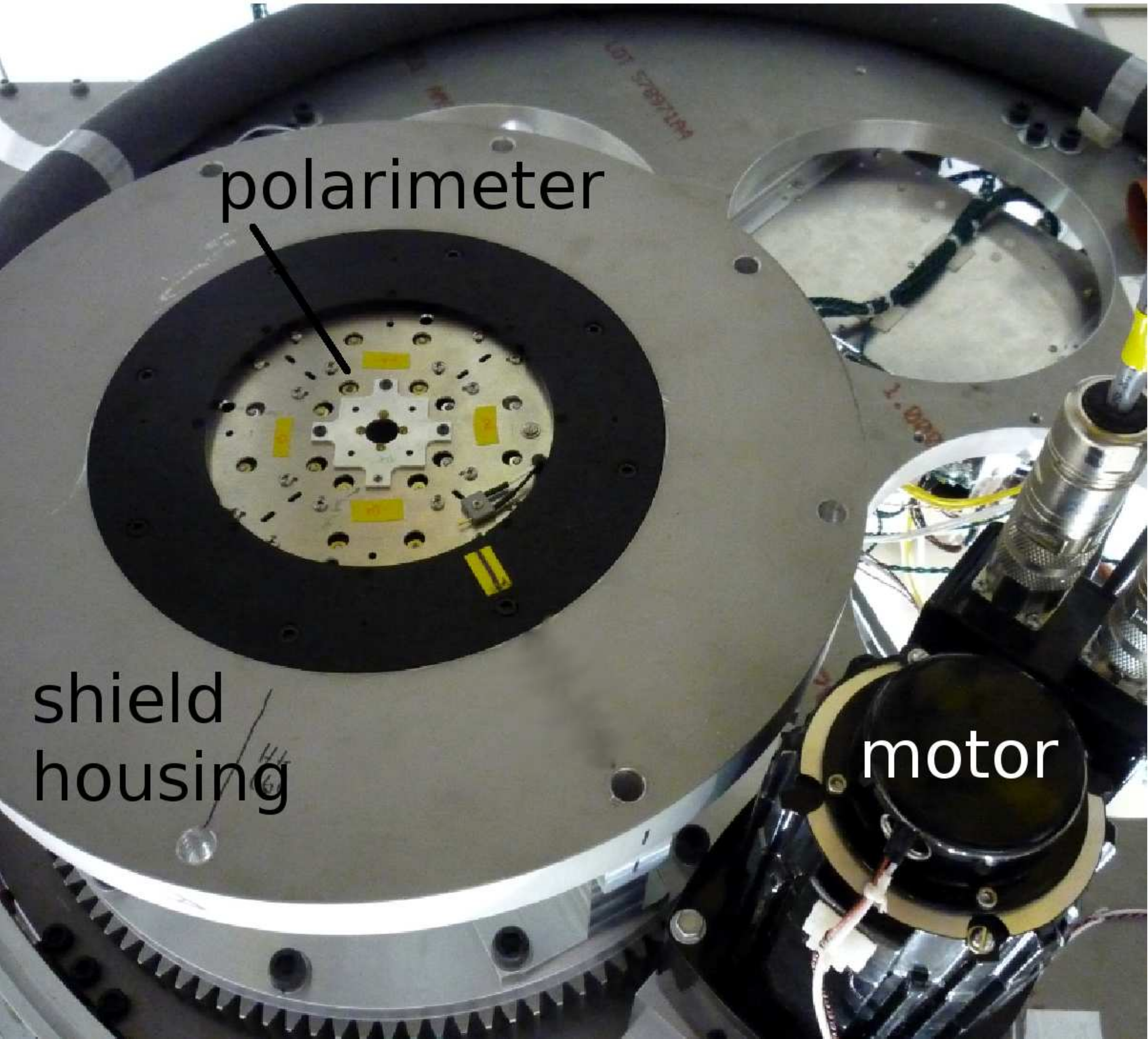}
\hfill
\includegraphics[height=0.31\textheight]{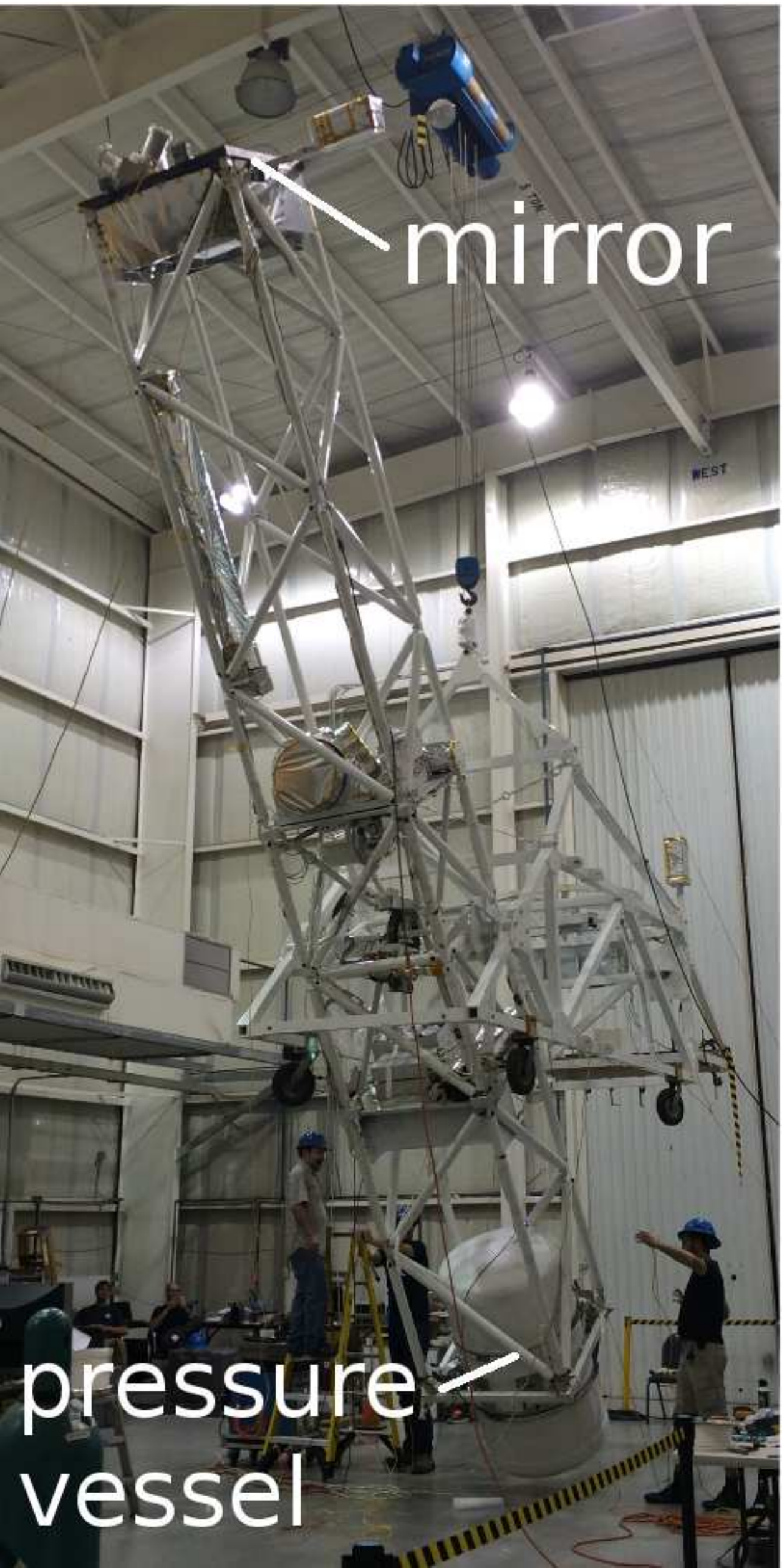}
\end{center}

\caption{Active/passive shield, rotation design, and \INFOCUS X-ray 
telescope. {\bf Left:} X-Calibur polarimeter, embedded by the CsI active 
shield, as well as the electronic readout and azimuthal rotation 
bearing. {\bf 2$^{\rm{nd}}$ from left, top:} Tungsten cap with 
collimator, only allowing X-rays from the mirror to enter the shield. 
{\bf 2$^{\rm{nd}}$ from left, bottom:} Motor and ring bearing to rotate 
polarimeter/shield assembly. {\bf 2$^{\rm{nd}}$ from right:} Opened 
pressure vessel with shield/X-Calibur installed (tungsten cap removed). 
The X-ray beam enters the polarimeter through the hole in the top plate. 
{\bf Right:} The \INFOCUS balloon gondola/truss with active pointing 
control and the X-ray mirror installed. The pressure vessel is installed 
in the back.}

\label{fig:ActiveShieldAndTelescope}

\end{figure*}

\subsection{The Shield}

During the balloon flight, the polarimeter will be hit by charged and 
neutral particle backgrounds with different spectral signatures and 
intensities. These backgrounds reduce the signal-to-noise ratio and, in 
the case of non-isotropic fluxes, can even lead to a fake polarization 
signature. In order to suppress these backgrounds, the polarimeter and 
the front-end readout electronics are operated inside an active CsI(Na) 
anti-coincidence shield. The $2.7 \, \rm{cm}$ thick CsI crystal of the 
shield covers the sides and the bottom of the polarimeter and produces 
scintillation light when particles interact. The top is protected by a 
passive tungsten plate/collimator 
(Fig.~\ref{fig:ActiveShieldAndTelescope}, left), blocking X-rays and 
particles that do not come from the X-ray mirror. The (active) CsI 
scintillator of the shield is read out by four Hamamtsu PMTs R\,6233 
which are biased at $V_{\rm{bi}} = +800 \, \rm{V}$. The analog signal of 
all four PMTs is merged and in turn digitized. A programmable, digital 
discriminator decides on whether a shield flag $f_{\rm{shld}}$ is set on 
the CZT readout board (kept up for $6 \, \mu\rm{s}$) and is merged into 
the data stream. The values of the discriminator and the width of the 
flag were optimized using a radio-active source to maximize the shield 
efficiency and minimize chance coincidences (see 
Sec.~\ref{subsec:BG_Data}).


In order to reduce the systematic uncertainties of the polarization 
measurements, the polarimeter and the active shield will be rotated 
around the optical axis with $\sim 2 \, \rm{rpm}$ using a ring bearing 
(see Fig.~\ref{fig:ActiveShieldAndTelescope}). The angle between the 
polarimeter/shield and the mounting fixture is read out by a code wheel 
with the accuracy of $1^{\circ}$. A counter-rotating mass can be used to 
cancel the net angular momentum of the rotating polarimeter assembly 
during the balloon flight. The computer reading the PMT and CZT events 
is part of the rotating assembly, and referred to as polarimeter CPU.

\subsection{The \INFOCUS X-ray Telescope}

The X-Calibur polarimeter will be flown in a pressurized vessel located 
in the focal plane of the \INFOCUS X-ray telescope 
\cite{InFocus_FirstFlight}. The telescope is shown in the right panel of 
Fig.~\ref{fig:ActiveShieldAndTelescope}. A Wolter grazing incidence 
mirror focuses the X-rays onto the polarimeter. The X-Calibur 
scintillator rod will be aligned with the optical axis of the \INFOCUS 
X-ray telescope (see Sec.~\ref{subsec:FtSumnerData}). The focal length 
of the mirror is $8 \, \rm{m}$ and the field of view is $\rm{FWHM} = 10 
\, \rm{arc min}$. The telescope truss of \INFOCUS is only coupled to the 
gondola by a ball joint in a support cup with floating oil, allowing for 
full inertial pointing of the telescope with an accuracy of $7 ''$ and 
$15 ''$ RMS in altitude and azimuth, respectively. To maintain the 
decoupling between truss and gondola, any communication between the two 
systems is done by wireless connections. In addition to the rotating 
polarimeter CPU (see above), a second CPU as part of the X-Calibur 
subsystem (motor CPU) is installed in the pressure vessel (non-rotating) 
and controls the motors, and a temperature system. Power and data 
communication between the polarimeter CPU and the motor CPU is achieved 
by a {\it Mercotac$^{\tiny \circledR}$} 830-SS rotating ring of mercury 
sliding contacts. Communication between the pressure vessel and the 
telescope gondola will be done via a wireless network. The data will be 
stored on solid state drives and will in parallel be down-linked to the 
ground.

\begin{figure}[th!]
\begin{center}
\includegraphics[width=0.44\textwidth]{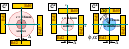} \\
\includegraphics[width=0.46\textwidth]{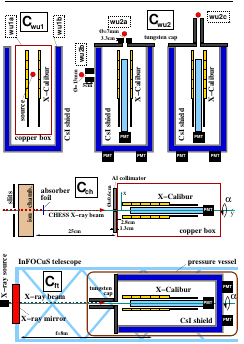}
\end{center}

\caption{Different X-Calibur configurations referred to via 
$C_{\rm{env}}^{\rm{stp}}$. {\bf Top row:} CZT detector geometries (see 
Fig.~\ref{fig:Design}, left) viewed along the optical axis looking into 
the X-Ray beam. Each detector is tangentially translated by $\Delta = 1 
\, \rm{mm}$. $C^{1}$: CZT detector cathodes being $c = 16 \, \rm{mm}$ 
away from the optical axis (regular-sized sealed radioactive source fits 
in between). $C^{2}$: Reduced distance of $c =11 \, \rm{mm}$ (only 
compact sources). $C^{3}$: same as $C^{2}$ but with the scintillator rod 
installed. The definitions of $\Phi$ and $\Delta \Phi$ as used in 
Sec.~\ref{sec:XCLB_Characterization} are illustrated. {\bf Bottom:} 
Schematic sketches (not to scale) of X-Calibur installed in the 
different environments. $C_{\rm{wu}}$: Washington University; detector 
calibration in copper box ($C_{\rm{wu1a}}$) or in passive CsI shield 
($C_{\rm{wu1b}}$); performance measurements in the active shield with 
different source positions ($C_{\rm{wu2a}}$, $C_{\rm{wu2b}}$, and 
$C_{\rm{wu2c}}$). $C_{\rm{ch}}$: X-Calibur/CHESS setup in hutch {\it 
C1}. The beam intensity is reduced by absorption foils; platinum 
($C_{\rm{ch1}}$), or platinum and lead ($C_{\rm{ch3}}$). The X-Calibur 
polarimeter can be rotated around its optical axis (angle $\alpha$) 
which is aligned with the CHESS beam. $C_{\rm{ft}}$: X-Calibur installed 
in the \INFOCUS X-ray telescope during a flight preparation campaign in 
Ft.~Sumner.}

\label{fig:X-Calibur_Configurations}

\end{figure}

\subsection{X-Calibur Configurations and Data Sets}

In order to characterize the different components and aspects of the 
X-Calibur polarimeter, different types of measurements were performed 
with different geometrical configurations of the instrument (e.g. with 
and without the shield, measurements without the scattering rod to 
calibrate the CZT detector response itself, illumination of the 
instrument with different X-ray sources from different angles, etc.). 
The measurements were performed at different facilities/locations (which 
we refer to as environments). The energy calibration and 
characterization of the CZT detectors was performed in the laboratory at 
Washington University (Sec.~\ref{sec:DetectorCharacterization}). 
Measurements of a polarized X-ray beam to study the performance of the 
polarimeter were conducted at the CHESS synchrotron facility at Cornell 
University (Sec.~\ref{subsec:CHESS}). Data were also taken with the 
fully integrated X-Calibur/\INFOCUS telescope in a field campaign in 
Ft.Sumner, NM (Sec.~\ref{subsec:FtSumnerData}). Measurements of the 
background (Sec.~\ref{subsec:BG_Data}) were performed at Washington 
University, CHESS, and in Ft.Sumner. Throughout the paper, each 
configuration and location is referred to as $C_{\rm{env}}^{\rm{stp}}$ 
(`env' referring to the environment/location and `stp' referring to the 
experimental setup). These different configurations are shown in 
Fig.~\ref{fig:X-Calibur_Configurations} and will be referred to in the 
sections to follow in which the corresponding results are presented.

It should be noted, that some of the data presented in this paper were 
taken without the CsI shield (e.g. the data taken at the CHESS 
facility), not allowing one to use the shield veto for background 
suppression. Since separate background runs were taken and subtracted 
from the data, and the majority of measurements is signal-dominated, 
this does not affect any result or conclusion presented in this paper.

\section{Reconstruction of Polarization Properties} \label{sec:Analysis}

The events recorded by X-Calibur consist of the digitized pulse height 
of one (or more) detector pixel(s), a time stamp, and flags describing 
whether the shield and/or the scintillator triggered. These raw events 
are first transformed into measured energies using the detector 
calibration. The reconstructed events are further processes in order to 
derive energy spectra and the polarization properties of the measured 
X-ray beam. This section outlines the corresponding procedures.

\subsection{Definitions}

The modulation in the measured azimuthal scattering distribution is the 
defining signature from which the polarization properties are derived. 
The modulation factor $\mu$ describes the polarimeter response to a 
polarized beam and is used to reconstruct the polarization properties. 
Assuming a $100 \%$ linearly polarized X-ray beam, the minimum 
($C_{\rm{min}}$) and maximum ($C_{\rm{max}}$) number of counts of the 
azimuthal scattering distribution define:

\begin{equation}
\label{eq:ModulationFactor}
\mu = \frac{C_{\rm{max}} - C_{\rm{min}}}{C_{\rm{max}} + C_{\rm{min}}}.
\end{equation}
It represents the modulation amplitude of a $100 \%$ polarized beam and 
depends on the polarimeter design and the physics of Compton-scattering. 

The performance of a polarimeter can be characterized by the minimum 
detectable polarization (MDP) as the minimum fraction of polarization 
that can be detected at the $99 \%$ confidence level for a given time of 
observation $T$. Assuming a polarimeter that detects all 
Compton-scattered photons with an ideal angular resolution~-- in this 
case $\mu$ becomes the modulation amplitude averaged over all solid 
angles and the Klein-Nishina cross section~-- one can estimate the MDP 
by integrating the scattering probability distribution 
\cite{Weisskopf2011, StokesAnalysis} ($R_{\rm{src}}$ and $R_{\rm{bg}}$ 
are the source and background count rates, respectively):

\begin{equation}
\label{eqn:MDP}
\rm{MDP} \simeq \frac{4.29}{\mu R_{\rm{src}}} \sqrt{\frac{R_{\rm{src}} +
R_{\rm{bg}}}{T}}.
\end{equation}

\subsection{X-Calibur Event Reconstruction and Selection}

Each recorded X-Calibur event contains an event number, a GPS time stamp 
$T_{\rm{gps}}$, the orientation angle $\theta_{\rm{w}}$ of the 
shield/polarimeter with respect to the mounting plate (read by a code 
wheel), and a list of CZT detector pixels that were hit (up to nine) 
including their digitized pulse heights. Furthermore, two flags are 
merged into the data stream: (i) a flag $f_{\rm{shld}}$ that indicates 
whether the PMTs reading the active CsI shield got a signal exceeding 
the defined discriminator threshold, and (ii) a flag $f_{\rm{sci}}$ that 
indicates if the PMT reading the central scintillator rod of the 
polarimeter (see Fig.~\ref{fig:Design}, left) exceeded its discriminator 
threshold. The average analog rise/fall times 
$\tau_{\rm{r}}$/$\tau_{\rm{f}}$ (time for a signal rise from $10-90\%$ 
of the amplitude) of the folded scintillator/PMT response were measured 
for the shield and for the scintillator. For the shield we find 
$\tau_{\rm{r}}^{\rm{shld}} \approx 70 \, \rm{ns}$ and 
$\tau_{\rm{f}}^{\rm{shld}} \approx 2.6 \, \mu\rm{s}$, respectively. The 
response of the scintillator rod of the polarimeter is much faster: 
$\tau_{\rm{r}}^{\rm{sci}} \approx 3.0 \pm 1.3 \, \rm{ns}$ and 
$\tau_{\rm{f}}^{\rm{sci}} \approx 13.7 \pm 5.1 \, \rm{ns}$ for cosmic 
rays, and $\tau_{\rm{r}}^{\rm{sci}} \approx 4.7 \pm 2.7 \, \rm{ns}$ and 
$\tau_{\rm{f}}^{\rm{sci}} \approx 13.5 \pm 5.0 \, \rm{ns}$ for a 
Cs$^{137}$ source placed at configuration $C_{\rm{wu2c}}^{3}$, 
respectively. Upon a CZT trigger, the event readout is delayed by 
$2.5-3.3 \, \mu\rm{s}$ (jitter) to allow the shield and scintillator PMT 
signals to built-up and being converted into the corresponding flags. 
The flags $f_{\rm{shld}}$ and $f_{\rm{sci}}$ are kept up for a duration 
of $6 \, \mu\rm{s}$.

\begin{figure}[t]
\begin{center}
\includegraphics[width=0.49\textwidth]{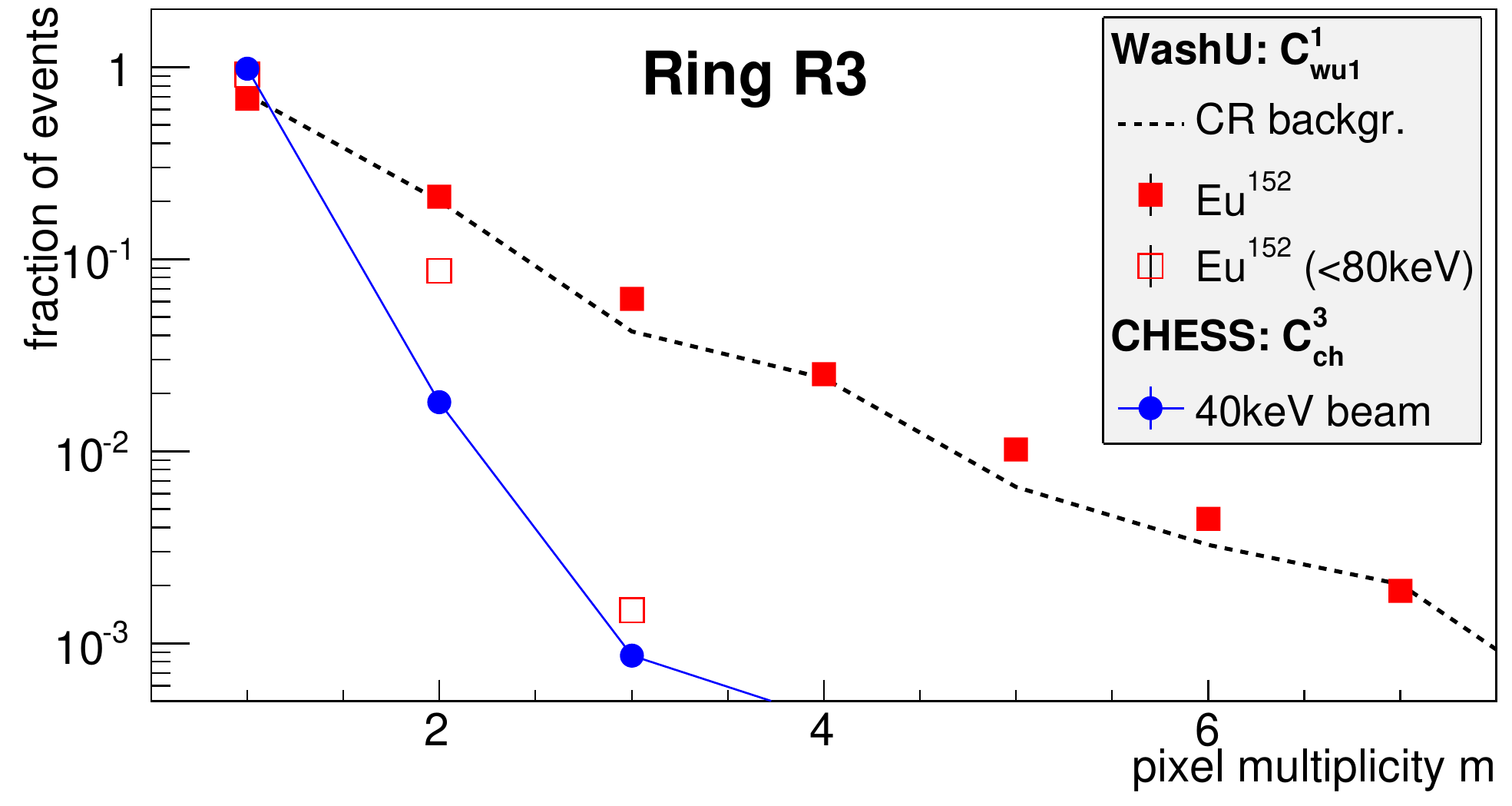}
\end{center}

\caption{Distribution of pixel multiplicities $m$ for events taken with 
different sources (recorded in detector ring {\it R3}): Cosmic ray (CR) 
background in the laboratory, energy calibration using Eu$^{152}$, 
Eu$^{152}$ calibration restricting event energies to $E < 80 \, 
\rm{keV}$, and data taken at the $40 \, \rm{keV}$ CHESS beam (see 
Sec.~\ref{subsec:CHESS}).}

\label{fig:MultiplicityDistribution}
\end{figure}

For each event, the pulse heights of all contributing channels/pixels 
$i=1...m$ are transformed into energies $E_{i}$ using the channel 
calibration, see Eq.~(\ref{eq:EnergyCalibration}) in 
Sec.~\ref{subsec:CZT_Calibration}. The total energy of the CZT event is 
$E = \sum_{i} E_{i}$. The number of pixels $m$ participating in the 
event is referred to as the pixel multiplicity. Selection cuts can be 
applied to the data based on the event properties mentioned above: 
$T_{\rm{gps}}$, $\theta_{\rm{w}}$, $E$, $m$, $f_{\rm{shld}}$, and 
$f_{\rm{sci}}$. Figure~\ref{fig:MultiplicityDistribution} shows the 
distribution of $m$ for different source types (measured in detector 
ring {\it R3}). It can be seen that sources with high energy 
contributions (such as the cosmic ray background or Eu$^{152}$ with its 
$E> 100 \, \rm{keV}$ energy lines) have $m=1$ pixel event contributions 
of $\simeq 70 \%$. Sources with energies concentrated in the $20-80 \, 
\rm{keV}$ interval, relevant for X-Calibur, have $m=1$ contributions of 
$> 90 \%$ (e.g. the $40 \, \rm{keV}$ CHESS beam, see 
Sec.~\ref{subsec:CHESS}). This is explained by the fact that low energy 
X-rays deposit smaller and more concentrated charge clouds in the CZT 
with a reduced chance of charge sharing between pixels (that would cause 
$m \geq 2$ events). Therefore, an event selection cut on $m=1$ is a 
reasonable way for background subtraction without loosing signal events 
at low energies. Cuts on the energy $E$ and the shield flag 
$f_{\rm{shld}}$ will further reduce the background (see 
Sec.~\ref{subsec:BG_Data}). A selection cut on the scintillator flag 
$f_{\rm{sci}}$ selects a very clean sample of events that 
Compton-scattered in the scintillator. This has the potential to further 
reduce the background~-- however, with a loss in efficiency at low 
energies (see Sec.~\ref{subsec:ScintillatorEfficiency}). However, a cut 
on $f_{\rm{sci}}$ is optional and not a requirement for sensitive 
polarization measurements with X-Calibur.

\subsection{Energy Spectra}

Energy spectra are used to study the scattering properties of the 
polarimeter and mark an intermediate step to derive the energy-dependent 
polarization properties. Energy spectra can be derived for individual 
pixels or for groups of pixels (e.g. a CZT detector or a detector ring 
{\it R}). The energy spectra shown in this paper are normalized to the 
acquisition time (dead time corrected), the anode detector surface 
covered by the corresponding pixel group and the width of the energy 
bins. Note, the limited dynamical range of the charge digitization of 
individual pixel channels will lead to discrete energies. Given the 
different energy calibrations of the channels, these will differ from 
pixel to pixel. This difference can lead to binning artifacts in energy 
spectra that are obtained from a small group of pixels.

\subsection{Polarization Properties} \label{subsec:AnalysisPolarization}

The signature of a polarized beam will be imprinted in the azimuthal 
scattering distribution of recorded events (see for example 
Figs.~\ref{fig:AzimuthDistribution} or \ref{fig:CHESS_2DAzimuthData} in 
Sec.~\ref{subsec:CHESS}). Either the scattering distribution itself, or 
a method involving the Stokes parameters, can be used to extract the 
polarization fraction $r$ and polarization direction $\Omega$. First, 
the details of the detector geometry have to be carefully considered in 
the analysis, as outlined below.

\begin{figure*}[t!]
\begin{center}
\includegraphics[width=0.5\textwidth]{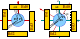}
\hfill
\includegraphics[width=0.48\textwidth]{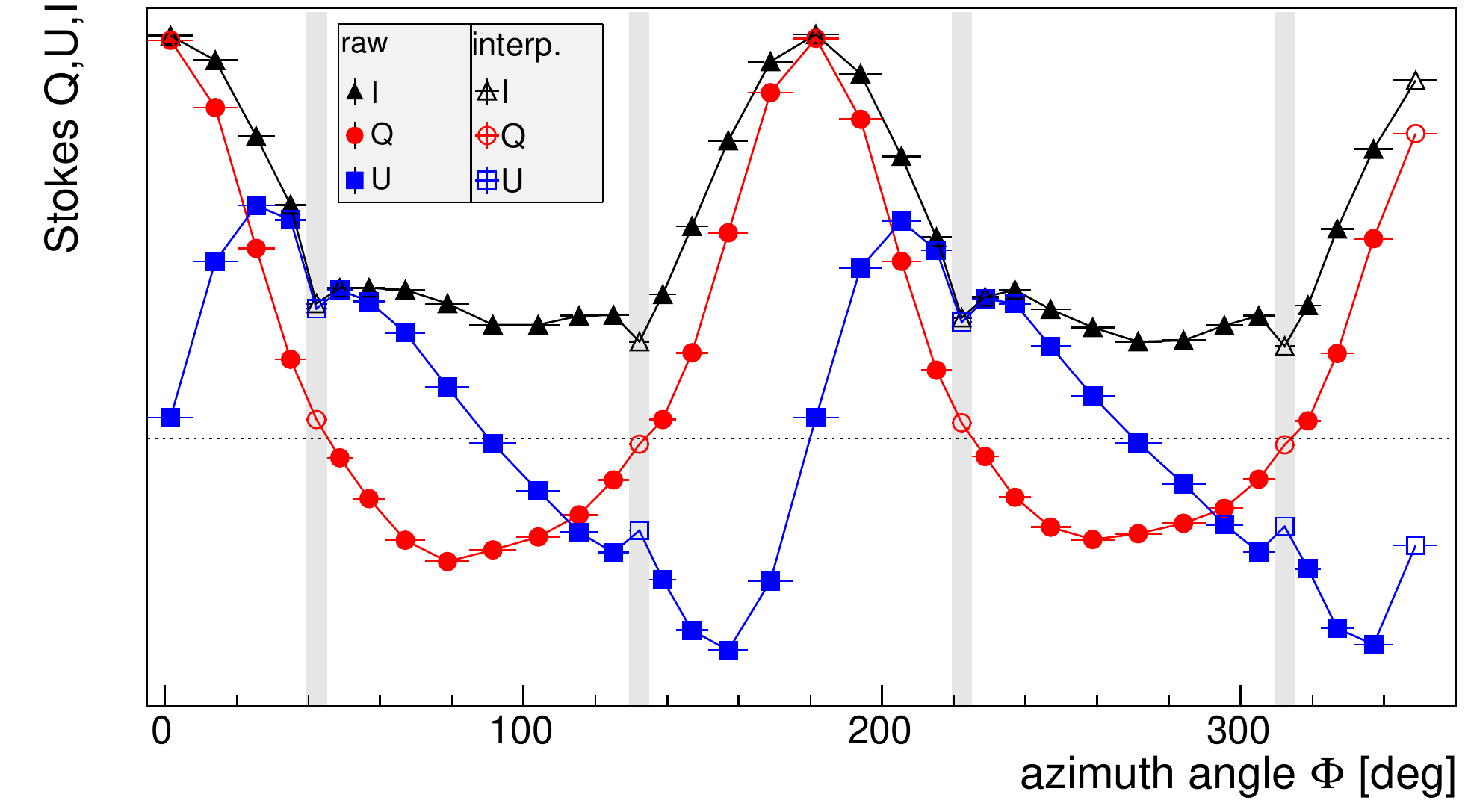}
\end{center}

\caption{{\bf Left:} Angular coverage of two selected CZT pixels within 
a \SPRing for different beam positions $P$. For a non-rotating detector 
(left) the angular coverage for each pixel only needs to be calculated 
once. In the rotating system (right) the beam offset in the horizon 
system (solid arrow) leads to varying angular coverage per pixel 
depending on the current rotation $\alpha$ that have to be updated on an 
event-by-event basis. Note, the pixel contacts are physically deposited 
on the backside of the detectors but are shown on the front side (facing 
the scintillator) for a better illustration of the angular measurement. 
The cyan dashed arrows indicate the $x/z$ detector coordinate system. 
{\bf Right:} Stokes parameters $Q$, $U$, and $I$ as measured within a 
\SPRing for a polarized beam with $\Omega = 90^{\circ}$. The pixels $j$ 
have varying angular coverage $\Delta \Phi_{j}$ (indicated by the 
horizontal error bars). The gray bands indicate the gaps between 
detector boards (see left panel). The filled data points show the 
measured values. The open data points show the interpolated Stokes 
parameters to recover the contributions missed by the gaps. Note, the 
entries in the figure are normalized to the solid angle $\Delta \Phi$ 
covered by each pixel. Therefore, the normalization leads to apparent 
jumps from the pixels to the values shown for the gaps. The open symbol 
at $\Phi \simeq 360^{\circ}$ corresponds to a dead pixel.}

\label{fig:SchematicBeamOffset_AndStokes}
\end{figure*}

{\it Azimuthal pixel coverage.} The $360^{\circ}$ range in azimuth is 
covered by the $4 \times 8 = 32$ pixels per \SPRing (see 
Fig.~\ref{fig:SchematicBeamOffset_AndStokes}, left). Different pixels 
cover different azimuthal ranges $\Delta \Phi$ with respect to the 
center of the scattering rod which defines the optical axis. Therefore, 
the four sides of detector boards lead to a 4-fold symmetry in the 
scattering distribution (Fig.~\ref{fig:AzimuthDistribution}). This 
purely geometrical effect can be corrected for by dividing the counts 
$C_{j}$ in each pixel $j$ by its azimuthal coverage $\Delta \Phi_{j}$.

{\it Azimuthal pixel coverage for beam offsets.} A special situation 
arises if the optical axis of the X-ray beam is not aligned with the 
geometrical axis of the polarimeter (see $P_{1}$ in 
Fig.~\ref{fig:SchematicBeamOffset_AndStokes}, left). Such an offset, if 
not corrected, will introduce asymmetries in the azimuthal scattering 
distributions and can mimic a wrong polarization signature (see 
Sec.~\ref{subsec:Systematics} for a detailed study). Assuming a known 
offset vector $\vec{P}1$, the azimuthal coverage $\Delta \Phi_{p1, j}$ 
can be recalculated for each pixel $j$. This correction re-aligns the 
origin of the detector system with the beam axis and strongly reduces 
the systematic effect introduced by the offset. In the case of 
sufficient event statistics, first moments can be used to estimate the 
beam offset from the data itself (see Sec.~\ref{subsec:Systematics}). If 
the polarimeter is rotating, the beam offset in the horizon system will 
rotate in the detector system (see 
Fig.~\ref{fig:SchematicBeamOffset_AndStokes}, left). In this case 
$\Delta \Phi_{p1,j}$ has to be updated on an event-by-event basis, 
taking into account the current polarimeter orientation $\alpha$ 
measured by the code wheel.

{\it Pixel acceptance and flat fielding.} Individual pixels have 
different trigger efficiencies, energy thresholds and energy resolutions 
affecting the number of counts derived from a given energy interval. To 
correct for these differences, the pixels are flat fielded using 
Compton-scattered events recorded from a non-polarized X-ray beam that 
results in a flat azimuthal scattering distribution. For a \SPRing the 
event counts per azimuthal coverage $C_{j} / \Delta \Phi_{j}$ are 
averaged for all $j=1...32$ pixels and are used to determine a relative 
azimuthal acceptance $a_{j}$:

\begin{equation}
w_{j} = \frac{1}{a_{j}} = \frac{\Delta \Phi_{j}}{C_{j}} \frac{1}{N}\sum_{i=1}^{N}\frac{C_{i}}{\Delta \Phi_{i}}.
\label{eqn:Acceptance}
\end{equation}

The pixel acceptance $a_{j}$ can in turn be used to weight individual 
events with $w_{j}=1/a_{j}$. Implicitly, $a_{j}$ depends on the event 
selection cuts~-- so that it has to be computed for the particular set 
of cuts applied to the data. Dead pixels cannot be recovered by a 
corresponding weight since they contribute zero events. However, once 
the polarimeter/shield assembly is rotating with respect to the 
polarization plane, the pixel acceptances can be ignored since they 
average out throughout the measurement~-- this includes the treatment of 
dead pixels.

{\it Polarization properties derived from the azimuthal scattering 
distribution.} Integrating $C_{j} / \Delta \Phi_{j}$ for each pixel $j$ 
in a certain energy range and weighting the individual counts with 
$w_{j}$ will result in the azimuthal scattering distribution. Here, the 
angular coverage $\Delta \Phi_{j}$ determines the horizontal error bar 
of the data point. The scattering distribution can be derived for 
individual detector rings. Fitting a sinusoidal function to the 
distribution (e.g. left panel in Fig.~\ref{fig:AzimuthDistribution}, 
Sec.~\ref{subsec:CHESS}) allows one to reconstruct (i) the orientation 
of the polarization plane (minimum), as well as (ii) the modulation of 
the data $\mu_{\rm{data}}$ following Eq.~(\ref{eq:ModulationFactor}). 
The corresponding modulation factor $\mu_{\rm{sim}}$ is derived from 
simulations of a $100 \%$ polarized beam (see 
Sec.~\ref{sec:Simulations}) that are analyzed using the same set of 
event selection cuts. The polarization fraction $r$ of the measured beam 
is in turn calculated to be:

\begin{equation}
r = \frac{\mu_{\rm{data}}}{\mu_{\rm{sim}}}.
\label{eqn:PolFracPhiDistri}
\end{equation}

The effect of dead detector pixels and gaps between the detector boards 
(see Fig.~\ref{fig:SchematicBeamOffset_AndStokes}, left) is accounted 
for automatically, since the corresponding points do not show up in the 
distribution and will not affect the fit.

{\it Polarization properties derived from Stokes parameters.} An 
alternative approach to reconstruct the polarization properties is based 
on the Stokes parameters \cite{Chandrasekhar1960, StokesAnalysis} that 
are calculated for each event $k$:

\begin{equation}
q_{k} = \cos(2 \Phi_{k}), \quad u_{k} = \sin(2 \Phi_{k}), \quad i_{k} = 1.
\label{eqn:StokesParams}
\end{equation}

The Stokes parameters can be summed for a subset of the data consisting 
of $N$ events (e.g. over a specific energy interval and detector ring):

\begin{equation}
Q = \sum_{k=1}^{N} w_{j(k)} q_{k}, \quad U = \sum_{k=1}^{N} w_{j(k)} u_{k}, \quad I = \sum_{k=1}^{N} w_{j(k)}.
\label{eqn:StokesSum}
\end{equation}
Here, each event $k$ (originating from pixel $j$) is weighted with 
$w_{j(k)} = 1/a_{j(k)}$. Since each pixel $j$ covers a range in azimuth 
$\Delta \Phi_{j} = \Phi_{j,\rm{max}} - \Phi_{j,\rm{min}}$, the values 
$q_{k}$ and $u_{k}$ derived from the mean angle $\Phi_{j(k)}$ in a 
non-rotating system may lead to inaccurate and/or biased results. 
Therefore, the mean Stokes parameters $\left< q \right>_{j(k)}$ and 
$\left< u \right>_{j(k)}$ are calculated based on the covered range in 
azimuth of the corresponding pixel $j$:

\begin{eqnarray}
\begin{split}
\left< q \right>_{j(k)} & = \frac{1}{\Delta \Phi_{j(k)}} \int_{\Phi_{j(k),\rm{min}}}^{\Phi_{j(k),\rm{max}}} \cos(2 \Phi) \, \rm{d}\Phi, \\
\left< u \right>_{j(k)} & = \frac{1}{\Delta \Phi_{j(k)}} \int_{\Phi_{j(k),\rm{min}}}^{\Phi_{j(k),\rm{max}}} \sin(2 \Phi) \, \rm{d}\Phi.
\end{split}
\label{eqn:MeanStokes}
\end{eqnarray}

The values of $\left< q \right>_{j(k)}$ and $\left< u \right>_{j(k)}$ 
are in turn used in Eq.~(\ref{eqn:StokesSum}). The proper treatment of 
dead pixels and gaps between the detector boards {\it Bd0}-{\it Bd3} 
(see Fig.~\ref{fig:SchematicBeamOffset_AndStokes}, left) is crucial when 
working with the Stokes parameters in a non-rotating coordinate system, 
since a lack of events from a particular azimuthal direction 
$\tilde{\Phi}$ will lead to an increase/decrease in $\sum_{k} \left< q 
\right>_{j(k)}$ and/or $\sum_{k} \left< u \right>_{j(k)}$, 
systematically affecting the reconstructed polarization properties (see 
Fig.~\ref{fig:SchematicBeamOffset_AndStokes}, right). The amplitude and 
sign of the resulting effect depends on the orientation between the 
polarization plane and the detector plane (assuming the gaps are located 
at angles of $45^{\circ}$, $135^{\circ}$, $225^{\circ}$, and 
$315^{\circ}$, see Fig.~\ref{fig:SchematicBeamOffset_AndStokes}, left). 
If the planes are exactly parallel or exactly perpendicular, the effect 
leads to a maximal underestimation of $r$. An angle of $45^{\circ}$ 
(polarization plane being aligned with two of the four gaps) leads to 
the maximal overestimation of $r$. At angles of $45/2 = 27.5^{\circ}$ 
(and multiples thereof) the effect of detector gaps cancels. A 
simulation for the X-Calibur detector layout was performed and resulted 
in a systematic effect of $\pm6 \%$ for a $100\%$ polarized beam. 
Therefore, a correction has to be applied: (i) the $m=1,2,...$ azimuthal 
ranges $\delta \tilde{\Phi}_{m}$ covered by the gaps have to be 
identified in which the detector is not sensitive (ii) The distributions 
of $\delta Q / \delta \Phi$, $\delta U / \delta \Phi$, and $\delta I / 
\delta \Phi$ have to be collected as a function of $\Phi$ 
(Fig.~\ref{fig:SchematicBeamOffset_AndStokes}, right). (iii) These 
distributions are in turn used to interpolate the data gaps $\delta 
\tilde{\Phi}_{m}$ from neighboring pixel to recover the `missing' 
contributions in Eq.~(\ref{eqn:StokesSum}):

\begin{eqnarray}
\begin{split}
Q & \rightarrow Q + \sum_{m} \Delta \tilde{Q}_{m} \delta 
\tilde{\Phi}_{m}, \\ U & \rightarrow U + \sum_{m} \Delta \tilde{U}_{m} 
\delta \tilde{\Phi}_{m}, \\ I & \rightarrow I + \sum_{m} \Delta 
\tilde{I}_{m} \delta \tilde{\Phi}_{m}.
\end{split} 
\label{eqn:StokesGapCorrection} 
\end{eqnarray}
The errors on the added sums are calculated using error propagation 
during the interpolation. Given the scalar nature of the Stokes 
analysis, there is no simple way of representing the azimuthal 
modulation of the scattering distribution. Therefore, it is useful to 
compare the Stokes results with the results obtained from the azimuthal 
scattering distribution (previous paragraph) in order to identify 
possible systematic effects. Note again, that a rotating polarimeter 
will not require a correction for dead pixels or detector gaps.

The Stokes sums in (\ref{eqn:StokesSum}) or 
(\ref{eqn:StokesGapCorrection}) can be used to reconstruct the 
polarization fraction $r$ and polarization angle~$\Omega$:

\begin{eqnarray}
\begin{split}
r & = \frac{2}{\mu_{\rm{sim}}} \frac{\sqrt{Q^{2} + U^{2}}}{I}, \\
\Omega & = \frac{1}{2} \rm{atan} (U / Q).
\end{split}
\label{eqn:PolarizationFromStokes}
\end{eqnarray}

Results from $l$ independent measurements (e.g. from different detector 
rings) can be combined in a weighted average. Since the polarization 
fraction is always positive it can lead to an overestimation if the true 
polarization fraction is in the MDP regime of the data set, see 
Eq.~(\ref{eqn:MDP}), where the error bars are highly asymmetric. This 
systematic effect can be avoided by averaging a set of modified Stokes 
parameters:

\begin{eqnarray}
\begin{split}
Q' & = \sum_{l} \frac{2 w_{l}}{\mu_{\rm{sim,l}}} \frac{Q_{l}}{I_{l}}, \quad U' = \sum_{l} \frac{2 w_{l}}{\mu_{\rm{sim,l}}} \frac{U_{l}}{I_{l}}, \\
r' & = \frac{1}{L'} \sqrt{Q'^{2} + U'^{2}} \quad {\rm with} \quad L' = \sum_{l} w_{l}.
\end{split}
\label{eqn:Stokes_Q_U_Average}
\end{eqnarray}
The weights $w_{l} = 1/\sigma_{l}^{2}$ account for the statistical 
uncertainties of the individual measurements $l$.

{\it Unfolding analysis.} An unfolding analysis that takes into account 
the energy-dependent detector response and photon detection efficiencies 
and that can be used to reconstruct polarization fraction and angle as a 
function of true photon energy will be described in a separate paper 
\cite{XCLB_LogLikelihoodAnalysis}.

{\it Forward folding.} Recording the azimuthal scattering distributions 
with planar detectors will lead to projection effects that depend on the 
polar angle of the scattering in the scintillator. For most parts of the 
polarimeter these effects are negligible, since (i) a particular 
detector ring sees the superposition of different polar scattering 
angles canceling the effect, and (ii) the same effect is present in the 
simulations that are used to determine the polarization fraction 
following Eq.~(\ref{eqn:PolFracPhiDistri}). Only for polarization 
fractions $r \ll 1$ measured in detector ring {\it R1}, which sees only 
back-scatter events, a second order correction may be 
needed\footnote{Here, the $r \propto \mu_{\rm{data}}$ relation in 
Eq.~(\ref{eqn:PolFracPhiDistri}) will no longer be exactly linear.}. To 
fully take into account these effects, the data can be analyzed with a 
forward folding method (which is beyond the scope of this paper). The 
modeling of the pixel acceptance in {\it R1}, only affecting 
measurements with the non-rotating polarimeter, will also be slightly 
affected by the effect.

\section{Simulations} \label{sec:Simulations}


Simulations of the energy-dependent response of the X-Calibur 
polarimeter are needed in order to reconstruct the polarization 
properties from measured data (see 
Sec.~\ref{subsec:AnalysisPolarization}). The detector geometry is 
modeled and the X-ray flux/spectrum was simulated for the different 
experimental setups in which the data presented in this paper were 
taken. The simulations were performed in the following steps.

\begin{enumerate}

\item Physics interactions, scatterings, and energy depositions in the 
scintillator and the CZT detectors were simulated using {\it 
GEANT4}\footnote{{\tt http://geant4.cern.ch/}} with the Livermore 
low-energy electromagnetic model list.

\item The charge collection efficiency as a function of 
depth-of-interaction in the CZT detectors was determined using an 
in-house developed software to (i) calculate the 2D electric potential 
inside the detector crystals followed by (ii) the integration of the 
weighting potential \cite{Jung2007} along the charge transport tracks, 
resulting in the collected charge for each energy deposition (using a 
dielectric constant for CZT of $10$). The simulations are valid for 
detectors with strip anode contacts but will roughly resemble the 
response for pixelated detectors, as well. A mobility of electrons/holes 
of $\mu_{\rm{e}} = 1000 \, \rm{cm}^{2} / \rm{V} / \rm{s}$, and 
$\mu_{\rm{h}} = -120 \, \rm{cm}^{2} / \rm{V} / \rm{s}$ is assumed, as 
well as life times of $\tau_{\rm{e}} = 10^{-6} \, \rm{s}$ and 
$\tau_{\rm{h}} = 4 \cdot 10^{-6} \, \rm{s}$, respectively. This 
corresponds to a mobility-lifetime product of $\mu_{\rm{e}}\tau_{\rm{e}} 
= 10^{-3}$ which is in reasonable agreement with the values measured for 
a selection of the detectors used in X-Calibur (see 
Fig.~\ref{fig:TempStudies}, bottom).

\item The energy resolution (asymmetric Gaussian function, implicitly 
including the electronic readout noise) and energy threshold were 
measured from real data (Sec.~\ref{subsec:ThreshAndResolution}) and were 
folded into the simulations on a pixel-by-pixel basis. However, 
differences in channel trigger efficiencies were not simulated. Using 
the reversed energy calibration in Eq.~(\ref{eq:EnergyCalibration}), the 
simulated events were in turn converted into the X-Calibur data format 
(ASIC/channel ID and digitized raw pulse height) and can be analyzed in 
the same way as the measured data.

\item Since we find that the simulations do not properly account for low 
energy tails in the detector response (see next paragraph), an empirical 
model was used to scatter an exponential tail $T(E) = A \cdot 
\exp(E/E_{\rm{t}})$ into the simulations with a relative fraction of 
$f=0.45$ per energy deposit. The parameter $E_{\rm{t}} = 15 \, \rm{keV}$ 
(at $E=40 \, \rm{keV}$) and $E_{\rm{t}} = 30 \, \rm{keV}$ (at $E = 120 
\, \rm{keV}$) was interpolated in $\log E$ for the energy range covered.

\item The scintillator trigger flag $f_{\rm{sci}}$ was generated for 
each event based on the scintillator trigger efficiency determined from 
real data (see Fig.~\ref{fig:Scintillator_TriggerEfficiency} in 
Sec.~\ref{subsec:ScintillatorEfficiency}).

\end{enumerate}

Different scenarios/setups were simulated, reflecting the measurements 
and studies presented in this paper.

{\it CZT detector response.} To test the validity of the simulation 
chain, the direct illumination of a single CZT detector with a 
Eu$^{152}$ point source was simulated\footnote{All lines with 
intensities above $2 \%$ were generated according to their relative 
emission intensities.}~-- corresponding to the experimental setup used 
for the detector calibration measurements presented in 
Sec.~\ref{subsec:CZT_Calibration}. The comparison between the 
simulations and data (Fig.~\ref{fig:EnergyCalibration}, right) shows 
reasonable agreement in terms of line positions, widths, and threshold 
effects. However, if ignoring step (4) of the simulation chain (`Sim' in 
the legend of Fig.~\ref{fig:EnergyCalibration}), a lack in continuum 
emission can be seen in the simulations. This motivated the introduction 
of step (4) which leads to a reasonable agreement between data and 
simulations over the whole energy band relevant for X-Calibur (`Sim+' in 
the figure legend). Possible reasons for the continuum in the data may 
be related to details in the detector response or back-reflection of 
emitted X-rays from the source off the surrounding fixture that is not 
simulated\footnote{For similar CZT detectors we find a photo-peak 
detection efficiency of order unity \cite{VarPitch}.}. It should be 
noted that the goal of the CZT simulations is not to find an accurate 
model for the detector response~-- but rather a suited parameterization 
that reproduces the integral spectral response for the different energy 
bins.

{\it CHESS beam.} X-Calibur performance measurements were performed at 
the highly polarized synchrotron X-ray beam at the CHESS facility~-- 
providing a strong, mono-energetic X-ray beam (Sec.~\ref{subsec:CHESS}). 
A corresponding set of simulations was performed using steps (1)-(5), 
resembling the CHESS setup of pencil-beam X-rays (polarized and 
non-polarized) at $40 \, \rm{keV}$, $80 \, \rm{keV}$, and $120 \, 
\rm{keV}$ entering the polarimeter along the optical axis of the 
scintillator. Detector pixels that were excluded during the CHESS data 
runs were also excluded in the simulations to resemble a configuration 
close to the one used for the measurements. The trigger efficiency of 
the scintillator as a function of energy deposition was derived from the 
CHESS data (see Fig.~\ref{fig:Scintillator_TriggerEfficiency} in 
Sec.~\ref{subsec:ScintillatorEfficiency}) and was fed into the 
simulations in step (5) to generate the trigger flag $f_{\rm{sci}}$ on 
an event-by-event basis. No backgrounds were simulated in the case of 
the CHESS measurements since the measurements were completely 
signal-dominated.

\begin{figure}[t]
\begin{center}
\includegraphics[width=0.49\textwidth]{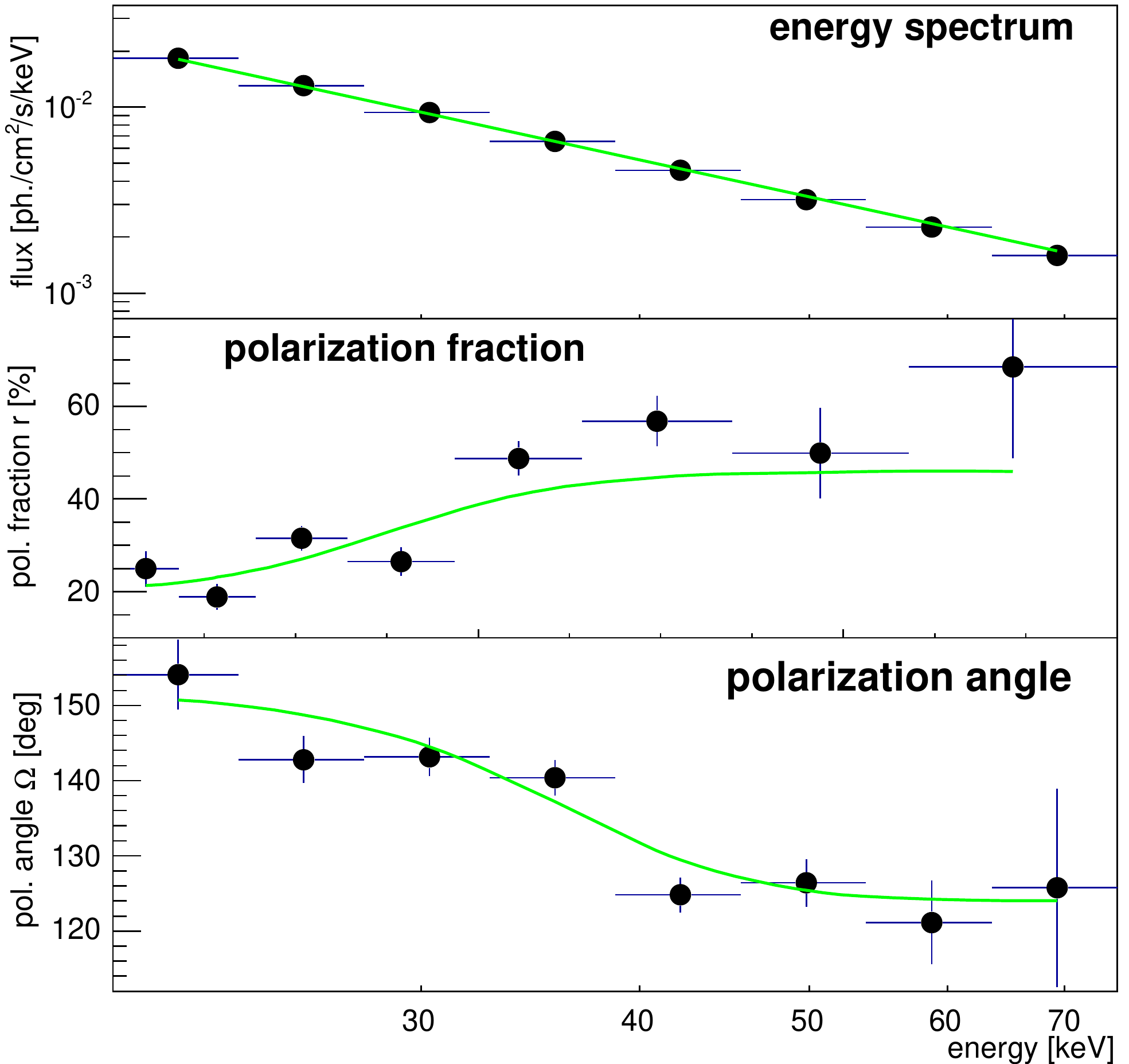}
\end{center}

\caption{Simulated X-Calibur observations of the Crab nebula ($5.6 \, 
\rm{h}$), taken from \citet{Guo2010}. The data points show the 
reconstructed flux (top), the fraction of polarization (middle), and 
polarization angle (bottom). The lines show the assumed flux, 
polarization fraction and polarization direction.}

\label{fig:XCLB_Sims}
\end{figure}

{\it Balloon flight.} A balloon flight in the focal plane of the 
\INFOCUS mirror assembly was assumed in an earlier simulation 
\cite{Guo2010} that only involved step (1) in the above chain. The 
effective detection areas of the X-ray mirror are $95/60/40 \, 
\rm{cm}^{2}$ at $20/30/40 \, \rm{keV}$, respectively. We accounted for 
atmospheric absorption at a floating altitude of $130,000$ feet using 
the NIST XCOM attenuation coefficients\footnote{{\tt 
http://www.nist.gov/pml/data/xcom/index.cfm}} and an atmospheric depth 
of $2.9 \, \rm{g/cm}^{2}$ (observations performed at zenith); the 
atmospheric transmissivity rapidly increases from $0$ to $0.6$ in the 
$20 - 80 \, \rm{keV}$ range. The trigger efficiency of the scintillator 
scatterer was assumed to be $f_{\rm{sci}} = 1$ above an energy 
deposition of $2 \, \rm{keV}$ and $f_{\rm{sci}} = 0$ below.

We simulated the most important backgrounds such as the cosmic X-ray 
background \cite{Ajello2008}, albedo photons and cosmic ray protons and 
electrons \cite{Mizuno2004}. The neutron background was not modeled 
since a detailed study of \citet{Parsons2004} showed that the 
contribution in CZT can be neglected. Different shield configurations 
and shield thicknesses were simulated. The configuration shown in 
Fig.~\ref{fig:ActiveShieldAndTelescope} (left) represents an optimized 
compromise balancing the background rejection power and the 
mass/complexity of the shield. A Crab-like source was simulated for a 
$5.6 \, \rm{hr}$ balloon flight. We assumed a power law energy spectrum, 
and a continuous change of the polarization fraction and angle between 
the values measured at $5.2 \, \rm{keV}$ with OSO-8 \cite{Weisskopf1978} 
and at $E>100 \, \rm{keV}$ with \Integral \cite{Dean2008} by modeling a 
transition following a Fermi distribution.

For a Crab-like source the simulations predict an event rate of $1.1 
(3.2) \, \rm{Hz}$ with (without) requiring a triggered scintillator 
coincidence ($f_{\rm{sci}} = 1$). Figure~\ref{fig:XCLB_Sims} compares 
the simulation results with the assumed model curves; the errors were 
computed in a similar way as described by \citet{Weisskopf2010}. 
Simulations performed at different zenith angles $\theta$ show that the 
source rate scales with $(\cos \theta)^{1.3}$ which is taken into 
account for simulating astrophysical observations. More details about 
these simulations are discussed in \citet{Guo2010} and \citet{Guo2013}.

\section{CZT Detector Characterization} \label{sec:DetectorCharacterization}

The performance of the individual CZT detectors used in X-Calibur is 
coupled to the performance of the polarimeter as a whole~-- including 
its energy threshold and its energy resolution. This section describes 
the energy calibration of the individual CZT detectors 
(Sec.~\ref{subsec:CZT_Calibration}), as well as measurements of the 
energy threshold and energy resolution 
(Sec.~\ref{subsec:ThreshAndResolution}). The CZT performance serves as 
important input for the simulations described in 
Sec.~\ref{sec:Simulations}. In contrast to the laboratory, the 
polarimeter will be operated in an environment of varying temperature 
conditions during the balloon flight which motivates the study of the 
temperature dependence of the CZT detector performance which will be 
described in Sec.~\ref{subsec:TempStudies}.

In order to quantify the characteristics of individual pixels, the 
emission lines in the calibrated energy spectra are fitted with a 
Gaussian function. The peak position is described by the mean 
$E_{\rm{p}}$. To account for the asymmetric shape of the peaks (see for 
example Fig.~\ref{fig:EnergyCalibration}), the fitted function allows 
for asymmetric spectral continua levels ($c_{1}$, $c_{2}$) and 
asymmetric peak widths ($\sigma_{1}$, $\sigma_{2}$) for the $E < 
E_{\rm{p}}$ (1) and $E > E_{\rm{p}}$ (2) regimes, respectively. The fit 
parameters are used to characterize the measured peak. The energy 
resolution is calculated as the full width half maximum, $\Delta E = 
\rm{FWHM} = 2 \sqrt{2 \ln 2} \cdot \frac{1}{2} (\sigma_{1} + 
\sigma_{2})$. The peak rate is determined by counting the events in the 
interval $\pm 2 \, \rm{FWHM}$ centered around $E_{\rm{p}}$, normalized 
by the observation time.

A Eu$^{152}$~source is used as calibration/test source in a variety of 
studies presented in this paper. Eu$^{152}$~emits X-ray lines at 
$39.5\,\rm{keV}$ ($\rm{K}_{\alpha2}$), $40.12\,\rm{keV}$ 
($\rm{K}_{\alpha1}$), $45.7 \, \rm{keV}$, $122.78 \, \rm{keV}$, $244.7 
\, \rm{keV}$, $344.28 \, \rm{keV}$, and at higher energies. The line at 
$40.12\,\rm{keV}$ is the strongest in the low energy triplet; relative 
to the $40.12\,\rm{keV}$ line, the $39.5 \, \rm{keV}$ line is emitted at 
$55 \%$ intensity and the $45.5 \, \rm{keV}$ line at $31 \%$ intensity. 
The $40.12\,\rm{keV}$, used as low-energy performance marker, is 
therefore fitted jointly with the two close-by lines that are set at 
fixed distance $\Delta E$ and fixed intensity relative to the 
$40.12\,\rm{keV}$ peak (with the free fit parameter $\sigma$ being the 
same for all three lines since the energy resolution of a detector pixel 
is not expected to change within a few keV). Including the two 
neighboring lines avoids systematic shifts and artificial broadening of 
the fitted peak (see Fig.~\ref{fig:EnergyCalibration}, left).

\subsection{Energy Calibration} \label{subsec:CZT_Calibration}

\begin{figure*}[t]
\begin{center}
\includegraphics[width=0.48\textwidth]{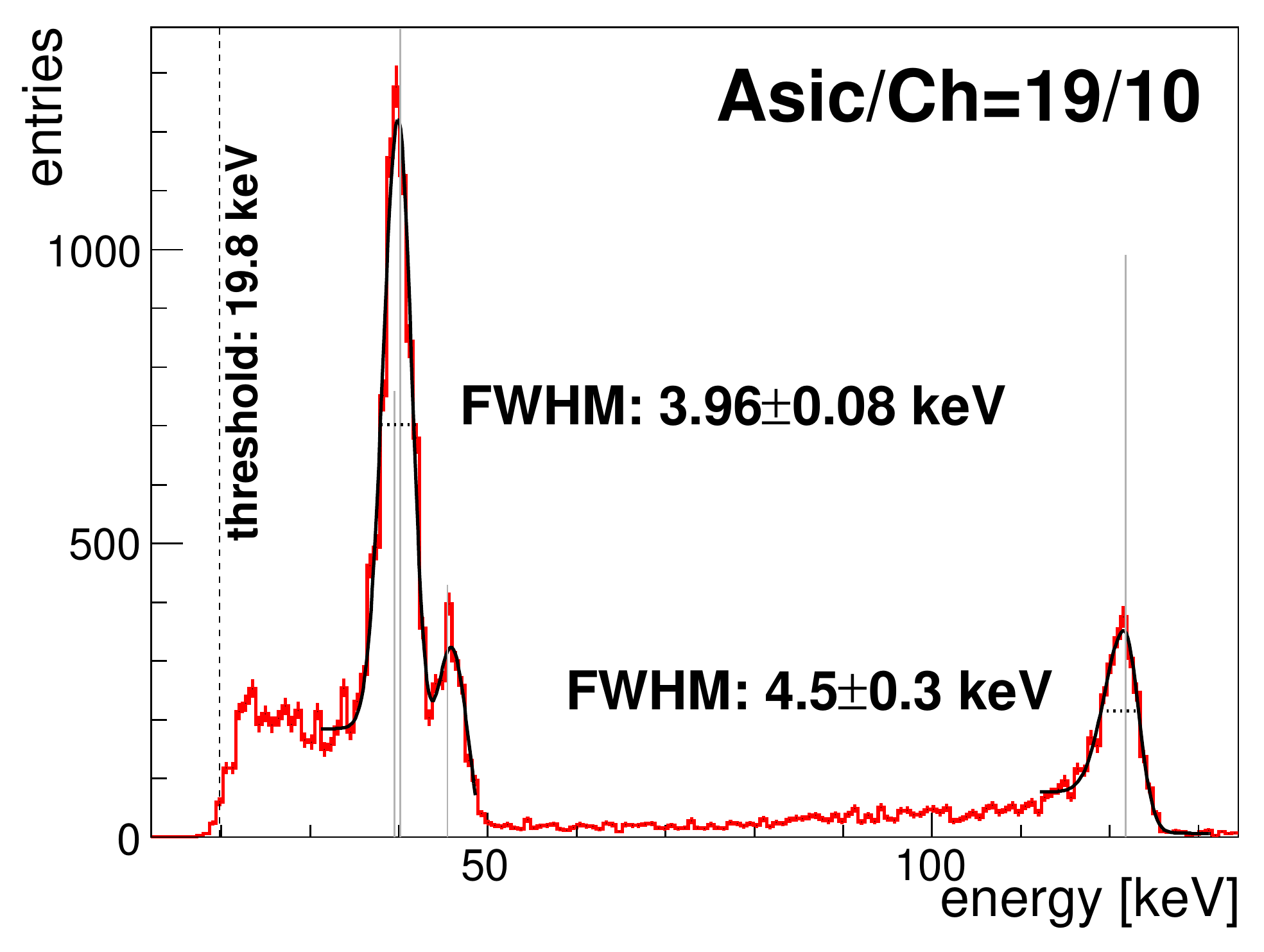}
\hfill
\includegraphics[width=0.50\textwidth]{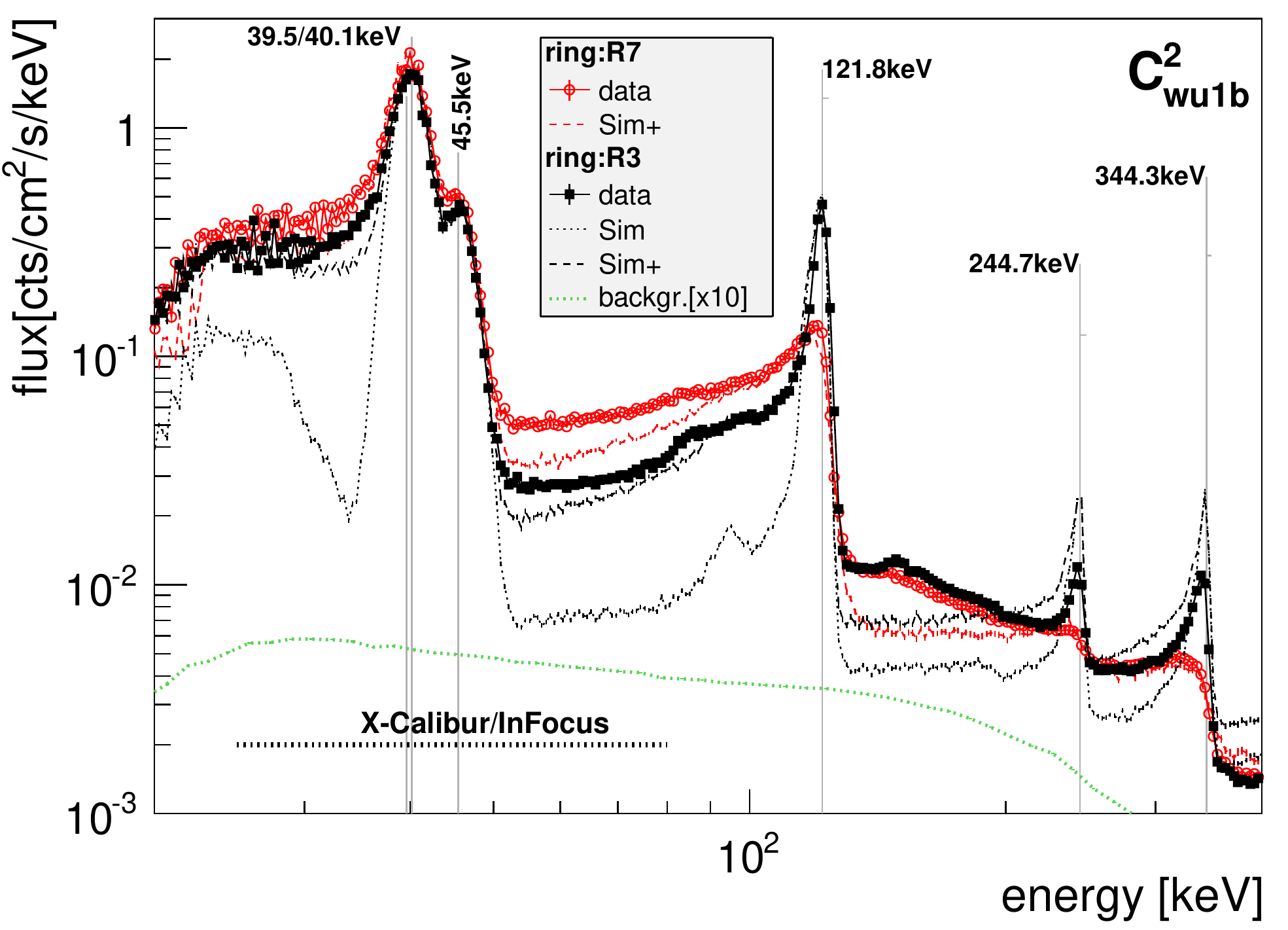}
\end{center}

\caption{Eu$^{152}$ energy spectra (direct CZT illumination). The 
vertical lines indicate the nominal X-ray line energies; relative 
heights indicate the emission intensity folded with the absorption 
probability in $5 \, \rm{mm}$ CZT and $2 \, \rm{mm}$ CZT (small tick), 
respectively. Data were taken with X-Calibur configuration 
$C_{\rm{wu1}}$ (see Fig.~\ref{fig:X-Calibur_Configurations}). {\bf 
Left:} Spectrum of an individual detector pixel (linear energy axis, 
$C_{\rm{wu1a}}^{1}$). The dashed vertical line indicates the trigger 
energy threshold. The fitted functions ($40.1 \, \rm{keV}$ and $121.8 \, 
\rm{keV}$) are used to determine the peak position and energy resolution 
(FWHM). {\bf Right:} Energy spectra summed over all pixels of the 
detectors in ring {\it R3} ($5 \, \rm{mm}$ thickness) and ring {\it R7} 
($2 \, \rm{mm}$ thickness), respectively. Also shown are the simulated 
spectra with/without (Sim+/Sim) additional continuum (see 
Sec.~\ref{sec:Simulations}). A CR background spectrum (scaled by a 
factor of $10$) is shown for reference. The horizontal dotted line 
indicates the energy range relevant for X-Calibur.}

\label{fig:EnergyCalibration}

\end{figure*}

The energy calibration of the individual CZT detector pixels is done 
with a compact Eu$^{152}$ source (cylindrical emitting volume with a 
diameter of $\simeq$$3 \, \rm{mm}$) in the X-Calibur configuration 
$C_{\rm{wu1b}}^{2}$ in which the the scintillator rod is not installed 
(see Fig.~\ref{fig:X-Calibur_Configurations}). The source was 
successively placed at the centers of the detector rings~-- allowing to 
calibrate rings {\it R1} to {\it R8}, one at a time. In a first step, an 
automatic routine is used to optimize the pixel trigger thresholds: data 
are taken in a special acquisition mode that adjusts ASIC/channel 
discriminators based on measured event rates for each pixel, such that 
the trigger threshold is as low as possible (maximizing the integral 
trigger rate), but at the same time is safely above the electronic noise 
regime. The noise regime leads to very high (artificial) trigger rates 
and differs from channel to channel. The routine works reliably for most 
channels. However, a visual inspection of all 2048 recorded energy 
spectra was performed to assure that channels with trigger thresholds 
set too high or too low (failed automatic detection of the noise regime) 
were adjusted manually.

About 5 million events were taken for each CZT detector (20 million 
events per ring {\it R}), and the known energy lines at $40.1 \, 
\rm{keV}$ and $122 \, \rm{keV}$ were used to determine the pedestal 
$p_{0}$ and amplification slope $a$ for each channel\footnote{The 
linearity of the channels was confirmed using the internal test pulse 
generator of the ASIC.}. The energy of a measured pulse height $p$ is in 
turn calculated using

\begin{equation}
\label{eq:EnergyCalibration} 
E(p) = (p-p_{0})/a.
\end{equation}

The left panel of Figure~\ref{fig:EnergyCalibration} shows the 
calibrated energy spectrum of a single pixel. The right panel shows the 
averaged calibration spectra of two chosen detector rings ($4 \times 64$ 
pixels each). Also shown are the energy spectra obtained from the 
simulations of the corresponding setup (see Sec.~\ref{sec:Simulations}). 
It can be seen, that the $2 \, \rm{mm}$ detectors (ring {\it R7}) loose 
performance at energies $E> 100 \, \rm{keV}$, as compared to the 
detectors with $5 \, \rm{mm}$ thickness (ring {\it R3}). However, in the 
$20-80 \, \rm{keV}$ energy band, relevant for X-Calibur, both types of 
detector thickness perform at a similar level.

In addition to the temperature-dependence of the detector performance 
discussed in Sec.~\ref{subsec:TempStudies}, another consideration has to 
be made when applying the calibration to the data. The CZT detectors 
used in X-Calibur are not setup for measuring the depth position of the 
X-ray interaction/absorption between the detector cathode and anode~-- 
referred to as the depth-of-interaction (DOI). The energy calibration in 
Eq.~(\ref{eq:EnergyCalibration}), however, depends on the average DOI of 
a given energy. The mean DOI, however, changes with the cosine of the 
inclination angle measured between the absorbed X-ray and the detector 
plane. The calibration was determined with the X-ray source located $c 
\simeq 11 \, \rm{mm}$ above the center of the detector cathode (see 
Fig.~\ref{fig:X-Calibur_Configurations}, top). Depending on their 
geometrical locations, the pixels are hit under angles between 
$0-45^{\circ}$. X-rays Compton-scattering in the X-Calibur scintillator, 
on the other hand, can hit the CZT detectors at angles between 
$0-90^{\circ}$, which adds a systematic error/uncertainty to the 
reconstructed energy for small incident angles. However, in the $20-80 
\, \rm{keV}$ band the DOI distribution is very narrow and localized 
close to the cathode. Therefore, the angle dependence is negligible~-- 
the systematic shift of the reconstructed $40.1 \, \rm{keV}$ line was 
experimentally constrained to be less than $1\%$ for shallow inclination 
angles.


A $< 3\%$ fraction of ASIC channels were found to be dead, too noisy, or 
did not make contact to the detector pixel. These channels were excluded 
from the analysis and are marked with `x' in the corresponding 2D plots 
shown in this paper (e.g. Fig.~\ref{fig:PixelThreshold}).

\subsection{Energy Resolution and Threshold} 
\label{subsec:ThreshAndResolution}

\begin{figure}[t]
\begin{center}
\includegraphics[width=0.49\textwidth]{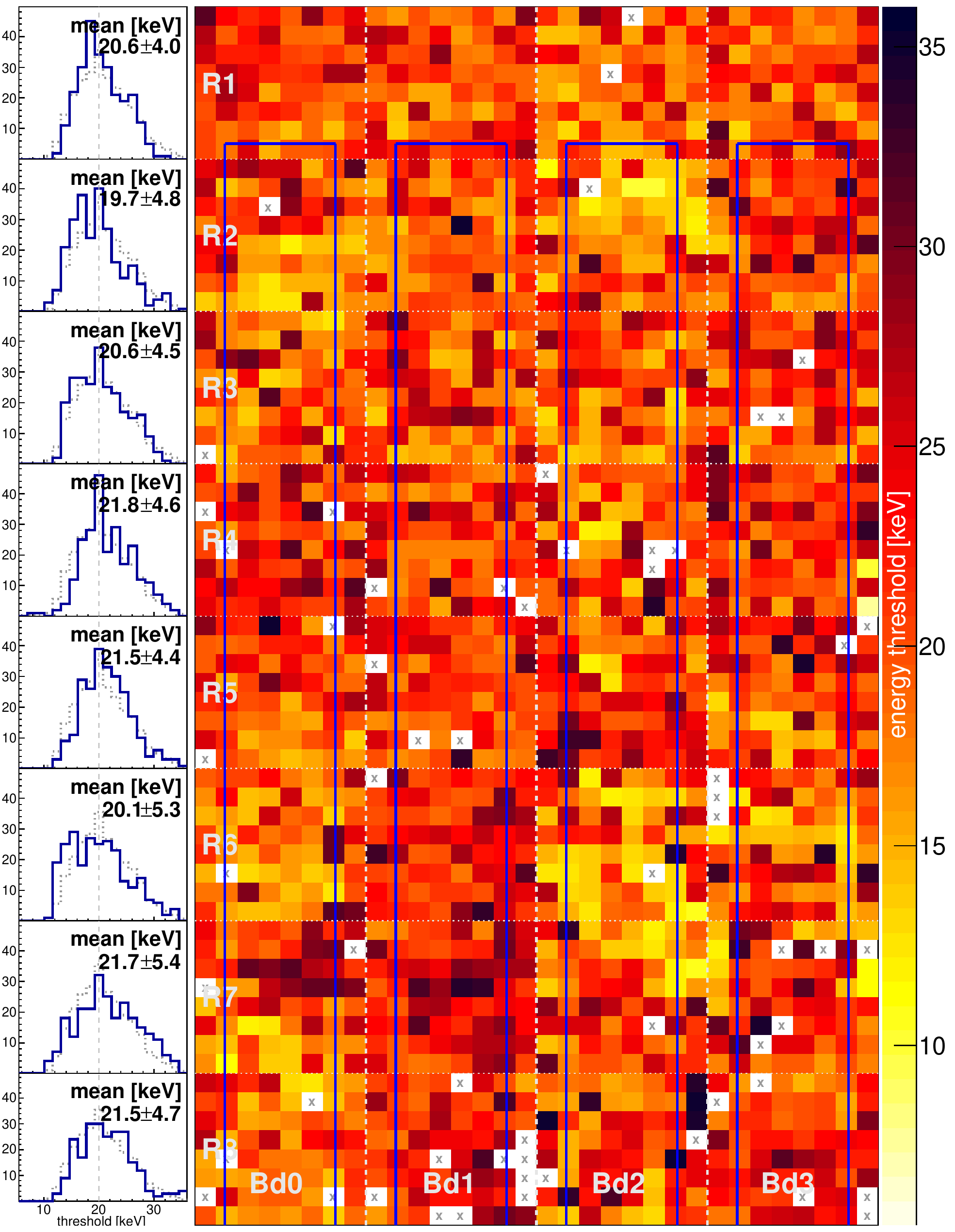}
\end{center}

\caption{2D distribution of energy thresholds of all X-Calibur pixels. 
In this representation, the four detector sides surrounding the 
scintillator ({\it Bd0} to {\it Bd3}, see Fig.~\ref{fig:Design}, left) 
are unfolded into a plane. The segments enclosed by dashed and dotted 
lines indicate the CZT detectors ($8 \times 8$ pixels each). The 
elongated box (solid line in {\it Bd0} to {\it Bd3}, each) indicates the 
position of the scintillator, even though it was not installed for this 
particular data set (configuration $C_{\rm{wu1b}}^{2}$, see 
Fig.~\ref{fig:X-Calibur_Configurations}, top). Pixels marked with `x' 
were not included in the analysis. The left panels show the threshold 
distributions per detector ring {\it R} (the gray distribution shows the 
scaled average of all pixels).}

\label{fig:PixelThreshold}

\end{figure}

To study the performance of the X-Calibur CZT detectors, the calibration 
data (Sec.~\ref{subsec:CZT_Calibration}) were used to characterize the 
energy spectra of individual pixels (see left panel of 
Fig.~\ref{fig:EnergyCalibration} for reference). The relevant properties 
studied in this section are the energy threshold, the fitted line/peak 
position, and the energy resolution (FWHM). As for the energy 
calibration, the data were taken with the configuration 
$C_{\rm{wu1b}}^{2}$ (see Fig.~\ref{fig:X-Calibur_Configurations}).

\begin{figure}[t!]
\begin{center}
\includegraphics[width=0.48\textwidth]{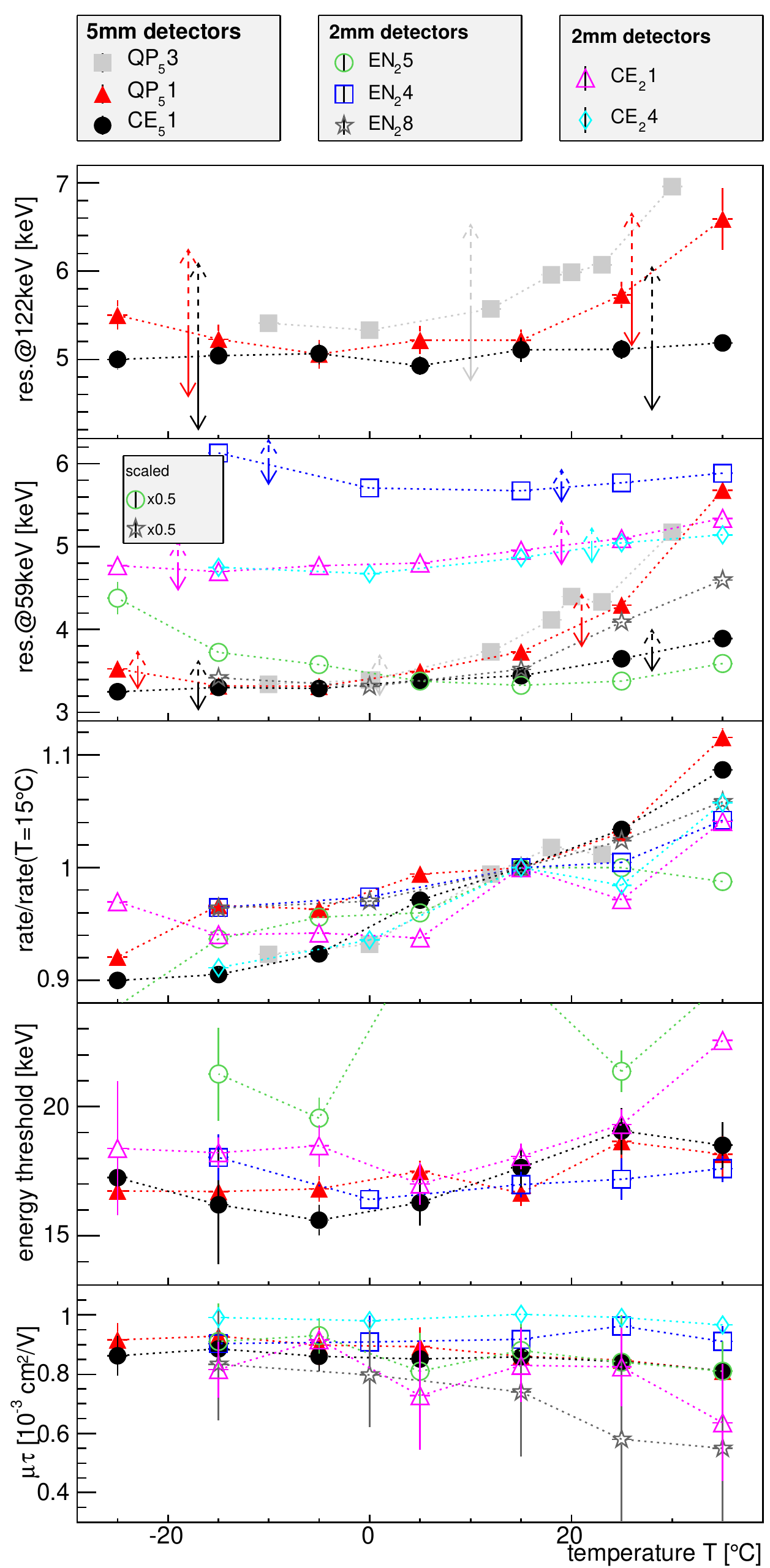}
\end{center}

\caption{Temperature-dependent performance of individual CZT detectors. 
Each data point represents the average of $64$ pixels of a detector. 
Vertical arrows indicate the shift of the curve if considering only edge 
pixels of a detector (dashed direction), or only central pixels of a 
detector (solid direction). {\bf Top:} Energy resolution at $122 \, 
\rm{keV}$ (only shown for the $5 \, \rm{mm}$ detectors). {\bf 
2$^{\rm{\bf{nd}}}$ from top:} Energy resolution at $59 \, \rm{keV}$. 
{\bf 3$^{\rm{\bf{rd}}}$ from top:} Peak detection rate of the $59 \, 
\rm{keV}$ line normalized to the rate measured at $T=15^{\circ} \, 
\rm{C}$. {\bf 4$^{\rm{\bf{th}}}$ from top:} Energy threshold as 
determined after threshold optimization performed at the corresponding 
temperature. {\bf Bottom:} Mobility-lifetime product 
$\mu_{\rm{e}}\tau_{\rm{e}}$ measured based on the $59 \, \rm{keV}$ 
line.}

\label{fig:TempStudies}
\end{figure}

The energy threshold of a pixel is defined as the reconstructed energy 
$E_{\rm{thr}}$ above which the corresponding ASIC channel starts to 
trigger on events (see Fig.~\ref{fig:EnergyCalibration}, left). The 
distribution of thresholds for all X-Calibur pixels is shown in 
Fig.~\ref{fig:PixelThreshold}. The thresholds vary from pixel to pixel, 
but no significant geometrical trends can be identified if comparing 
edge pixels versus central pixels, or pixels of detectors with different 
thickness ($5 \, \rm{mm}$, rings {\it R1-R5} versus $2 \, \rm{mm}$, 
rings {\it R6-R8}). About $50\%$ of all pixels have an energy threshold 
of $E<20 \, \rm{keV}$. It should be noted that one has to account for 
the energy resolution of a pixel in order to determine the analysis 
threshold that is about $3-4 \, \rm{keV}$ higher. The thresholds shown 
in Fig.~\ref{fig:PixelThreshold} reflect the status of the compact 
X-Calibur configuration. The thresholds of the detectors operated as a 
single unit are up to $5 \, \rm{keV}$ lower (see 
Fig.~\ref{fig:TempStudies}).

\begin{figure*}[t!]
\begin{center}
\includegraphics[height=0.42\textheight]{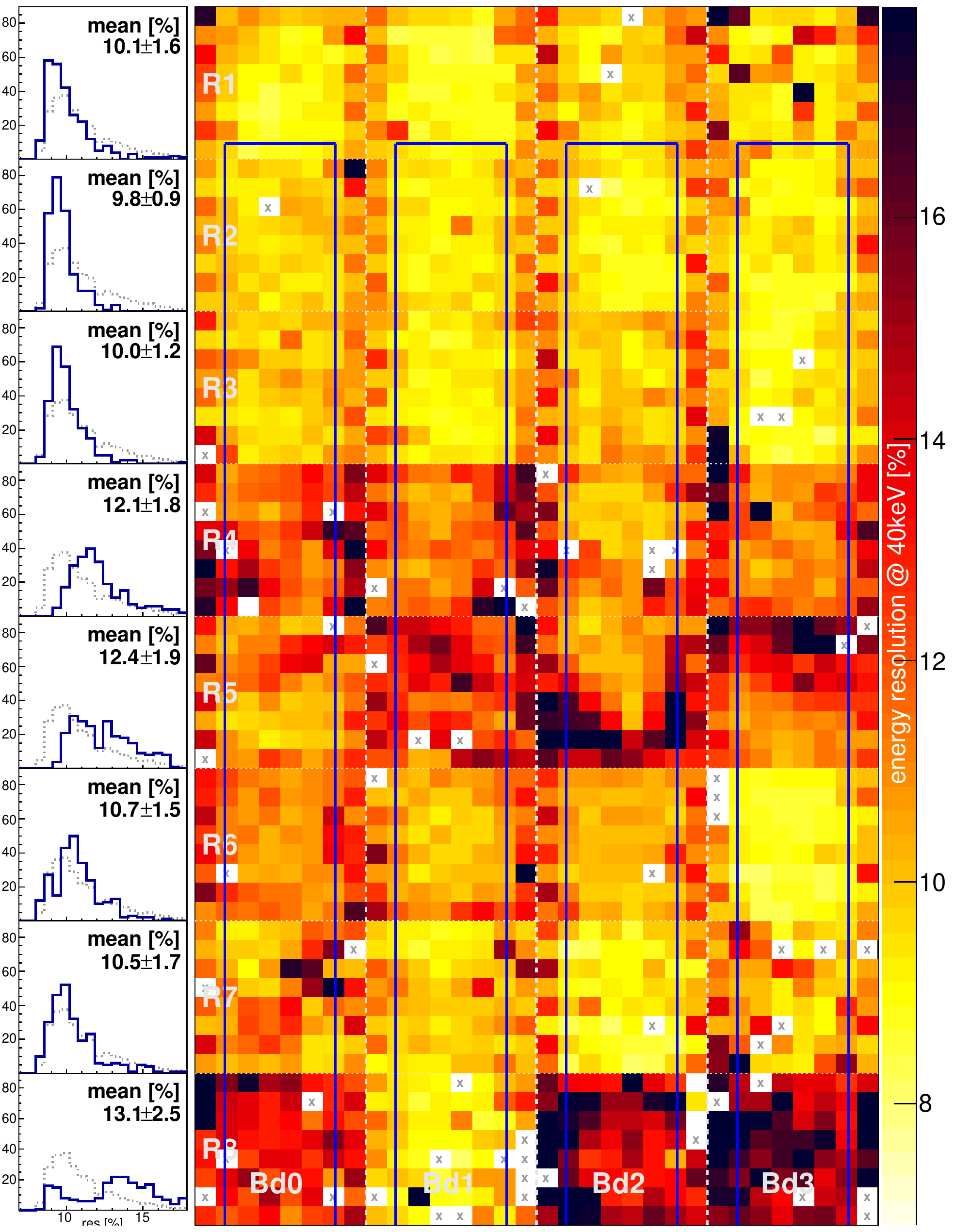}
\hfill
\includegraphics[height=0.42\textheight]{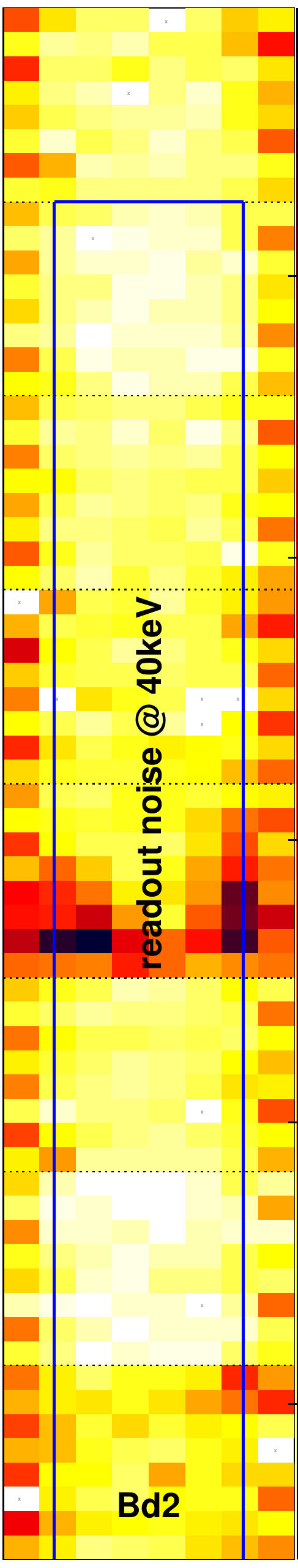}
\hfill
\includegraphics[height=0.42\textheight]{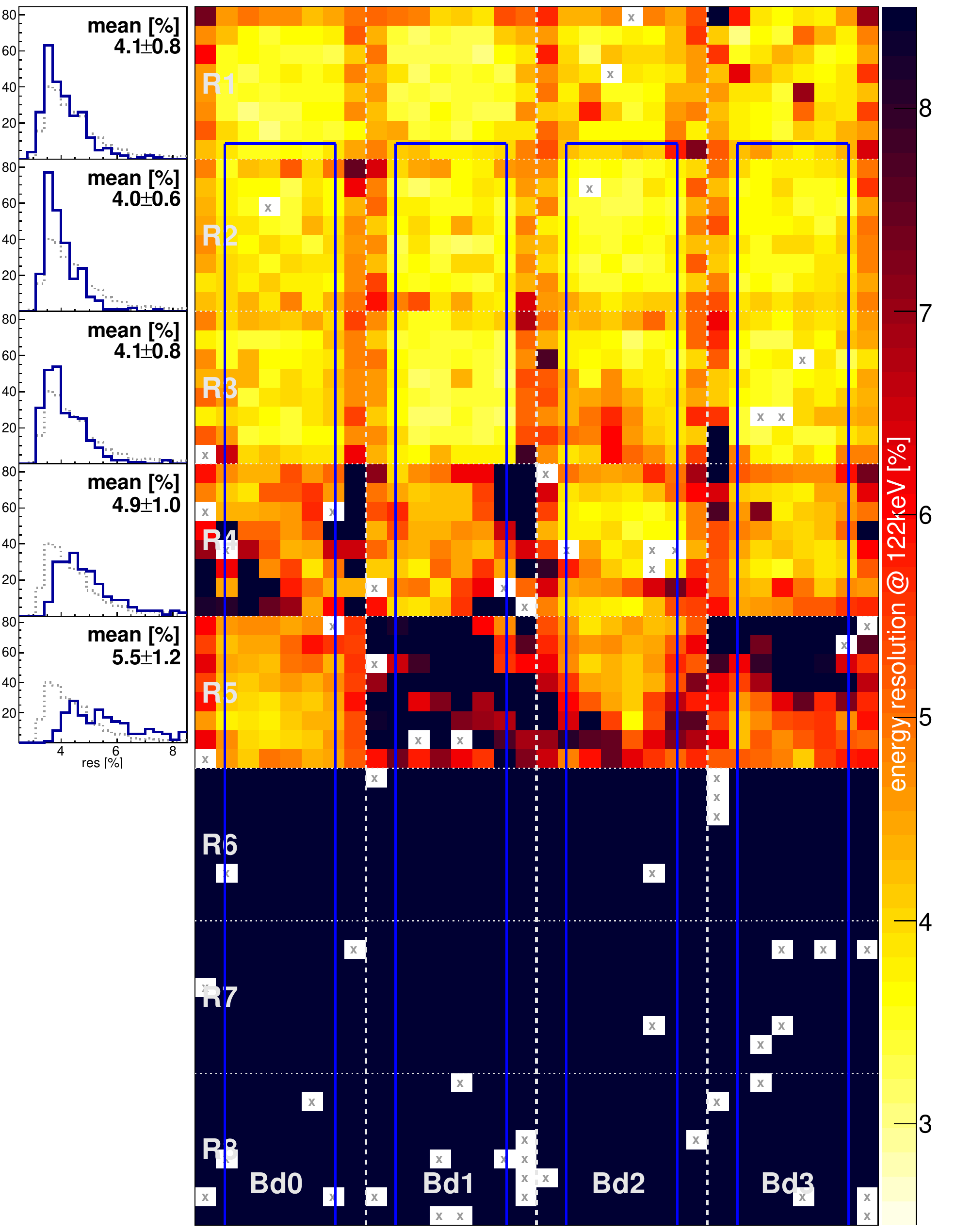}
\end{center}

\caption{2D distribution of the pixel-by-pixel energy resolution 
(compare with Fig.~\ref{fig:PixelThreshold}). {\bf Left:} Energy 
resolution at $40.1 \, \rm{keV}$. {\bf Middle:} Electronic readout noise 
at $40 \, \rm{keV}$ as determined using the ASICs internal test pulse 
generator (only shown for {\it Bd2}). The same range of the color scale 
as in the left panel is used. {\bf Right:} Energy resolution at $121.8 
\, \rm{keV}$. This energy line is not well-defined in the energy spectra 
measured with the $2 \, \rm{mm}$ detectors (see 
Fig.~\ref{fig:EnergyCalibration}) so that results are only shown for the 
$5 \, \rm{mm}$ rings {\it R1} to {\it R5}.}

\label{fig:PixelResolution}

\end{figure*}

Figure~\ref{fig:PixelResolution} shows the energy resolutions at $40.1 
\, \rm{keV}$ and $121.8 \, \rm{keV}$, respectively. Detector rings {\it 
R2} and {\it R3} are the most sensitive ones when it comes to detecting 
the Compton-scattered X-rays in the polarization measurements (see 
Sec.~\ref{subsec:CHESS}). Therefore, the best performing $5 \, \rm{mm}$ 
detectors were positioned in these rings accordingly. Some of the lower 
detector rings ({\it R4} to {\it R8}) show regions with clearly 
poorer-than-average energy resolution. However, any asymmetry in 
azimuthal detector performance will cancel out due to the rotation of 
the polarimeter in the final mode of operation. The energy line at 
$121.8 \, \rm{keV}$ is not well-defined in the spectra measured with the 
$2 \, \rm{mm}$ detectors (see Fig.~\ref{fig:EnergyCalibration}, right). 
Therefore, Fig.~\ref{fig:PixelResolution} only shows the $121.8 \, 
\rm{keV}$ energy resolutions for the $5 \, \rm{mm}$ detectors. The 
average energy resolution in rings {\it R1} to {\it R3} at $40.1 \, 
\rm{keV}$ amounts to $4.0 \, \rm{keV}$ ($10.0\%$); the corresponding 
value at $121.8 \, \rm{keV}$ is $4.9 \, \rm{keV}$ ($\simeq 4\%$). The 
average performance of rings {\it R4} to {\it R7} at $40.1 \, \rm{keV}$ 
amounts to $4.6 \, \rm{keV}$ ($\simeq 11\%$). The performance of the 
detectors in ring {\it R8} is modest.

The energy resolution of a detector pixel is mainly determined by two 
factors~-- the quality of the CZT crystal and the noise of the read-out 
electronics. The electronic readout noise was determined for all 
channels using the ASICs internal pulse generator (see 
Sec.~\ref{sec:XCLB}). The generator injects charge into the amplifier of 
the corresponding channel and allows one to test the 
trigger/digitization chain on a chennel-by-channel basis. 1000 events 
were taken per channel with the detectors connected and biased at 
nominal operation voltage (to also account for noise introduced by dark 
currents in the CZT). The results are shown in the middle panel of 
Fig.~\ref{fig:PixelResolution} for one of the four boards. More details 
on the electronic noise studies are presented in 
Sec.~\ref{subsec:TempStudies}. Note, that the internal ASIC capacitor 
does not allow to inject charges that correspond to energies lower than 
$\approx 200 \, \rm{keV}$. However, the noise versus energy trend seems 
to level off for energies lower than $\simeq 500 \, \rm{keV}$ (see 
Fig.~\ref{fig:TempStudiesNoise}). Therefore, the absolute noise 
resolution measured at $\simeq 200 \, \rm{keV}$ was used to estimate the 
relative noise contribution at $40 \, \rm{keV}$, as shown in the middle 
panel of Fig.~\ref{fig:PixelResolution}.

A comparison between the electronic noise and the energy resolution 
determined from the spectral lines shows that the low-energy resolution 
is dominated by the electronic noise. This kind of comparison can in 
general assist in localizing the cause for modestly performing 
detectors, e.g. by disentangling the contributions of electronic noise 
versus CZT crystal quality. The detector located in ring {\it R5} of 
{\it Bd2}, for example, shows noisy regions in both, the Eu$^{152}$ 
data, as well as in the electronic noise measurement~-- indicating a 
high leakage current as the reason for the sub-optimal performance. The 
ring {\it R8} detector in the same board, on the other hand, does not 
exhibit a poorer energy resolution than others in terms of noise~-- 
therefore, the poor performance visible in the Eu$^{152}$ data is 
probably related to a low CZT crystal quality.

On average, the side pixels of individual detectors exhibit a poorer 
than average energy resolution. At low energies, this shows up as a 
periodic structure on the left and right side of each detector. The same 
pattern can be seen in the electronic noise measurement 
(Fig.~\ref{fig:PixelResolution}, left versus middle), and can therefore 
be attributed to the readout noise. This hypothesis is supported by the 
fact that the leads between the corresponding edge pixels and the ASIC 
channels are located on the `outside' region of the printed circuit 
board~-- being more susceptible to noise pick up from the surrounding 
electronics. Subtracting the readout noise, these edge pixels do no 
longer show poorer energy resolution at low energies as compared to the 
other pixels. For energies $E > 100 \, \rm{keV}$ (not relevant for 
X-Calibur), the detector edge pixels show a reduced resolution in 
addition to the electronic noise component (`frame-like' structure 
surrounding each detector in Fig.~\ref{fig:PixelResolution}, right). 
This is a known issue with CZT detectors and can be explained by a less 
homogeneous electric field in the edge regions of a detector, affecting 
the charge collection. The effect is less prominent (if visible at all) 
for the horizontal edge pixels, for which the field is stabilized by the 
neighboring detectors located in the same plane.

\subsection{CZT Performance at different Temperatures} 
\label{subsec:TempStudies}

During a balloon flight the pressure vessel housing the polarimeter will 
undergo several changes in temperature that can potentially affect its 
performance. At float altitude, the vessel will be in an outside 
temperature environment of around $-20^{\circ} \rm{C}$. During daytime, 
the thermal radiation fields of the sun and the earth will provide 
additional sources of energy. During the night, the thermal radiation of 
the earth is the only external source of heat flow. The electronics 
inside the vessel generate heat at a rate of less than $100 \, \rm{W}$. 
The outside of the vessel will be insulated using a layer of aluminized 
mylar (reflection of sun light and high emittance of thermal radiation) 
that will be contact-separated from the surface of the vessel with a 
layer of Dacron mesh. Furthermore, heater bands with a total power of 
$175 \, \rm{W}$ are installed inside the vessel to guarantee a 
controllable temperature in the range of $0$ to $25^{\circ} \rm{C}$. 
This thermal design will prevent overheating during day-time and will 
avoid cold temperatures during the night. Nonetheless, variations in 
temperature of the polarimeter during flight are expected to some 
extent. Therefore, it is important to understand the 
temperature-dependent performance of the CZT detectors.

Individual CZT detectors were used to quantify the temperature 
dependence of the energy resolution, the energy threshold and the 
detection rate. Data were taken in a temperature chamber in the range of 
$T \in [-25; +35]^{\circ} \, \rm{C}$ with the detectors being 
illuminated with an Am$^{241}$ source ($59 \, \rm{keV}$) and a Co$^{57}$ 
source ($122 \, \rm{keV}$). The sources were located at a distance of 
$\simeq 2 \, \rm{cm}$ above the detector cathode. The data were used to 
re-calibrate each detector (pixel-by-pixel) for each environment 
temperature in order to cancel temperature-dependent calibration 
effects. Since these measurements were time-intensive and could only be 
done for one detector at a time, the study was limited to a 
representative subset of the detectors shown in 
Tab.~\ref{tab:Detectors}.

{\it Temperature-dependent detector characteristics.} The results of the 
measurements are shown in Fig.~\ref{fig:TempStudies}, averaged over all 
64 pixels per detector. The $122 \, \rm{keV}$ line is not well defined 
in the $2 \, \rm{mm}$ detectors, so that the high-energy results are 
only shown for the $5 \, \rm{mm}$ detectors. Given the limited sample of 
detectors, we cannot expect to attribute observed trends to a specific 
detector class (such as the brand); however, some general findings can 
be identified and are described in the following.

The energy resolution generally improves if reducing the temperature 
from $T = 35^{\circ} \, \rm{C}$ to $T = 0^{\circ} \, \rm{C}$. This 
effect is most prominent for the two tested Quikpak detectors and 
amounts to an improvement of up to $50\%$/$30\%$ at $59/122 \, 
\rm{keV}$, respectively. The effect is much weaker for the Creative 
Electron detectors. For temperatures $T < 0^{\circ} \, \rm{C}$, the 
resolution levels out. For some of the Endicott detectors ($2 \, 
\rm{mm}$) it even gets poorer again; however, the two detectors showing 
this effect (EN$_{2}$4 and EN$_{2}$5) show very poor performance in 
general. All three $5 \, \rm{mm}$ detectors show a much better energy 
resolution at $122 \, \rm{keV}$ in the central pixels (indicated by the 
downward arrow in the top panel of Fig.~\ref{fig:TempStudies}) as 
compared to the edge pixels (upward arrow). Here, the aspect ratio 
allows for E-field lines to bulge out of the sides of the detector (see 
also Fig.~\ref{fig:PixelResolution}). This difference is clearly less 
pronounces at $59 \, \rm{keV}$ where the charge collection is much more 
concentrated in the cathode region of the detector crystal.

The peak detection rate at $59 \, \rm{keV}$ (3$^{\rm{rd}}$ panel from 
the top in Fig.~\ref{fig:TempStudies}) shows a clear trend for all 
detectors: a reduced detection efficiency with decreasing temperature. 
This effect is not understood. Although the measurements were carefully 
set up, it cannot be excluded that a temperature-dependent contraction 
of the casing/fixture could have lead to a slight change in distance 
between the source and the detector as a function of environment 
temperature.

For each temperature, the energy thresholds were re-optimized. The 
results are shown in the 4$^{\rm{th}}$ panel of 
Fig.~\ref{fig:TempStudies}. No clear trends can be identified~-- the 
average energy threshold of the studied detectors lies between $15-20 \, 
\rm{keV}$.

Measurements at different bias voltages $V_{\rm{bi}}$ were used to 
determine the mobility lifetime product $\mu_{\rm{e}}\tau_{\rm{e}}$. 
Data were taken at $V_{\rm{bi}} = -200 \, \rm{V}$, $-500 \, \rm{V}$, and 
$-700 \, \rm{V}$ ($5 \, \rm{mm}$ detectors) and at $V_{\rm{bi}} = -100 
\, \rm{V}$, $-150 \, \rm{V}$, and $-200 \, \rm{V}$ ($2 \, \rm{mm}$ 
detectors), respectively. The shift of the line position with increasing 
$V_{\rm{bi}}$ was used to calculate $\mu_{\rm{e}}\tau_{\rm{e}}$ 
\cite{MuTauImproved}. The results are shown in the bottom panel of 
Fig.~\ref{fig:TempStudies}. No significant change in 
$\mu_{\rm{e}}\tau_{\rm{e}}$ can be identified for the temperature range 
studied. \citet{Jung2007} discuss the temperature dependence of Imarad 
High-Pressure Bridgman CZT detectors and find a decreasing trend of 
$\mu_{\rm{e}} \tau_{\rm{e}}$ for temperatures $T > 10^{\circ} \, \rm{C}$ 
and for $T < -25^{\circ} \, \rm{C}$.

\begin{figure*}[t!]
\begin{center}
\includegraphics[width=0.99\textwidth]{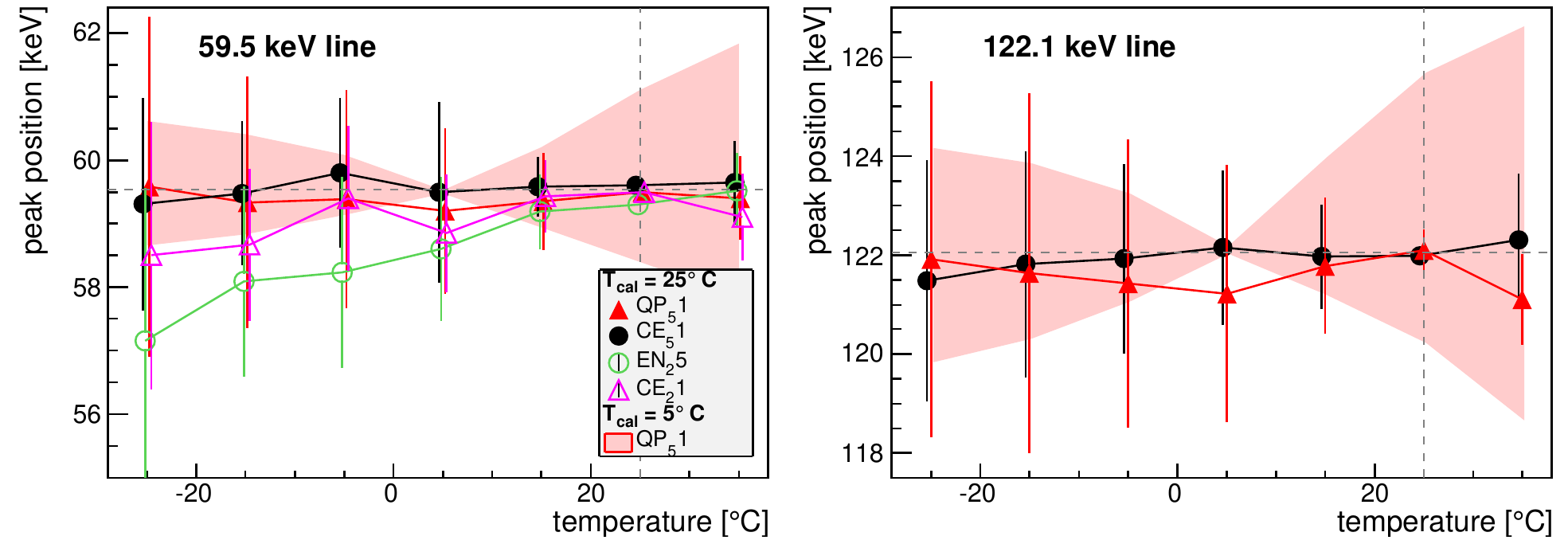}
\end{center}

\caption{Reconstructed line energy for different temperatures using the 
calibration obtained at $T_{\rm{cal}} = 25^{\circ} \, \rm{C}$ (dashed 
vertical line). The data points indicate the mean of the peak positions 
of all 64 pixels and the error bars represent the standard deviation of 
that distribution (separately derived for each temperature). The dashed 
horizontal line indicates the nominal line energy.}

\label{fig:PeakPosVsTemp}
\end{figure*}

{\it Temperature-stability of the calibration.} In the studies shown in 
Fig.~\ref{fig:TempStudies} all detector pixels were re-calibrated at 
each temperature. It is important to understand how the calibration 
itself changes with temperature. To study this effect, a reference 
calibration at $T_{\rm{cal}} = 25^{\circ} \, \rm{C}$ was applied to the 
measurements taken at the different temperatures. The line positions at 
$59.5 \, \rm{keV}$ and $122.1 \, \rm{keV}$ were determined for the 
individual pixels. Each distribution (one per temperature) of 
reconstructed line positions was in turn characterized by its mean and 
its standard deviation. The results are illustrated in 
Fig.~\ref{fig:PeakPosVsTemp}, where the standard deviation (spread of 
the corresponding distribution) is represented as error bar. For 
reference, the spectra of detector QP$_{5}$1 were corrected with the 
calibration obtained at $T_{\rm{cal}} = 5^{\circ} \, \rm{C}$. While the 
mean reconstructed line energy does not change significantly, the 
widening of the error band indicates that individual channels show a 
temperature-dependent upward/downward drift of the line position. This 
makes it difficult to globally correct for temperature-dependent changes 
in the calibration~-- unless calibration data are taken for all 2048 
X-Calibur channels at a variety of temperatures. The calibration changes 
by up to $\simeq 2\%$ for temperatures varying around $\pm 10^{\circ} \, 
\rm{C}$ relative to the reference calibration $T_{\rm{cal}}$.


\begin{figure}[t!]
\begin{center}
\includegraphics[width=0.49\textwidth]{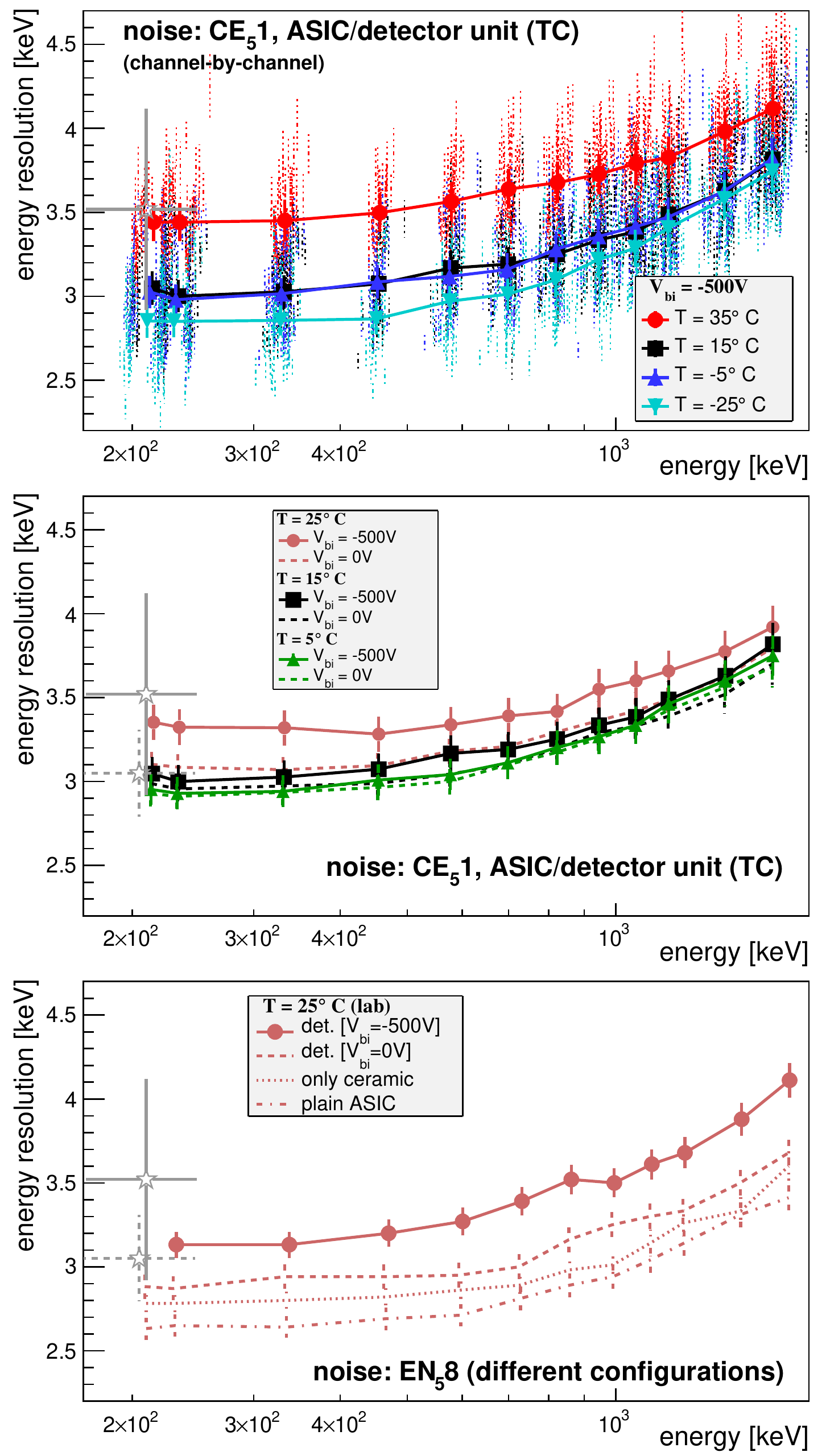}
\end{center}

\caption{Electronic readout noise measured with the ASIC's internal test 
pulser for different temperatures and different energies (detector 
CE$_{5}$1 and EN$_{5}$8). For reference, the gray asterisk marker 
indicates the 1~std.dev. range of all 2048 channels as operated in the 
X-Calibur assembly at room temperature of $T \approx 25^{\circ} \, 
\rm{C}$ with/without biased detectors (solid/dashed line). {\bf Top:} 
Channel-by-channel noise (vertical dotted lines) and detector average 
with the detector being biased at $V_{\rm{bi}}=-500 \, \rm{V}$. {\bf 
Middle:} Detector averages for three different temperatures with and 
without the cathode bias $V_{\rm{bi}}$. {\bf Bottom:} A series of $T 
\approx 25^{\circ} \, \rm{C}$ measurements with different configurations 
as described in the text.}

\label{fig:TempStudiesNoise}

\end{figure}

{\it Electronic noise.} The read-out electronic contributes a certain 
amount of jitter to the measured energy resolution of a detector pixel. 
In order to quantify this contribution, a series of measurements was 
taken with detectors CE$_{5}$1 and EN$_{5}$8 (see 
Tab.~\ref{tab:Detectors}) at different temperatures using the internal 
pulse generator of the ASIC (see Sec.~\ref{sec:XCLB}). The detectors 
were operated in an electrically shielded copper box. With the detector 
being plugged into the ASIC, the measurements actually reflect the 
readout noise of the ASIC/detector assembly, rather than the noise of 
the ASIC alone. Different amounts of charge were injected (corresponding 
to different energies)\footnote{Note, that given the differences in 
pixel acceptance and pixel calibration, a fixed amount of charge 
injected into the ASIC translates to slightly varying reconstructed 
energies (if comparing different channels).}. The measured pulse heights 
were transformed to energies using the corresponding energy calibration 
determined for each temperature. For each configuration and channel, a 
total of 1000 events were injected. The calibrated distribution was 
fitted in order to determine the corresponding mean energy and energy 
resolution.

The results are shown in Fig.~\ref{fig:TempStudiesNoise}. At energies of 
around $200 \, \rm{keV}$ the electronic noise of the ASIC/detector unit 
increases by $(21 \pm 6)\%$ in the studied temperature range of 
$-25^{\circ} \, \rm{C}$ to $+35^{\circ} \, \rm{C}$. The shape of the 
energy-dependent noise curve depends on the temperature. At room 
temperature, the noise increases by $(20 \pm 4)\%$ if going from $200 \, 
\rm{keV}$ to $1.7 \, \rm{MeV}$ and levels out around $3 \, \rm{keV}$ at 
low energies. This suggests, that the energy resolution in the X-Calibur 
range ($\approx 4 \, \rm{keV}$ at $40 \, \rm{keV}$, see 
Fig.~\ref{fig:PixelResolution}) is dominated by the electronic readout 
noise, rather than charge transport properties in the CZT crystal. The 
gray asterisk marker in Fig.~\ref{fig:TempStudiesNoise} shows the 
1~std.dev.~range of the noise distribution of all 2048 data channels as 
measured at room temperature ($\approx 25^{\circ} \, \rm{C}$) in the 
final X-Calibur configuration, compare with 
Fig.~\ref{fig:PixelResolution} (middle).

The middle panel of Fig.~\ref{fig:TempStudiesNoise} illustrates the 
effect of the bias voltage of the detector. A biased cathode at $T = 
25^{\circ} \, \rm{C}$ increases the low-energy noise by $(8 \pm 5)\%$, 
whereas no noticeable change can be measured for temperatures lower than 
that. Therefore, the cathode bias only seems to systematically affect 
the readout noise for temperatures higher than $\simeq 15^{\circ} \, 
\rm{C}$.


The bottom panel in Fig.~\ref{fig:TempStudiesNoise} shows the electronic 
noise measured at room temperature for different configurations: (i) the 
detector/ASIC unit with biased cathode, (ii) the detector/ASIC unit with 
unbiased cathode, (iii) the ASIC with a ceramic chip carrier but no 
detector bonded to it, and (iv) only the plain ASIC. It can be seen that 
steps (i)--(iii) each add $\approx 0.2 \, \rm{keV}$ readout noise to the 
single-detector system. The average readout noise of the whole X-Calibur 
assembly (gray asterisk in Fig.~\ref{fig:TempStudiesNoise}) is shown for 
reference. Another series of measurements was performed with the plain 
ASIC at different temperatures (not shown). No significant noise trend 
could be identified in the $T = -20^{\circ} \, \rm{C}$ to $+25^{\circ} 
\, \rm{C}$ temperature range which leads to the conclusion that the 
temperature dependence of the readout noise of the ASIC/detector unit 
(top panel of Fig.~\ref{fig:TempStudiesNoise}) is mostly a result of the 
temperature dependence of the dark currents in the CZT crystal.

{\it Caveats:} The electronic readout noise varies from ASIC to ASIC and 
depends on the electronic shielding environment in which the ASIC is 
operated. Therefore, the comparison between the absolute noise levels of 
the single ASIC system shown in Fig.~\ref{fig:TempStudiesNoise} and the 
average readout noise in the X-Calibur assembly should be treated with 
care; the relative noise trends found, however, can likely be applied to 
whole X-Calibur assembly. It should also be mentioned, that the cooling 
aggregates of the temperature chamber (increased activity at low 
temperatures) can potentially introduce external noise pick-up in the 
ASIC.

\subsection{Summary of the Detector Calibration and Tests}

Each detector pixel has been energy calibrated according to 
Eq.~(\ref{eq:EnergyCalibration}). With the current readout electronics 
and the compact X-Calibur configuration, the CZT detectors achieve a 
mean trigger threshold of $\simeq 21 \, \rm{keV}$. The mean energy 
resolution at $40 \, \rm{keV}$ in the three front-side detector rings 
{\it R1}-{\it R3} (detecting most of the scattered events in the 
polarization measurements) is found to be $\Delta E_{\rm{czt}} \simeq 4 
\, \rm{keV}$ FWHM, when operated at room temperature. The energy 
resolution of the polarimeter as a whole is determined by the energy 
resolution of the individual detectors and by the energy deposited/lost 
in the scintillator ($\Delta E_{\rm{sci}} = 0 - 5.4 \, \rm{keV}$ at $40 
\, \rm{keV}$). The energy resolution of the detectors is thus not 
entirely negligible and X-Calibur would benefit from an optimized 
readout ASIC. We are currently working on modifying and adopting the 
HD-3 ASIC \cite{DeGeronimo2003, Vernon2010}. Using a pre-amplifier chain 
optimized for the $2-100 \, \rm{keV}$ energy range, we expect a trigger 
threshold of $1.7 \, \rm{keV}$ and electronic readout noise of $\simeq 
550 \, \rm{eV}$ RMS. The noise contribution of the new ASIC to the 
energy resolution of the polarimeter would be negligible for all 
energies above $20 \, \rm{keV}$. The low energy threshold can 
potentially be used on a satellite-borne version of the polarimeter.

The effective energy threshold of the polarimeter is slightly higher 
than the energy threshold of the individual CZT detectors, as a $25 \, 
\rm{keV}$ photon looses up to $2.2 \, \rm{keV}$ in the scatterer. 
However, the polarimeter will detect a large fraction of the X-rays at 
125,000 feet flight altitude, as the residual atmosphere only transmits 
photons above $\simeq 25 \, \rm{keV}$.

Even though the temperature-dependent trends in energy resolution would 
favor operating the detectors at $T \leq 0^{\circ} \, \rm{C}$ 
(Fig.~\ref{fig:TempStudies}, top), the thermal design of the X-Calibur 
assembly and local internal heat built-up of the polarimeter during the 
balloon flight makes an operation at $T \simeq (15 \pm 10)^{\circ} \, 
\rm{C}$ a more likely scenario~-- still guaranteeing a reasonable energy 
resolution for most of the detectors. A change in temperature, which 
will be monitored during flight, leads to a shift in reconstructed 
energy~-- which can go either way (Fig.~\ref{fig:PeakPosVsTemp}). 
Ideally, one should use a data base of temperature-dependent calibration 
values on a pixel-by-pixel basis to correct for the temperature trends. 
If ignoring the temperature-dependence of the calibration, a systematic 
error on the reconstructed energy of a few percent has to be accounted 
for in the temperature interval of $\pm 10^{\circ} \, \rm{C}$ around the 
calibration temperature. For the first X-Calibur flight, we will choose 
the second option.

\section{X-Calibur: Instrument Characterization} \label{sec:XCLB_Characterization}

This section describes measurements of the fully assembled polarimeter 
installed in the CsI shield. The goal of the measurements is to 
characterize the efficiency of the shield, and to estimate the 
background levels in the different (ground-based) environments the 
polarimeter was operated in (Sec.~\ref{subsec:BG_Data}). The reduction 
of the background is crucial in order to perform sensitive measurements 
of the polarization properties of astrophysical sources, see 
Eq.~(\ref{eqn:MDP}).

The X-rays that enter the polarimeter along the optical axis will 
produce a certain amount of scintillation light when scattering in the 
scintillator rod which is read out by a PMT. The efficiency curve, 
describing the trigger probability for different energy depositions in 
the scatterer, is discussed in Sec.~\ref{subsec:ScintillatorEfficiency}. 
As will be shown in Sec.~\ref{subsec:BG_Data}, a high trigger efficiency 
of the scintillator will allow a further reduction of the background. 
The efficiency curve is fed into the simulations that were discussed in 
Sec.~\ref{sec:Simulations}. The measurements described in this chapter 
were done with disk-like radioactive sources with a diameter of $\simeq 
0.5 \, \rm{cm}$.

\subsection{Shield Performance and Cosmic Ray Background} \label{subsec:BG_Data}

The shielding and suppression of backgrounds is a crucial task for 
sensitive polarimetry measurements. The background in the ground-based 
measurements presented in this paper (non-flight) results mostly from 
secondary particles produced in air showers in the earth's atmosphere, 
induced by cosmic rays (CRs). The flux of the secondary particles 
depends on the geographical location at which the measurement is 
performed, and on the structure/materials of the building in which the 
polarimeter is operated (partly shielding the secondary particles). As 
can be seen in Fig.~\ref{fig:MultiplicityDistribution}, the CR 
background has a higher fraction of multiplicity $m \geq 2$ pixel events 
as compared to Compton-scattered X-rays in the $E < 100 \, \rm{keV}$ 
regime, relevant for the X-Calibur polarimetry measurements. In general, 
the distribution of $m$ depends on the energy and the kind of 
interaction; high-energy muons (ionization), for example, trigger events 
along a row of pixels (unless they cross the detector perpendicular to 
the pixel plane). Low energy X-rays, on the other hand, are 
photo-absorbed with a charge deposition usually contained well within 
one pixel. Therefore, a requirement of $m = 1$ pixel events already 
suppresses the background for the polarization measurements by a certain 
amount.

\begin{figure}[t!]
\begin{center}
\includegraphics[width=0.49\textwidth]{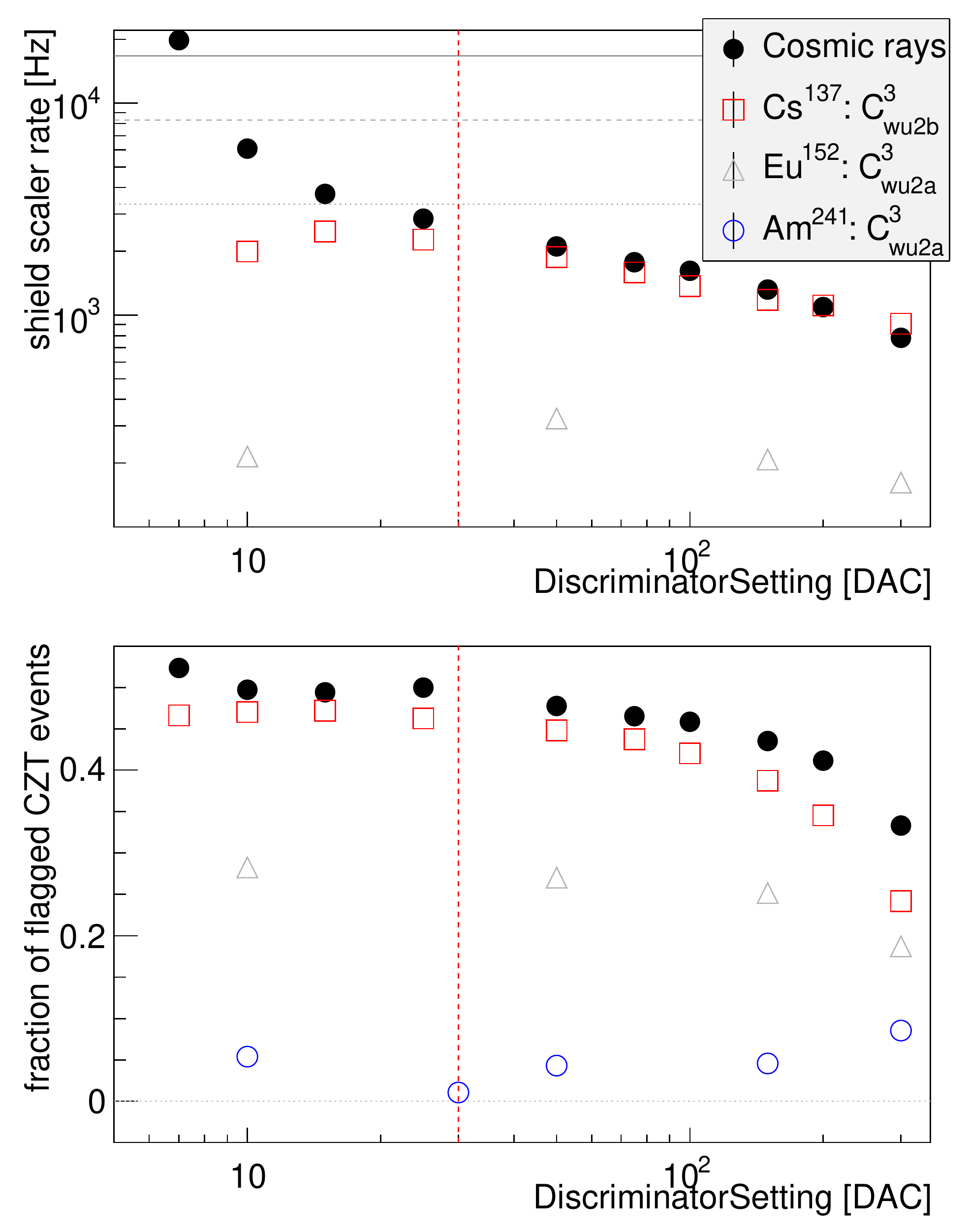}
\end{center}

\caption{Optimization of the shield discriminator using three different 
data runs (a)-(c), as described in the text. The background contribution 
is subtracted for the runs with Cs$^{137}$ outside shield (b), and for 
Am$^{241}$/Eu$^{152}$ on the optical axis (c). The dashed vertical line 
indicates the optimal discriminator setting. {\bf Top:} Shield trigger 
rate for different threshold settings. The horizontal lines indicate the 
rates corresponding to $10/5/2\%$ (from top to bottom) dead time 
produced by the shield vetoes. The noise regime starts to dominate 
around DAC $<20$; this in turn leads to an underestimation of the 
Cs$^{137}$/Eu$^{152}$ trigger contribution since the CR background is 
subtracted. In the case of the Am$^{241}$ source, no significant trigger 
rate above background was measured. {\bf Bottom:} Fraction of detected 
CZT events that are vetoed with $f_{\rm{shld}} = 1$.}

\label{fig:ShieldOptimization}
\end{figure}

{\it Optimization of the shield trigger threshold.} A particle crossing 
the active shield (see Fig.~\ref{fig:ActiveShieldAndTelescope, left) 
produces scintillation light in the CsI crystal. The crystal is read by 
four PMTs which signals are summed and digitized.} A programmable 
discriminator decides whether the shield veto flag $f_{\rm{shld}}$ is 
activated ($f_{\rm{shld}}$ is kept active for $6 \, \mu\rm{s}$ per 
trigger) and merged into the data stream. Events with the corresponding 
flag can in turn be filtered out in the data analysis. In order to 
optimize the shield trigger efficiency, a series of measurements was 
performed with different settings of the discriminator. The measurements 
comprise: (a) cosmic ray background only, (b) a collimated Cs$^{137}$ 
source aimed from outside the shield at the X-Calibur CZT detector 
assembly (configuration $C_{\rm{wu2b}}^{3}$ in 
Fig.~\ref{fig:X-Calibur_Configurations}), and (c) collimated 
Am$^{241}$/Eu$^{152}$ sources placed on the optical axis of the 
polarimeter (configuration $C_{\rm{wu2a}}^{3}$) to simulate X-rays from 
the X-ray mirror entering the polarimeter without interaction in the 
shield\footnote{Obviously, the lead used to collimate the sources leads 
to indirect scatterings and shield contamination, in particular in the 
case of the high energy lines of Eu$^{152}$.}. Two shield 
characteristics were measured: (i) the raw trigger rate of the shield, 
and (ii) the fraction of triggered CZT events with $f_{\rm{shld}} = 1$. 
The results are shown in Fig.~\ref{fig:ShieldOptimization}. Since the 
cosmic ray background was present in all three runs, the X-ray data runs 
(b) and (c) were corrected for the rates measured in run (a). The 
setting of the discriminator was optimized according to the following 
criteria.

\begin{enumerate}

\item The dead time produced by noise triggers should not exceed a few 
percent (horizontal lines in Fig.~\ref{fig:ShieldOptimization}). The 
Am$^{241}$/Eu$^{152}$ sources located on the optical axis (c) should 
result in CZT events with a $f_{\rm{shld}} = 1$ contribution as low as 
possible (no signal suppression).

\item For the CR (a) and Cs$^{137}$ (b) runs, the fraction of CZT 
triggers with $f_{\rm{shld}} = 1$ should be as high as possible, 
reflecting a high rejection power.

\end{enumerate}

We determined an optimal shield discriminator setting of $\rm{DACQ} = 
30$ which is indicated by the vertical line in 
Fig.~\ref{fig:ShieldOptimization}. Note, the fraction of Eu$^{152}$ 
events with $f_{\rm{shld}} = 1$ is around $25 \%$, which is probably due 
to X-rays that Compton scatter and interact with the shield and CZT~-- a 
result of not having a well collimated X-ray beam entering the 
polarimeter in this measurement.

\begin{table}[t!]

\begin{tabular}{lrr}

Data set & $\left< T_{5 \, \rm{mm}} \right>$ &  $\left< T_{2 \, \rm{mm}} \right>$ \\
 & [mHz] &  [mHz] \\
\hline \hline


\noalign{\smallskip}
\multicolumn{3}{l}{CZT initial calibration (Washington Univ.): $C_{\rm{wu1a}}^{1}$} \\
\hline

Eu$^{152}$ illumination & $7100$ & $6200$ \\
Background & $59$ & $35$ \\

\noalign{\smallskip}
\multicolumn{3}{l}{Scintillator characterization (Wash. Univ.): $C_{\rm{wu2a}}^{3}$} \\
\hline

Eu$^{152}$ on optical axis$^{\rm{*}}$ & $31$ & $8.7$ \\
Background in shield (active) & $9.9(5.4)$ & $4.9(2.0)$ \\

\noalign{\smallskip}
\multicolumn{3}{l}{CHESS synchrotron beam (Cornell Univ.): $C_{\rm{ch}}^{3}$} \\
\hline

$40 \, \rm{keV}$ X-ray beam & $1100$ & $300$ \\
Background & $13$ & $8.4$ \\

\noalign{\smallskip}
\multicolumn{3}{l}{X-Calibur/shield/InFocus (Ft.~Sumner): $C_{\rm{ft}}^{3}$} \\
\hline

X-ray source$^{\rm{*}}$ & $853$ & $190$ \\
Background in shield (active) & $8.5(3.9)$ & $4.7(1.6)$ \\

\noalign{\smallskip}
\multicolumn{3}{l}{X-Calibur flight (simulations)} \\
\hline

Crab$^{\rm{*}}$ & $1.25$ & $0.3$ \\

\end{tabular}

\caption{Average event trigger rates per CZT detector pixel for $5 \, 
\rm{mm}$ detectors $\left< T_{5 \, \rm{mm}} \right>$ and for $2 \, 
\rm{mm}$ detectors $\left< T_{2 \, \rm{mm}} \right>$. Rates are shown 
for the different measurements/environments presented in this paper (see 
Fig.~\ref{fig:X-Calibur_Configurations}). No event selection cuts are 
applied. Rates marked with a `*' are background subtracted. The 
corresponding background spectra for the different data sets are shown 
in Fig.~\ref{fig:BG_Spectra}.}

\label{tab:PixelRates}

\end{table}

\begin{figure*}[t!]
\begin{center}
\includegraphics[width=0.49\textwidth]{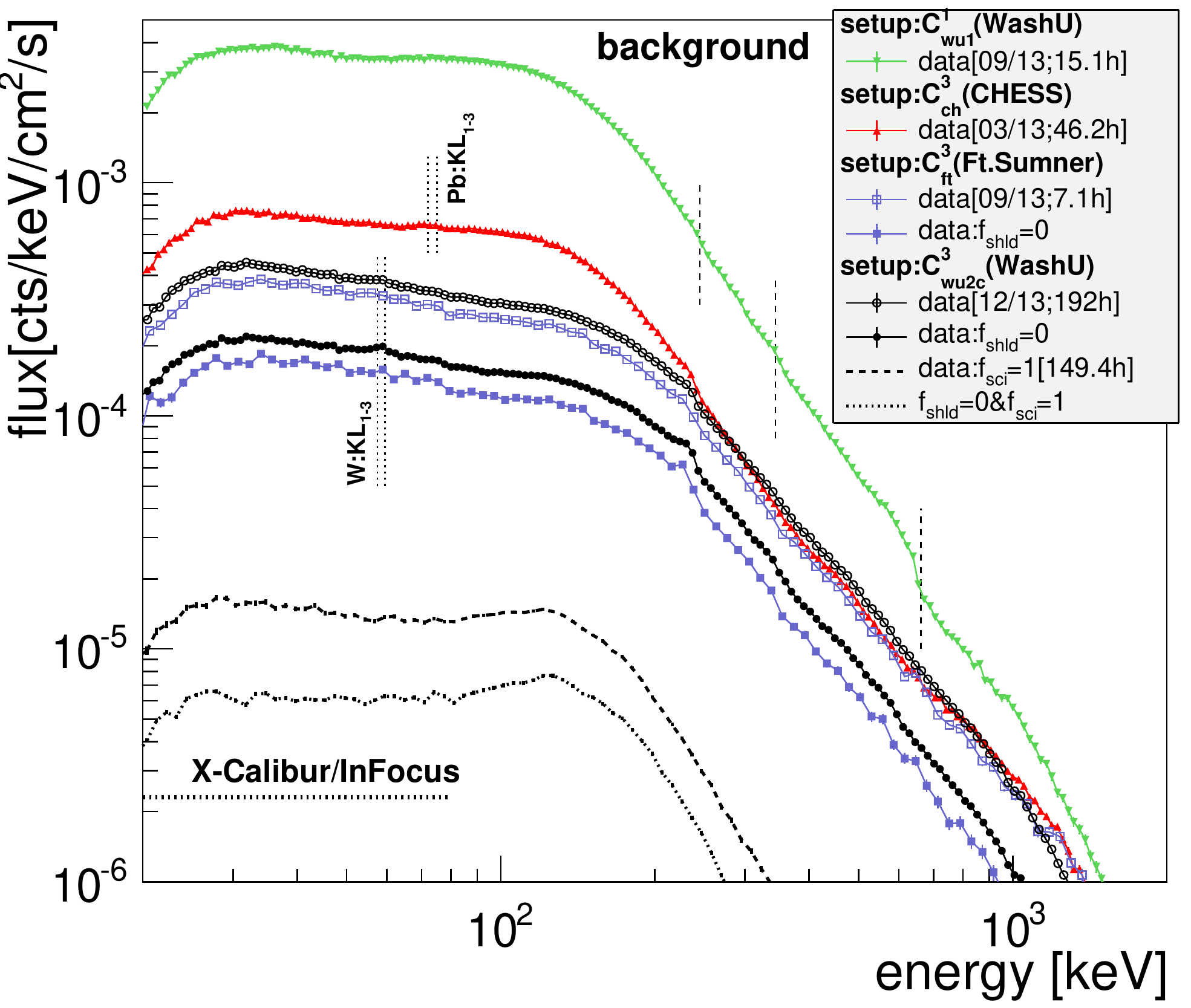}
\hfill
\includegraphics[width=0.49\textwidth]{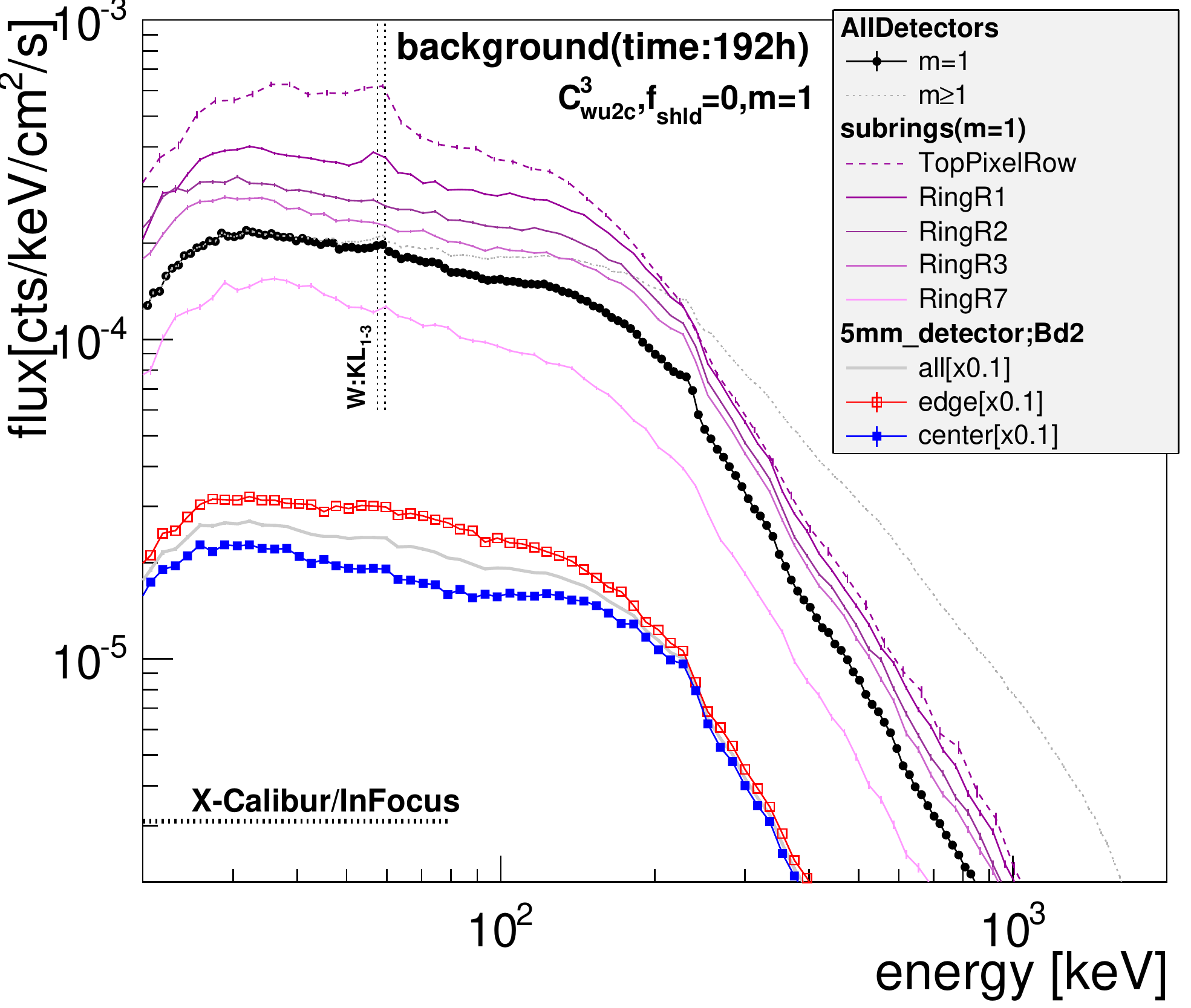}
\end{center}

\caption{Cosmic ray background spectra (events with pixel multiplicity 
$m=1$, if not mentioned otherwise). The dotted vertical lines indicate 
the ranges of the $\rm{KL}_{1-3}$ transition energies of tungsten (W) 
and lead (Pb) which show up in some of the spectra. The 
X-Calibur/\INFOCUS sensitive energy range is indicated, as well. {\bf 
Left:} Spectra measured in different environments (see 
Fig.~\ref{fig:X-Calibur_Configurations}). Date and duration of the runs 
is given in square brackets. The (passive/active) CsI shield strongly 
reduces the background. The three dashed vertical lines on the 
$C_{\rm{wu1a}}^{1}$ spectrum represent the high-energy lines of 
Eu$^{152}$ ($244.7 \, \rm{keV}$ and $344.3 \, \rm{keV}$) and Cs$^{137}$ 
($661.7 \, \rm{keV}$) for which an onset indication can be seen; the 
sources were stored in a lead bunker $\sim$$2 \, \rm{m}$ away during the 
background measurements and were probably contributing to the spectrum 
at a low level. {\bf Right:} Energy spectra for individual detector 
rings measured at Washington University in the active shield (after 
shield veto, $f_{\rm{shld}} = 0$). We also show energy spectra (scaled 
by a factor of 0.1) from different pixels of the detectors on the 
readout board {\it Bd2} to illustrate the difference between central 
pixels and pixels located on the edge of detectors, see also 
Fig.~\ref{fig:BG_Maps2D} for reference.}

\label{fig:BG_Spectra}

\end{figure*}

{\it Cosmic ray background levels.} Figure~\ref{fig:BG_Spectra} shows 
the CR energy spectra ($m=1$ pixel multiplicity) measured in the CZT 
detectors at the different locations: the laboratory at Washington 
University ($C_{\rm{wu}}$), the CHESS X-ray beam facility 
($C_{\rm{ch}}$, Sec.~\ref{subsec:CHESS}), and in Ft.~Sumner 
($C_{\rm{ft}}$, Sec.~\ref{subsec:FtSumnerData}). For reference, the 
corresponding event trigger rates ($m \geq 1$) per pixel are summarized 
in Tab.~\ref{tab:PixelRates}. Note, that the actual pixel rates can vary 
quite substantial, since different pixels see different income fluxes 
depending on the type of measurement (e.g. an X-ray beam Compton 
scattered in the scintillator leads to a strongly depth-dependent 
illumination of CZT detectors, compare with 
Fig.~\ref{fig:CHESS_2DAzimuthData}). However, it can be seen that for 
most measurements presented in this paper the background can be 
neglected. During the balloon flight, however, the expected 
source-to-background ratio will be much lower and a proper understanding 
and the background suppression will be crucial \cite{Guo2010, Guo2013}. 

Even though the background at flight altitude will be different as 
compared to the ground-based CR background, some characteristics in the 
detector response can be discussed qualitatively based on the spectra 
shown in Fig.~\ref{fig:BG_Spectra}. While $\alpha$-particles interact 
close to the surface of the detector, muons as well as primary and 
secondary high-energy gamma rays will penetrate deeper and their energy 
deposition is proportional to the detector volume. This volume-dependent 
background rate can be seen by comparing the spectra (and trigger rates) 
measured with $5 \, \rm{mm}$ versus $2 \, \rm{mm}$ detectors. Note, that 
the spectra shown in Fig.~\ref{fig:BG_Spectra} represent particle fluxes 
folded with the energy-dependent response of the CZT detectors (with an 
energy calibration derived from, and valid for, X-rays). The drop in 
event rate below $30 \, \rm{keV}$, for example, is an effect of the 
superposition of the different trigger thresholds of the pixels 
contributing to the spectrum. The dynamical energy range covered by a 
single pixel saturates around $2000 \, \rm{keV}$ (differing from pixel 
to pixel), so that the combined $m=1$ pixel spectra shown in 
Fig.~\ref{fig:BG_Spectra} drop off around this energy. Allowing events 
with $m \geq 1$ multiplicities will extend the energy range (see 
Fig.~\ref{fig:BG_Spectra}, right) which is, however, not relevant for 
the operation of X-Calibur.

\begin{figure}[t!]
\begin{center}
\includegraphics[width=0.49\textwidth]{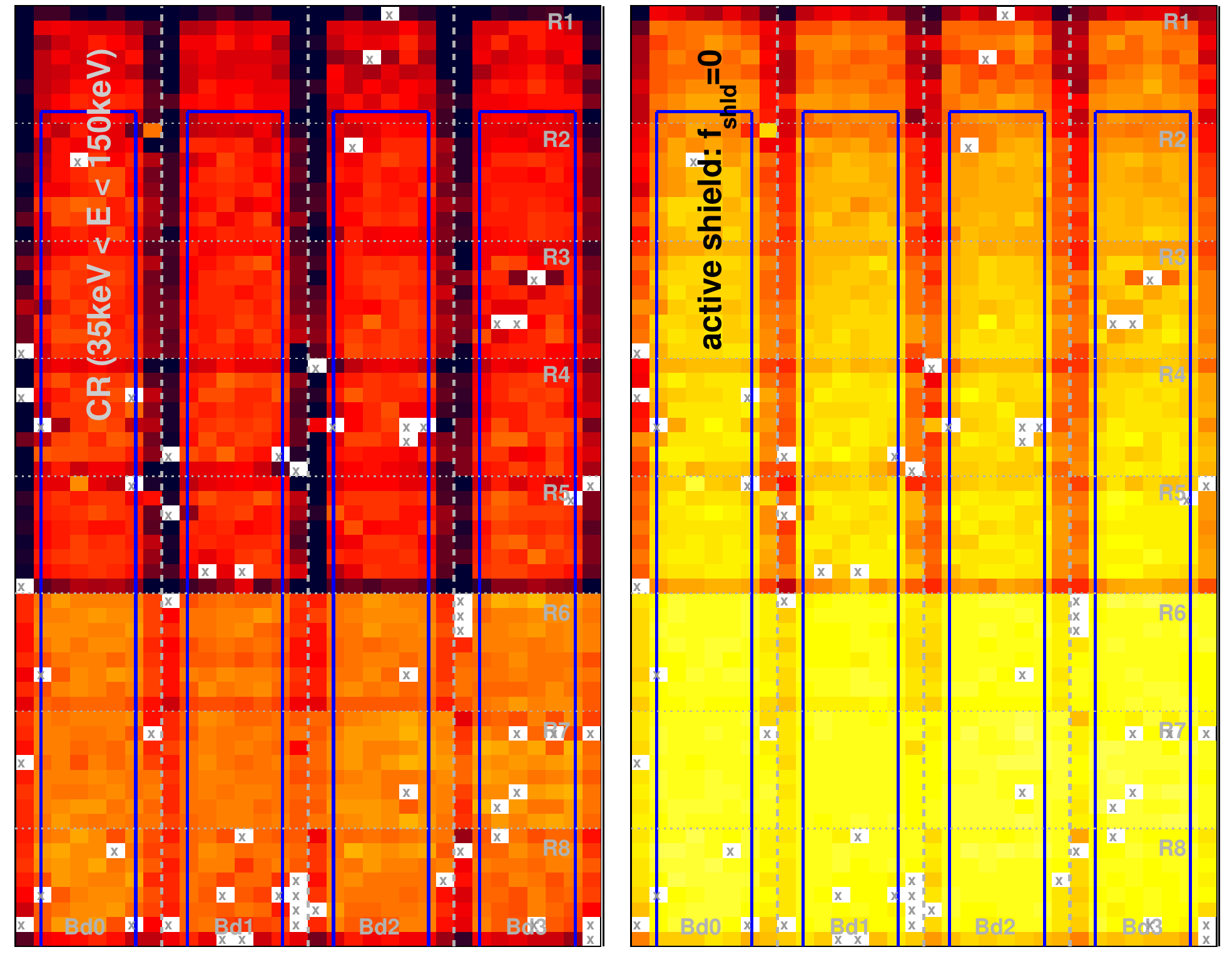} \\
\includegraphics[width=0.49\textwidth]{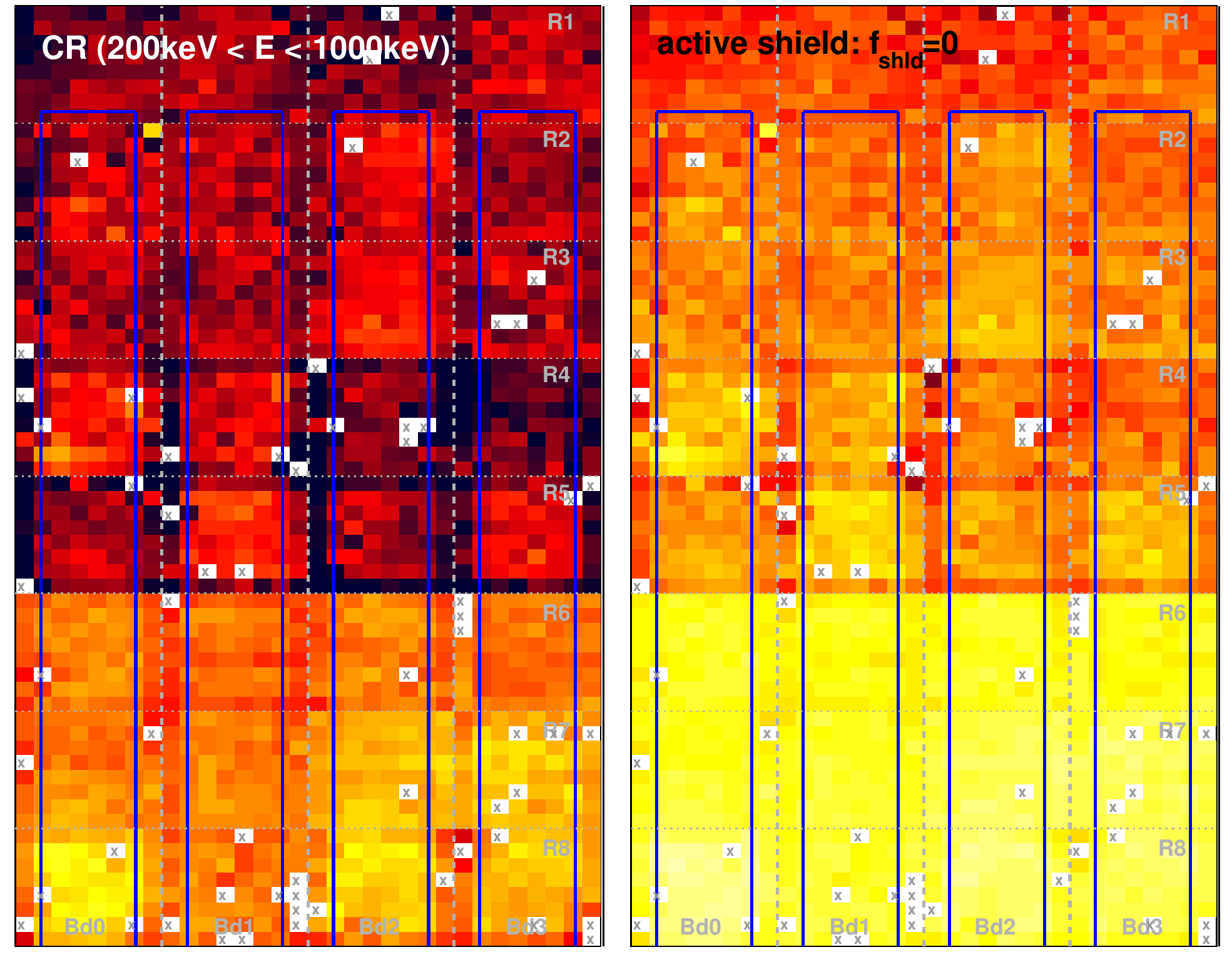}
\end{center}

\caption{2D maps of the CR count rates (pixel-by-pixel, representation 
same as in Fig.~\ref{fig:PixelThreshold}) of a $192 \, \rm{h}$ 
background run with X-Calibur being installed in the active CsI shield 
($C_{\rm{wu2c}}^{3}$ in Fig.~\ref{fig:X-Calibur_Configurations}). The 
tungsten cap is located at the top of each panel. The count rate is 
derived by integrating the corresponding energy spectra ($m=1$ events) 
in the given energy range: $35-150\, \rm{keV}$ (top) and $200-1000 \, 
\rm{keV}$ (bottom). The left panel shows the measured rate (passive 
shield rejection only) and the right panel shows the non-vetoed events 
with $f_{\rm{shld}} = 0$ (active and passive rejection). Left and right 
panels are shown with the same axis/color ranges, each.}

\label{fig:BG_Maps2D}
\end{figure}

{\it The CsI shield efficiency.} Figure~\ref{fig:BG_Maps2D} shows the 2D 
distribution of event count rates from background data taken with 
X-Calibur installed in the CsI shield at Washington University. The 
rates are shown for two different energy bands. The left panels show the 
raw rates and the right panels show the rates after rejecting events 
that triggered the active shield, only allowing non-vetoed events 
($f_{\rm{shld}} = 0$). It again becomes obvious that the $5 \, \rm{mm}$ 
detectors ({\it R1}-{\it R5}) collect more background compared to the $2 
\, \rm{mm}$ detectors ({\it R6}-{\it R8}). Furthermore, the edge pixels 
of the detector rows on the individual boards {\it Bd0-Bd3} see a higher 
background rate as compared to central pixels (likely because of the 
higher exposed surface area detecting charged particles and low energy 
X-rays). The energy-resolved difference between edge and central pixels 
can be seen in Fig.~\ref{fig:BG_Spectra}, right ($5 \, \rm{mm}$ 
detectors of boards {\it Bd2}): the central pixels show an almost two 
times lower background compared to edge pixels in the energy regime 
relevant for X-Calibur.

The background distribution after shield veto (Fig.~\ref{fig:BG_Maps2D}, 
right) exhibits a spatial gradient with more background events being 
detected closer to the front side of the experiment, as that side is 
only shielded by the passive tungsten cap (see 
Fig.~\ref{fig:ActiveShieldAndTelescope}, left). In particular, the 
front-side (top) pixel row of ring {\it R1}, with its exposed detector 
side walls, suffers strongly from primary radiation leaking through the 
tungsten, as well as secondary particles being produced in the tungsten. 
Here, spectral signatures of the tungsten KL transitions can be 
identified in the measured spectra shown in Fig.~\ref{fig:BG_Spectra} 
(right). Since this single pixel row is not crucial for the polarimetry 
sensitivity it can be excluded from the analysis.

The passive/active rejection efficiency of the CsI shield becomes 
obvious if comparing the different background spectra in 
Fig.~\ref{fig:BG_Spectra}, left. Moving the polarimeter from its copper 
housing (used in the initial test measurements, $C_{\rm{wu1}}^{1}$) into 
the CsI shield ($C_{\rm{wu2c}}^{3}$) reduces the background by roughly 
one order of magnitude due to passive shielding. Applying the 
$f_{\rm{shld}} = 0$ veto from the active shield leads to an additional 
background rejection by a factor of $\sim$$2$. Note, that the 
recombination of the tungsten and lead $\rm{KL}_{1,2,3}$ transition 
energies\footnote{http://www.nist.gov/pml/data/xraytrans/} (probably 
activated by CRs) can be identified in the background spectra. This 
feature is most prominent in the spectrum of ring {\it R1} 
(Fig.~\ref{fig:BG_Spectra}, right) which is located closest to the 
tungsten cap.

It should be noted that an additional cut on the scintillator rod 
coincidence flag $f_{\rm{sci}} = 1$ reduces the background by another 
$1.5$ orders of magnitude (Fig.~\ref{fig:BG_Spectra}, left)~-- strongly 
rejecting events that did not enter the polarimeter along the optical 
axis and interacted in the scintillator, see 
Sec.~\ref{subsec:ScintillatorEfficiency}.

\subsection{Compton-scattering and Scintillator Trigger Efficiency} \label{subsec:ScintillatorEfficiency}

\begin{figure}[th!]
\begin{center}
\includegraphics[width=0.47\textwidth]{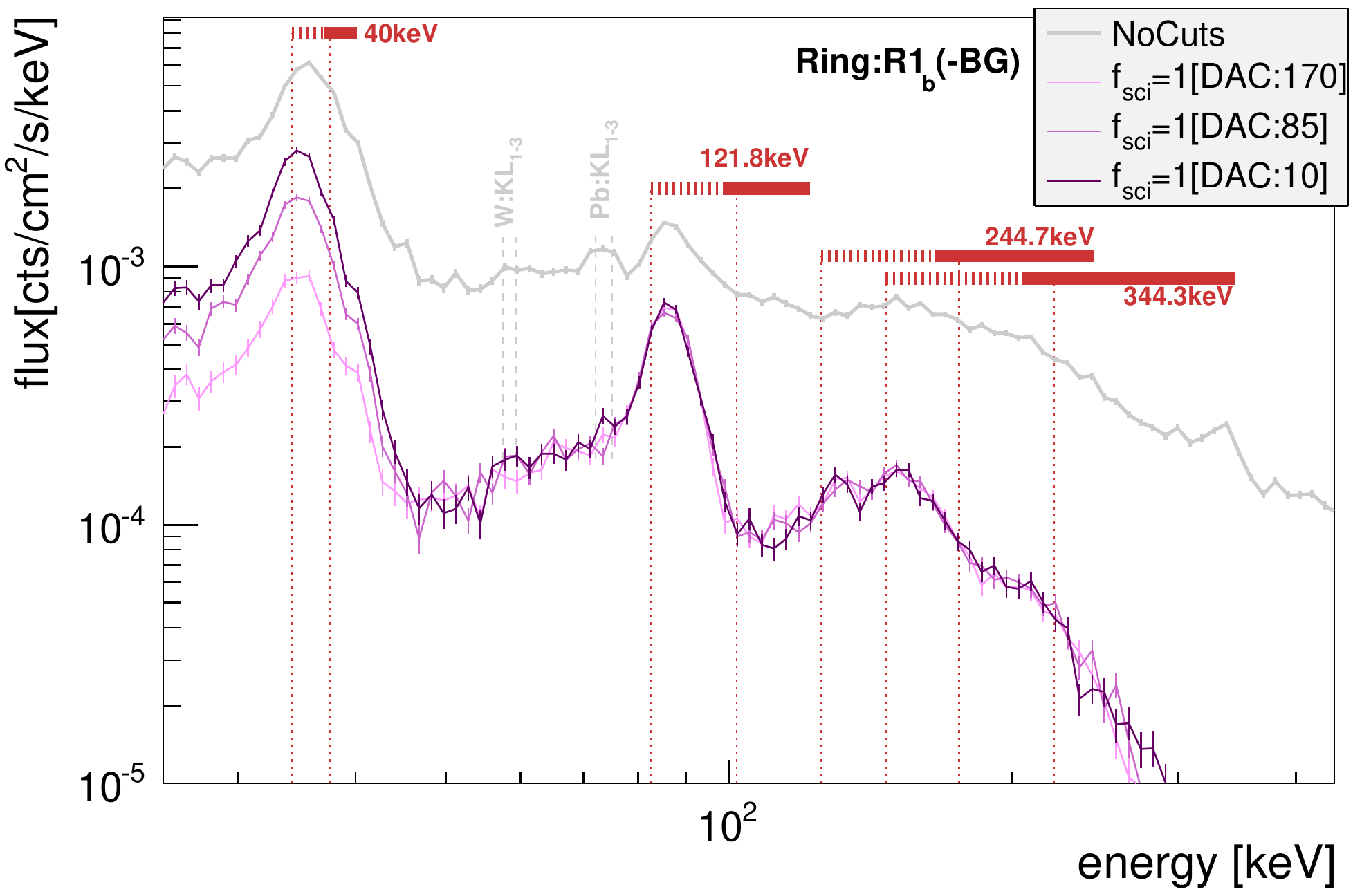} \\
\includegraphics[width=0.47\textwidth]{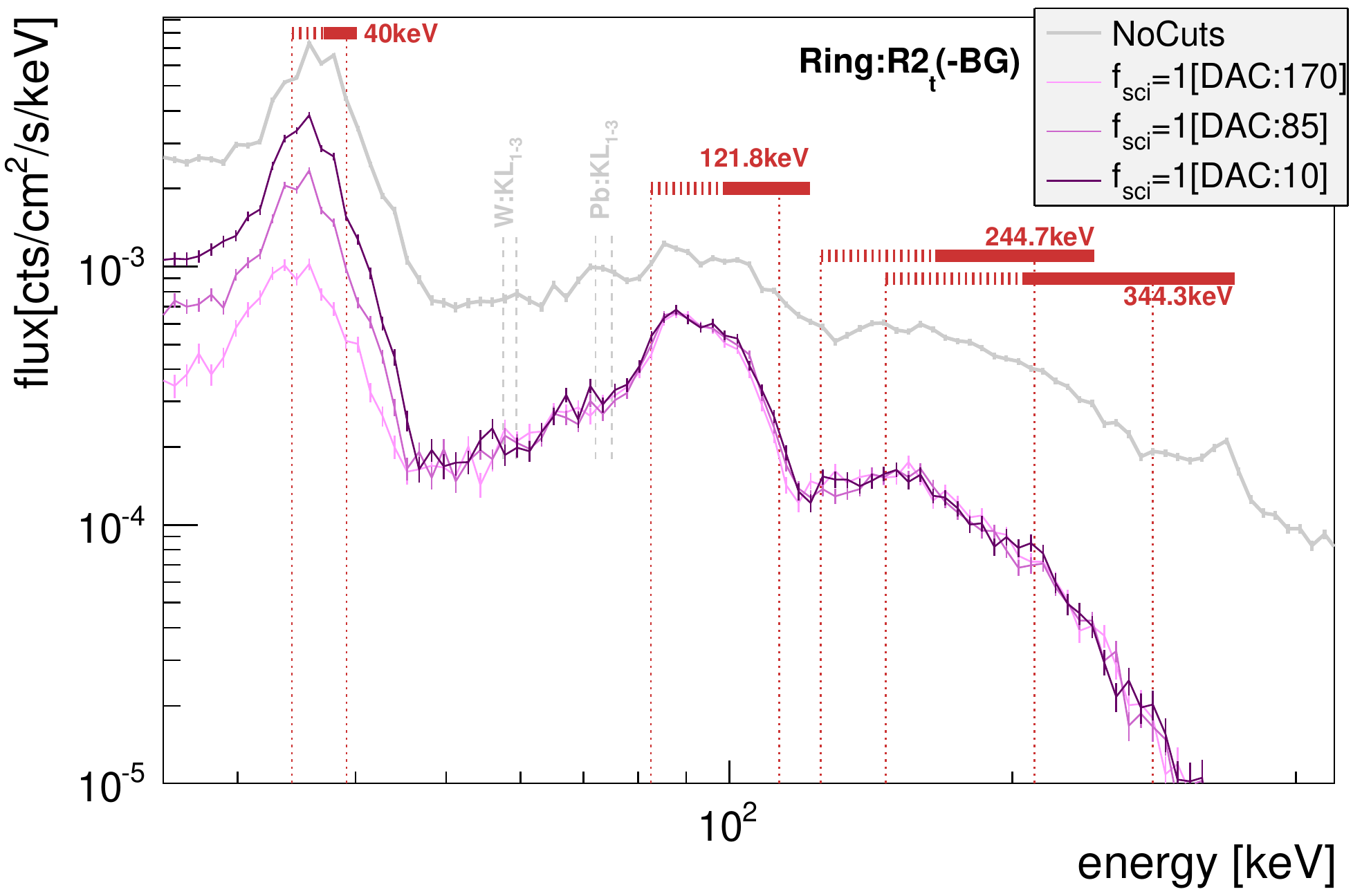} \\
\includegraphics[width=0.47\textwidth]{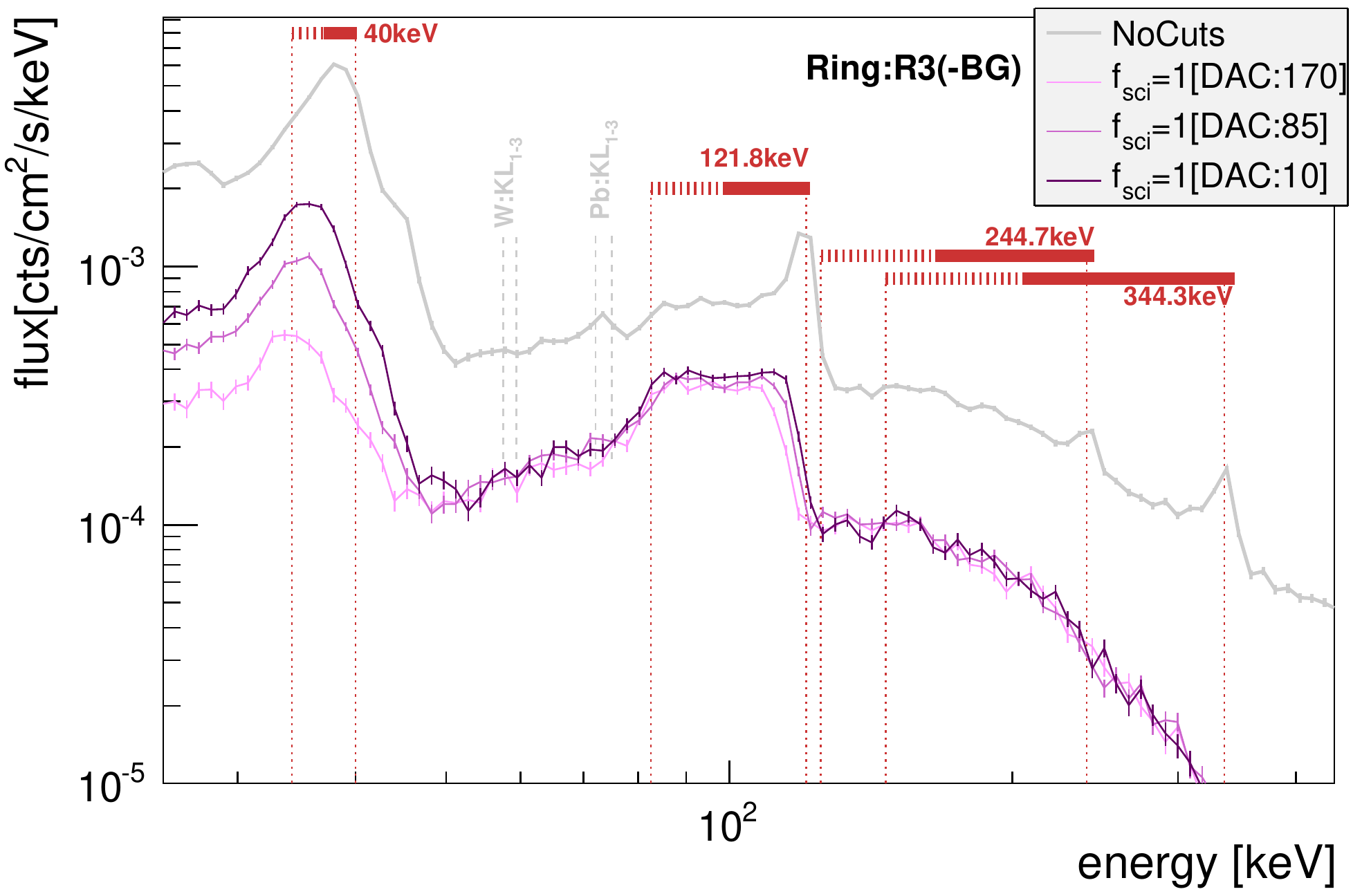}
\end{center}

\caption{A lead-collimated Eu$^{152}$ source enters the polarimeter 
along the optical axis and Compton-scatterers in the scintillator rod 
($C_{\rm{wu2a}}^{3}$ in Fig.~\ref{fig:X-Calibur_Configurations}). The 
resulting CZT spectra ($m=1$, background subtracted) are shown for 
different rings. The horizontal lines indicate the complete range of 
possible energies $E_{\rm{czt}}$ after Compton-scattering (dotted: 
$180-90^{\circ}$, solid: $90-0^{\circ}$). The attached vertical dotted 
lines indicate the sub range for the particular ring~-- given its 
geometrical coverage of the scintillator. Spectra are measured with 
different scintillator trigger thresholds settings (Digital to Analog 
Converter, DAC) and are compared to the spectra without coincidence 
requirement (no cuts). The lead collimator prevents direct source hits 
(rings {\it R1-R3}), but scattered X-rays and/or activation of the 
$\rm{KL}_{1-3}$ transitions in tungsten (W) and lead (Pb) partly 
contaminate the spectra.}

\label{fig:Scintillator_Spectra}
\end{figure}

The central scintillator rod of the polarimeter (Fig.~\ref{fig:Design}) 
acts as a Compton-scatterer for the incoming X-ray beam. At the same 
time, it produces scintillation light that is read out by a PMT. In 
order to characterize the Compton-scattering and the trigger response of 
the scintillator, a collimated Eu$^{152}$ source was placed at the 
polarimeter entrance, illuminating the scintillator along the optical 
axis (see setup $C_{\rm{wu2a}}^{3}$ in 
Fig.~\ref{fig:X-Calibur_Configurations}). The Compton-scattered X-rays 
are recorded by the surrounding CZT detector rings. An X-ray enters the 
polarimeter with a certain energy $E_{\rm{line}}$, deposits the energy 
$\Delta E_{\rm{sci}}$ in the scintillator and is absorbed in a CZT 
detector with its post-scattering energy of $E_{\rm{czt}} = 
E_{\rm{line}} - \Delta E_{\rm{sci}}$. The amount of scintillation light 
can be assumed to be proportional to $\Delta E_{\rm{sci}}$.

Data were taken with different settings of the scintillator/PMT 
discriminator threshold (DAC). As can be seen in 
Tab.~\ref{tab:PixelRates}, the background level is not negligible in 
this measurement, so that background spectra without a X-ray source were 
recorded and subtracted. It should be noted, however, that X-rays 
scattered in the lead collimator and activation/recombination of the 
$\rm{KL}_{1,2,3}$ electron transitions in the lead collimator and the 
tungsten cap lead to a slight contamination of the measured spectra. 
These corresponding spectral features do not cancel out after CR 
background subtraction since they are (indirectly) introduced by the 
Eu$^{152}$ source itself. Since those features will over-proportionally 
affect the data without the $f_{\rm{sci}}=1$ coincidence (X-ray paths 
that do not necessarily cross the scintillator), this may lead to an 
underestimation of the trigger efficiency (which in the ideal case would 
only compare Compton-scattered events in the scintillator with and 
without the $f_{\rm{sci}} = 1$ trigger coincidence). The scintillator 
trigger efficiency for different energies $\Delta E_{\rm{sci}}$ is 
derived as follows.

{\it Compton spectra.} The resulting Compton-scattered energy spectra 
$E_{\rm{czt}}$ are shown in Fig.~\ref{fig:Scintillator_Spectra} for 
different CZT detector rings. Data are shown with and without the 
scintillator coincidence requirement. The energy $\Delta E_{\rm{sci}}$ 
deposited in the scintillator depends on the scattering kinematics, 
including the primary energy of the X-ray $E_{\rm{line}}$ and its 
scattering angle. The theoretical range of energies after 
$0-180^{\circ}$ Compton scattering is indicated by horizontal lines for 
the four main Eu$^{152}$ energy lines. Due to geometrical reasons, some 
of the detector rings can only detect X-rays originating from a limited 
range of Compton-scattering angles. With the scintillator starting at 
ring {\it R2} (Fig.~\ref{fig:Design}, left), the sub rings {\it 
R1$_{\rm{t}}$}, and {\it R1$_{\rm{b}}$}, for example, can only detect 
back-scattered X-rays which deposit/loose the highest amount of energy 
$\Delta E_{\rm{sci}}$ in the scintillator~-- leading to a higher trigger 
probability. For a given incoming energy $E_{\rm{line}}$, the range of 
possible scattering angles translates into a range of possible 
Compton-scattered energies $E_{\rm{czt}}$. These ring-dependent sub 
ranges in scattered energy are indicated by the vertical lines in 
Fig.~\ref{fig:Scintillator_Spectra}. Note, however, that the scattering 
angles/energies will not be equally distributed within these boundaries. 
One can qualitatively discuss the X-Calibur response to 
Compton-scattering for the different scattered energy lines.

\begin{itemlist}

\item $E_{\rm{line}} = 40 \, \rm{keV}$: The Compton-scattered energy 
range of the Eu$^{152}$ line doublet ($39.52\,\rm{keV}$ and 
$40.12\,\rm{keV}$) only covers $E_{\rm{czt}} = 34.2 - 40.1\,\rm{keV}$ at 
an energy resolution of $\Delta E_{40} \simeq 4 \, \rm{keV}$. This makes 
it difficult to resolve structures in the scattered energy spectrum. 
However, the geometrical coverage of ring {\it R1$_{\rm{b}}$} 
constraints the energy deposition in the scintillator to $\Delta 
E_{\rm{sci}} = 2.4-5.4 \, \rm{keV}$ since only back-scattering angles of 
$80-180^{\circ}$ are possible (top panel in 
Fig.~\ref{fig:Scintillator_Spectra}). A clear improvement in trigger 
efficiency is obvious when lowering the scintillator DAC threshold from 
170 to 10.

\item $E_{\rm{line}} = 121.8 \, \rm{keV}$: This well-defined line at 
higher energy leads to a broader Compton-continuum which can be tested 
at different energies $\Delta E_{\rm{sci}}$. While ring {\it 
R1$_{\rm{b}}$} again only sees the back-scattered X-rays, the rings 
further down cover a broad continuum range. The trigger efficiency for 
the back-scattering of the $121.8 \, \rm{keV}$ line ($\Delta 
E_{\rm{sci}} = 23-39 \, \rm{keV}$) does not seem to depend on the 
scintillator DAC setting and is therefore already at its maximum. In the 
close-to forward scattering regime ($\Delta E_{\rm{sci}}$ a few keV) in 
ring {\it R3} (bottom panel in Fig.~\ref{fig:Scintillator_Spectra}) a 
lower DAC threshold leads to a broader Compton continuum in the 
$f_{\rm{sci}} = 1$ data, reflecting the increase in trigger efficiency. 
Starting at ring {\it R3}, however, the effective thickness of the lead 
collimator at the polarimeter entrance is not high enough to fully 
absorb all X-rays; this leads to some direct detector hits in the raw 
energy spectrum~-- and therefore an underestimation of the trigger 
efficiency at the corresponding value of $\Delta E_{\rm{sci}}$.

\item $E_{\rm{line}} = 244.7 \, \rm{keV}$ and $344.3 \, \rm{keV}$: These 
two high-energy (but lower intensity) lines lead to partially 
overlapping Compton-continua. This makes it difficult to study the 
response in more detail. It can also be seen, that $344.3 \, \rm{keV}$ 
direct CZT hits are not fully prevented by the lead collimator (peak in 
raw spectra).

\end{itemlist}

\begin{figure}[t!]
\begin{center}
\includegraphics[width=0.49\textwidth]{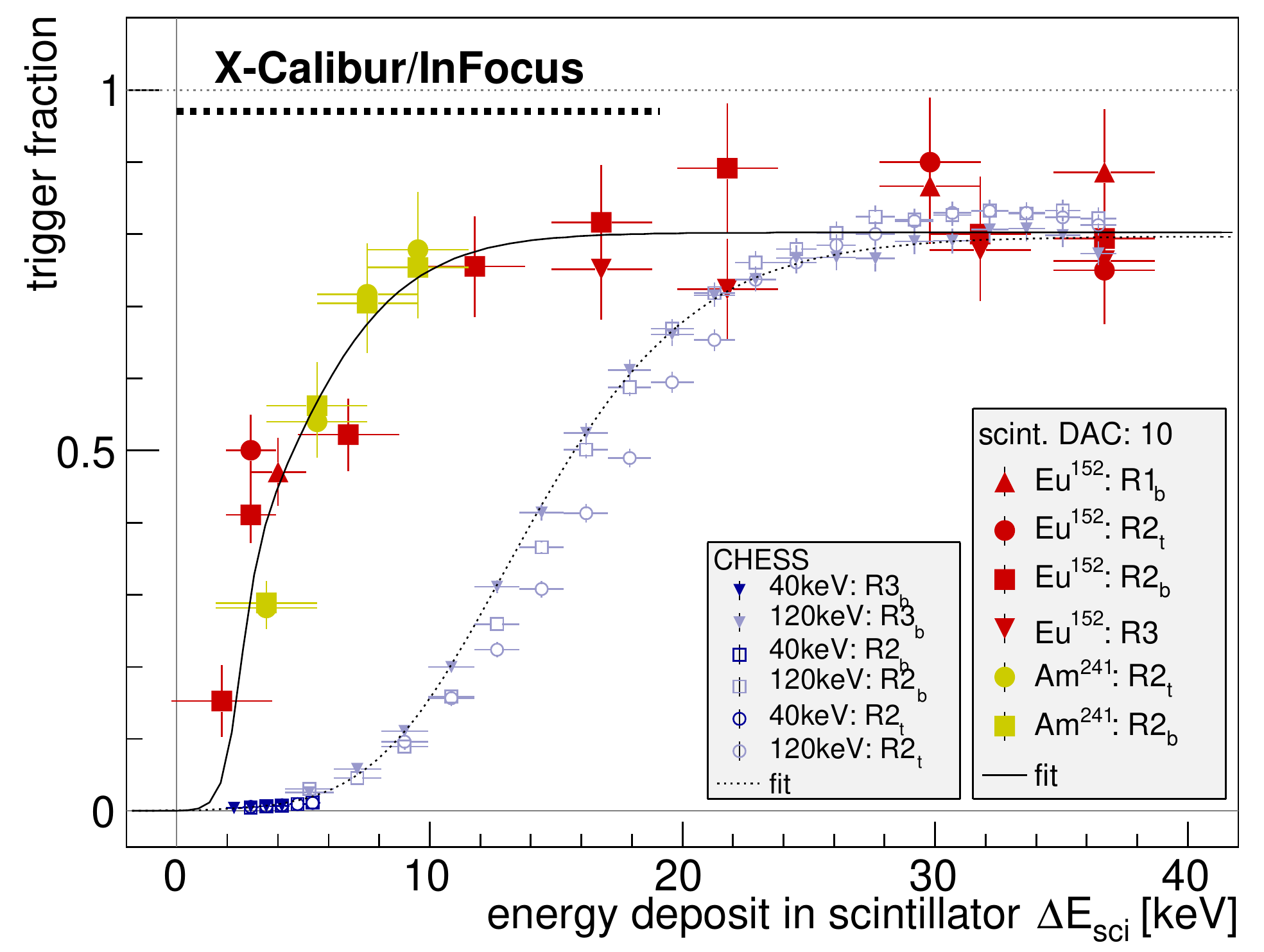}
\end{center}

\caption{Fraction of CZT events with a scintillator coincidence trigger 
($f_{\rm{sci}}=1$) as a function of energy deposition $\Delta 
E_{\rm{sci}}$. Data were taken with Am$^{241}$, Eu$^{152}$, and at the 
CHESS beam (Sec.~\ref{subsec:CHESS}, however with a much higher 
scintillator threshold). The Compton-scattered continua (above 
background, see for example Fig.~\ref{fig:Scintillator_Spectra}) were 
used to derive the trigger efficiency for different $\Delta 
E_{\rm{sci}}$. The distribution was fitted by a function shown in 
Eq.~(\ref{eqn:ScintTriggEffFit}). The dashed horizontal line indicates 
the range of scintillator energy depositions for incoming X-rays in the 
$20-80 \, \rm{keV}$ band.}

\label{fig:Scintillator_TriggerEfficiency}
\end{figure}

{\it Scintillator efficiency.} The scintillator trigger efficiency as a 
function of $\Delta E_{\rm{sci}} = E_{\rm{line}} - E_{\rm{czt}}$ was 
estimated from the flux ratio at energy $E_{\rm{czt}}$ with and without 
the $f_{\rm{sci}} = 1$ coincidence requirement (after continuum 
subtraction). This can be done independently for the different line 
energies and for different detector rings. Although the different 
Eu$^{152}$ energies are ideal to test different values of $\Delta 
E_{\rm{sci}}$, the reflection and re-processing of the high energy lines 
contaminates the raw spectrum hampering the quantitative determination 
of the trigger efficiency (underestimation). Therefore, additional data 
were taken with an Am$^{241}$ source which has a line at $59.5 \, 
\rm{keV}$, but no significant emission above. The trigger efficiencies 
determined from the different source lines and detector rings are 
summarized in Fig.~\ref{fig:Scintillator_TriggerEfficiency} for the 
optimized discriminator threshold of DAC=10. The trigger distribution 
was fitted by a function

\begin{equation}
f(\Delta E_{\rm{sci}}) = \frac{a}{1 + e^{-b (\Delta E_{\rm{sci}}-E_{0})} + e^{-c 
(\Delta E_{\rm{sci}}-E_{0})}}.
\label{eqn:ScintTriggEffFit}
\end{equation}

For the optimized setting of DAC=10, the trigger efficiency reaches 
$50\%$ at $\Delta E_{\rm{sci}} \simeq 4.6 \, \rm{keV}$. Note, that the 
trigger fraction does not level out at $a = 1$ ($100 \%$). This is 
potentially a result of the contamination of the spectrum (direct 
detector hits)~-- caused by the non-ideal measurement setup that leads 
to a systematic underestimation of the efficiency. For the simulations 
in Sec.~\ref{sec:Simulations}, we therefore assumed the same trigger 
function, however with a value of $a=1$. Note, that we currently do not 
model any depth-dependence of $f(\Delta E_{\rm{sci}})$ which becomes 
important when looking at the different detector rings. In a detailed 
{\it Monte Carlo} study, \citet{PogoScintSims} simulate the trigger 
efficiency of a similar scintillator material (EJ-204) and find a 
threshold of $2-3 \, \rm{keV}$. The authors also find a $13 \%$ drop in 
light yield along a $20 \, \rm{cm}$ long scintillator rod (from close to 
the PMT towards the distant end).

The raw scintillator trigger rate (independent of CZT event triggers) of 
the optimized threshold setting of DAC=10 was measured to be only a few 
hundred Hz~-- a further reduction in trigger threshold by updating the 
signal amplifier seems possible and is work in progress.

A cleaner setup to measure the scintillator response is the 
well-collimated X-ray beam at CHESS (see Sec.~\ref{subsec:CHESS}). The 
results are also shown in Fig.~\ref{fig:Scintillator_TriggerEfficiency} 
for a non-polarized beam at $40 \, \rm{keV}$ and $120 \, \rm{keV}$, 
respectively. However, the amplifier for the PMT reading the 
scintillator was not optimized at the time of the CHESS measurements, so 
that the $50\%$ trigger probability is only reached around $\Delta 
E_{\rm{sci}} \simeq 15 \, \rm{keV}$. The corresponding fitted trigger 
efficiency (again with the $a=1$ assumption)\footnote{Also the CHESS 
measurements suffered from indirect contamination due to beam absorber 
foils, see Sec.~\ref{subsec:CHESS}.} was fed into the simulations that 
were done for the CHESS measurements discussed in 
Sec.~\ref{subsec:CHESS}.

\subsection{Summary of the Instrument Characteristics}

The rejection power of the active shield was studied in the laboratory. 
It reduces the background by more than one order of magnitude in the 
energy range relevant for X-Calibur.

When we require a scintillator trigger ($f_{\rm{sci}} = 1$), the 
effective energy threshold of the polarimeter increases to $\simeq 25 \, 
\rm{keV}$ (with the current scintillator threshold of $\sim$$5 \, 
\rm{keV}$). We are working on a further reduction of the scintillator 
threshold \cite{Fabiani2013}. Above an energy of $30 \, \rm{keV}$, we 
can detect a high fraction of events with a coincident scintillator 
signal. For ground-based background measurements, the scintillator 
coincidence reduces the background by another factor of $\sim$$30$. On 
upcoming balloon flights, we will use events with (for best high-energy 
signal to noise ratio) and without (for best low-energy response) 
scintillator coincidences.

\section{X-Calibur: Polarization Measurements} \label{sec:XCLB_PolarimetryMeasurements}

This section discusses measurements performed with the fully assembled 
and calibrated X-Calibur polarimeter. Measurements of non-polarized and 
polarized X-ray beams at the CHESS facility at Cornell University are 
described in Sec.~\ref{subsec:CHESS} and are compared to simulations. 
The results illustrate the functionality of X-Calibur as an X-ray 
polarimeter. Section~\ref{subsec:FtSumnerData} describes X-Calibur 
measurements that were taken with the fully integrated 
X-Calibur/\INFOCUS X-ray telescope in Ft.~Sumner during a test 
integration of a flight-ready balloon setup. 

Systematic effects and asymmetries in the detector response can, under 
certain circumstances, cause measurements of aparent polatization 
fractions, even if the measured X-ray beam itself is non-polarized. It 
is therefore crucial to understand and control these kinds of systematic 
effects in order to correctly study the polarization properties of 
astrophysical sources. The systematic effects (and corrections thereof) 
caused by a X-ray beam hitting the polarimeter off-center are discussed 
in Sec.~\ref{subsec:Systematics}.

\subsection{Polarized X-rays from the CHESS Beam} \label{subsec:CHESS}

In order to measure the response to a polarized X-ray beam, the 
X-Calibur polarimeter was operated at the Cornell High Energy 
Synchrotron Source (CHESS)\footnote{http://www.chess.cornell.edu/} for 
one week in March 2013. CHESS provides a highly collimated and highly 
polarized beam of synchrotron X-rays. Using Bragg reflection from a 
2-bounce silicon (220) monochromator, a $40 \, \rm{keV}$ beam was 
generated with the 2$^{\rm{nd}}$/3$^{\rm{rd}}$ harmonics at $80 / 120 \, 
\rm{keV}$, respectively. The measurements were performed in hutch {\it 
C1}. The polarimeter was mounted on an adjustable X/Y/Z stage table with 
the scintillator being aligned with the X-ray beam (using X-ray 
fluorescence paper). The setup, referred to as $C_{\rm{ch}}$, is shown 
in Fig.~\ref{fig:X-Calibur_Configurations}. X-Calibur was mounted in a 
fixture that allowed us to rotate the polarimeter around the 
optical/beam axis, in order to test the response at different 
orientations between the polarization plane and the detector. The 
rotation angle is referred to as $\alpha$ with detector board {\it Bd0} 
located at the top position at $\alpha = 0^{\circ}$ (see $C^{3}$ in the 
top panel of Fig.~\ref{fig:X-Calibur_Configurations}). Looking into the 
beam, a positive angle $\alpha$ corresponds to a counter clock-wise 
rotation. The accuracy of setting the angle was estimated to be $\Delta 
\alpha \simeq 2^{\circ}$. It should be noted that we did not use a high 
precision rotation mechanism (as will be used during the balloon 
flight). Therefore, an $\alpha$ dependent mis-alignment/tilt of the 
scintillator during the measurements cannot be excluded.

{\it The CHESS X-ray flux.} A total of 30 synchrotron-emitting electron 
bunches cycle the CHESS accelerator ring. The relative intensity of the 
X-ray beam entering the {\it C1} hutch is controlled by a system of 
slits, as well as the tunable orientation of the Bragg reflection 
crystals. The intensity, however, varies with (i) decaying electron 
population in the ring during a run (re-charged every $\sim$$2 \, 
\rm{h}$), (ii) local heat built-up on the Bragg crystals, and (iii) the 
position stability of the electron beam in the ring. The intensity of 
the monochromatic beam entering the hutch was monitored using an argon 
ion chamber ($6 \, \rm{cm}$ in length along the beam, set to a counter 
range of $10^{-8} \, \rm{A}/\rm{V}$). The absolute flux calibration of 
X-Calibur was not a major objective of the measurements; therefore, we 
did not setup a system to automatically stabilize the beam intensity 
entering the hutch~-- but rather kept it manually in the ballpark of 
$1-2 \cdot 10^{8} \, \rm{cts}/\rm{s}$ (ion chamber) throughout the 
measurements and logged the average intensity for each data run. 
Absorber foils placed behind the ion chamber further reduced the beam 
intensity by several orders of magnitude. This brought the X-ray flux 
hitting the polarimeter down to the level of $1-2 \, \rm{kHz}$, a regime 
that can be well handled by the readout electronics of the polarimeter.

{\it Configurations.} An aluminum collimator plate of thickness $0.5''$ 
was placed directly in front of the X-Calibur polarimeter with an 
entrance hole of $0.25''$ in diameter. The CZT detector configuration 
used for the CHESS measurements is listed in Tab.~\ref{tab:Detectors}. 
CHESS data presented in this paper were taken with three configurations 
(variations of $C_{\rm{ch}}$ as depicted in 
Fig.~\ref{fig:X-Calibur_Configurations}).

\begin{itemlist}

\item {\bf $C_{\rm{ch1}}$}: the absorber foils consist of platinum (Pt), 
a $125 \, \mu\rm{m}$ layer placed directly after the ion chamber and a 
stack of four $90 \, \mu\rm{m}$ layers placed between the ion chamber 
and X-Calibur. This setup is optimized for a band-pass at $40 \, 
\rm{keV}$.


\item{\bf $C_{\rm{ch3}}$}: One layer of $125 \, \mu\rm{m}$ Pt absorber 
foil is placed directly behind (downstream) the ion chamber, and a $1.27 
\, \rm{mm}$ lead foil (Pb) is placed between the ion chamber and 
X-Calibur. This setup is optimized for a band-pass at $80 \, \rm{keV}$.

\item{\bf $C_{\rm{ch4}}$}: Same as $C_{\rm{ch3}}$, but the ion chamber 
was removed from the beam path (in order to extend the range of beam 
offsets without interfering with the housing of the ion chamber).

\end{itemlist}

X-Calibur detects individual events, but only the ASIC that triggered 
the event (corresponding to $1/64$ of the polarimeter) is dead during 
the read-out which takes $\sim$$130 \,\mu\rm{s}$. The dead-time during 
data runs is therefore usually not higher than $5\%$. In order to 
guarantee a homogeneous data set, some additional pixels (as compared to 
the calibration runs) were excluded from the analysis which had high 
thresholds or were too noisy~-- even though the azimuthal acceptance 
(see Sec.~\ref{sec:Analysis}) is in general capable of correcting for 
most of these effects. Only events with multiplicity $m=1$ pixels were 
selected. X-Calibur was not operated in its active CsI shield, and the 
scintillator PMT discriminator was not yet optimized at the time of the 
CHESS measurements.

{\it Data runs.} A series of measurements was taken with configuration 
$C_{\rm{ch1}}$ (optimized for $40 \, \rm{keV}$) and $C_{\rm{ch3,4}}$ 
(optimized for $80 \, \rm{keV}$). Data were taken with different 
polarimeter orientations $\alpha$. The $C_{\rm{ch1}}$ data runs span a 
range of $\alpha \in [-90; +90]^{\circ}$ in steps of $10^{\circ}$ with 
$4 \, \rm{Mio}$ events per orientation. A non-polarized beam was 
`generated' by the superposition of data from two perpendicular 
orientations. A higher number of events were taken for these sets: $20 
\, \rm{Mio}$ events for the $\alpha = 0/-90^{\circ}$ pair, and $10 \, 
\rm{Mio}$ events for the $\alpha = \pm 45^{\circ}$ reference pair. The 
$C_{\rm{ch3,4}}$ series scanned a range of $\alpha \in 
[-90;+10]^{\circ}$ with a non-polarized beam composed from $10 \, 
\rm{Mio}$ events derived from the $\alpha = 0/-90^{\circ}$ orientations.

\begin{figure}[t!]
\begin{center}
\includegraphics[width=0.49\textwidth]{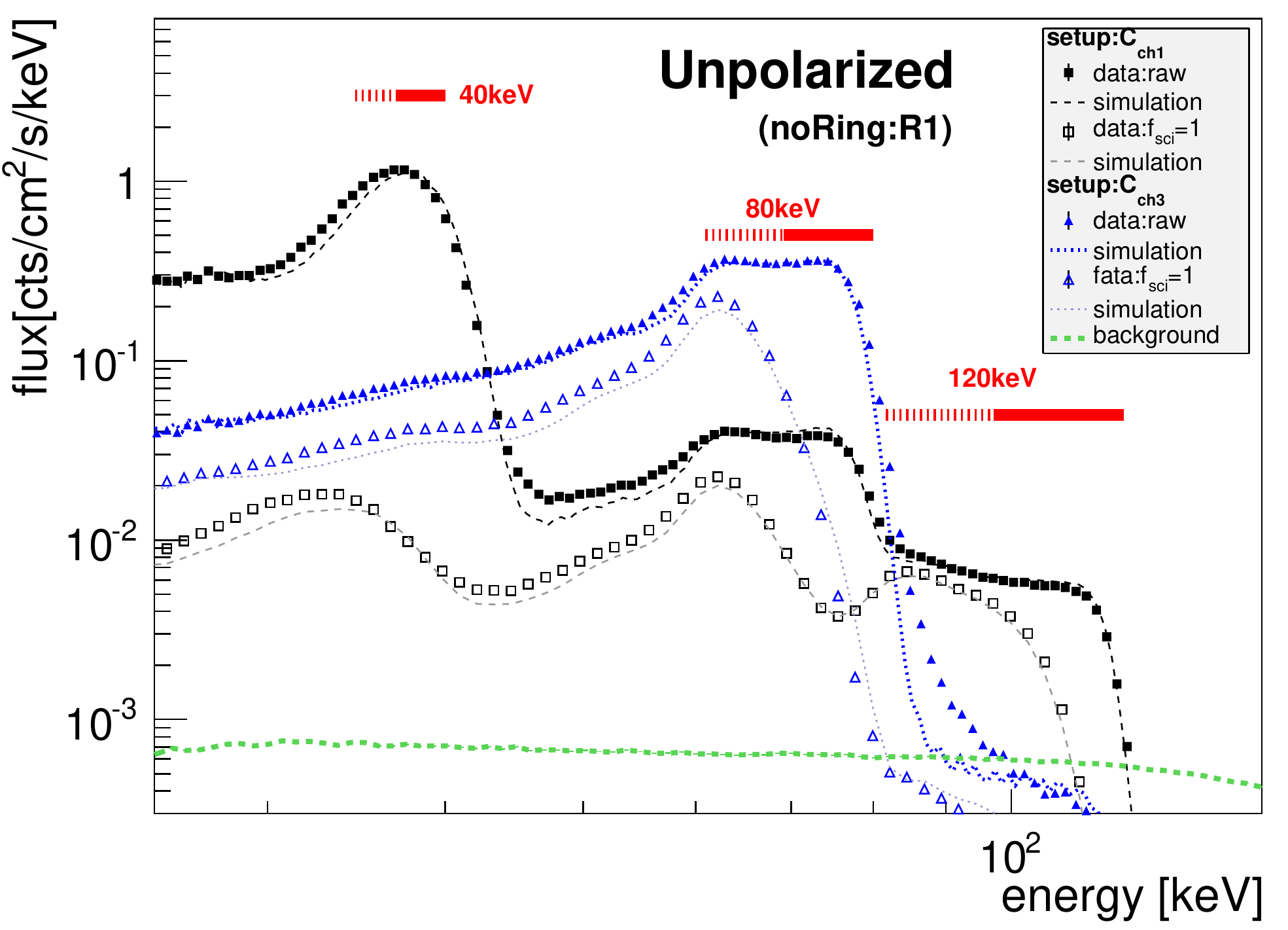}
\end{center}

\caption{Energy spectra of the Compton-scattered non-polarized CHESS 
X-ray beam (superposition of perpendicular polarization planes). The 
measurements were performed with the two configurations $C_{\rm{ch1}}$ 
and $C_{\rm{ch3}}$ (see text and 
Fig.~\ref{fig:X-Calibur_Configurations}). Shown are the background 
subtracted integral spectra of all detectors~-- only ring R1 was 
excluded due to flux contamination (see text and 
Fig.~\ref{fig:CHESS_ComtponSpectra_UnPol}, right, for the reason). The 
background level is shown for reference. The horizontal lines represent 
the nominal energies of the three harmonics for (left to right): $180 - 
90^{\circ}$ Compton-scattering (dotted), and $90 - 0^{\circ}$ 
Compton-scattering (solid). Spectra are shown with and without the 
$f_{\rm{sci}} = 1$ scintillator coincidence requirement. Simulated 
spectra are shown, as well.}

\label{fig:CHESS_ComtponSpectra_AllDetector_UnPol}
\end{figure}

\begin{figure*}[t!]
\begin{center}
\includegraphics[width=0.49\textwidth]{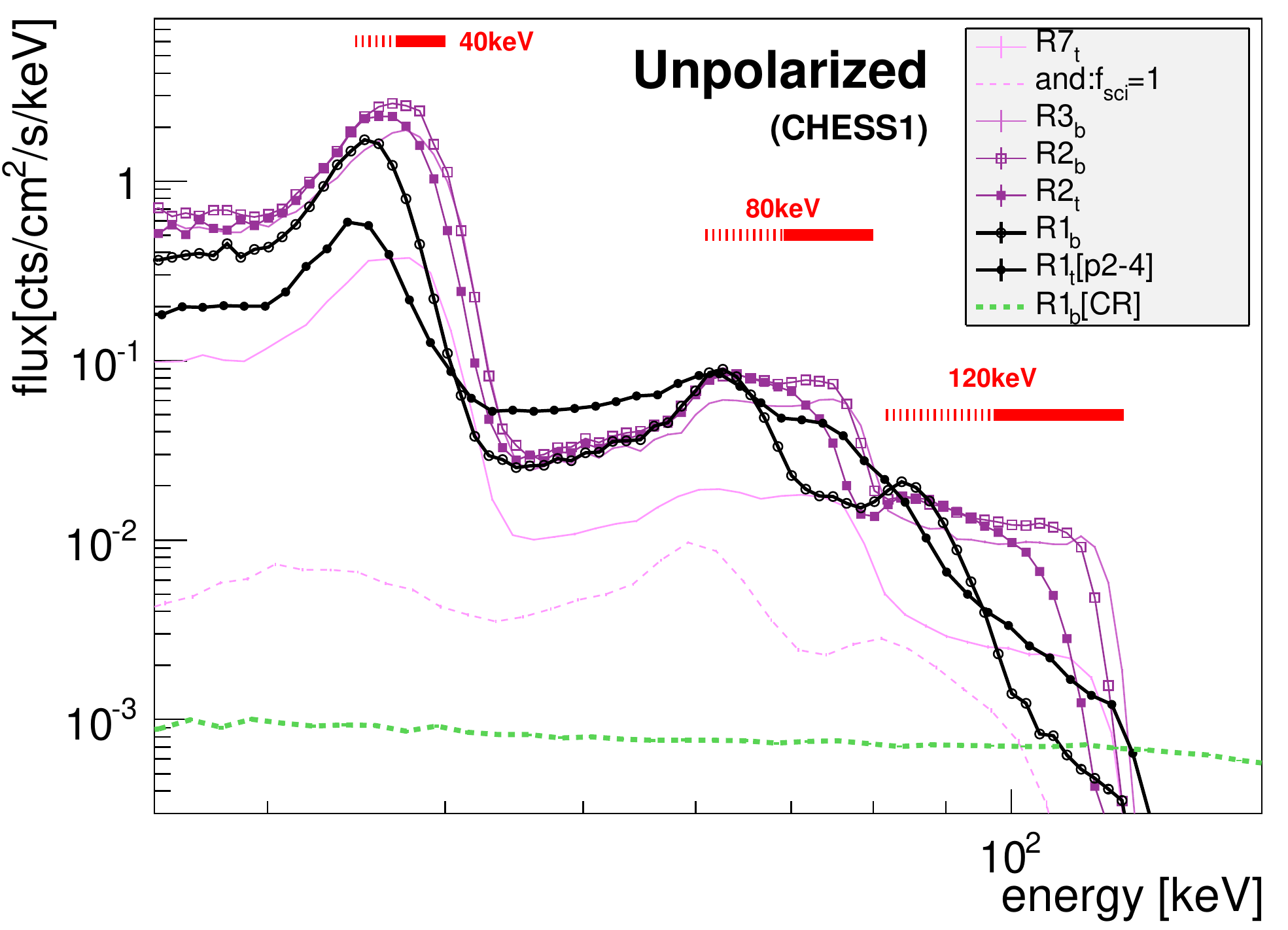}
\hfill
\includegraphics[width=0.49\textwidth]{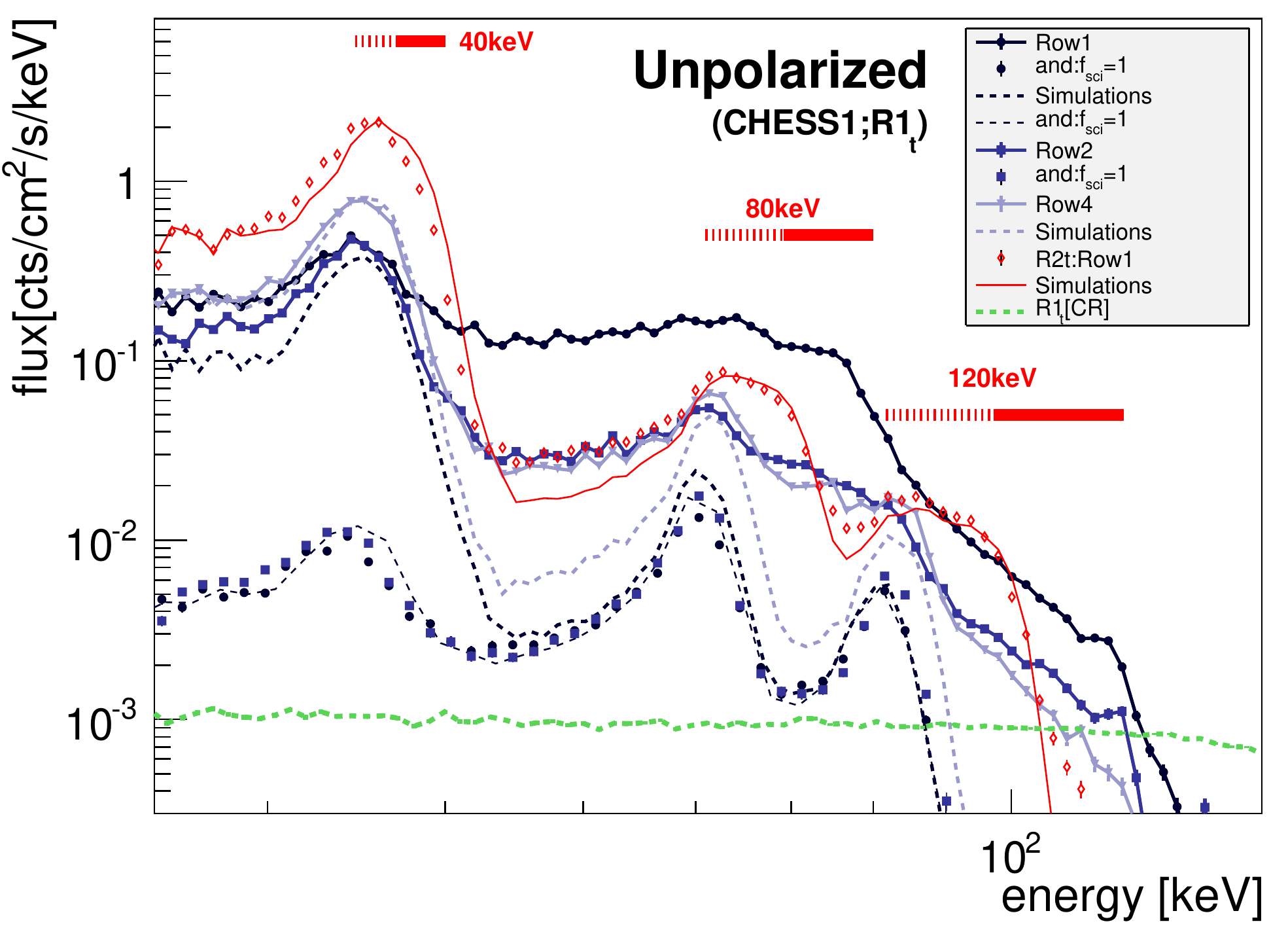}
\end{center}

\caption{Energy spectra of the Compton scattered, non-polarized $40/80/120 
\, \rm{keV}$ CHESS X-ray beam (averaged over the complete azimuthal 
scattering range, $m=1$ events). The horizontal lines represent the 
nominal energies after Compton-scattering. For reference, the CR 
spectrum is shown for one of the $5 \, \rm{mm}$ rings. {\bf Left:} 
Spectra from selected sub rings. In the case of ring {\it 
R1$_{\rm{t}}$}, the first pixel row was removed (see text and right 
panel). {\bf Right:} Special case of ring {\it R1$_{\rm{t}}$} which is 
located at the polarimeter entrance; individual pixel rows are shown. It 
can be seen that the first pixel row has significantly higher flux at 
$E>40 \, \rm{keV}$ as compared to the other rows~-- likely due to 
external X-ray contamination. The same spectra are shown with the 
scintillator trigger condition ($f_{\rm{sci}} = 1$). Simulated spectra 
are shown for the pixel rows 1 and 4. The spectrum obtained from the 
first pixel row in Ring {\it R2$_{\rm{t}}$} is shown for reference.}

\label{fig:CHESS_ComtponSpectra_UnPol}
\end{figure*}

\begin{figure}[t!]
\begin{center}
\includegraphics[width=0.49\textwidth]{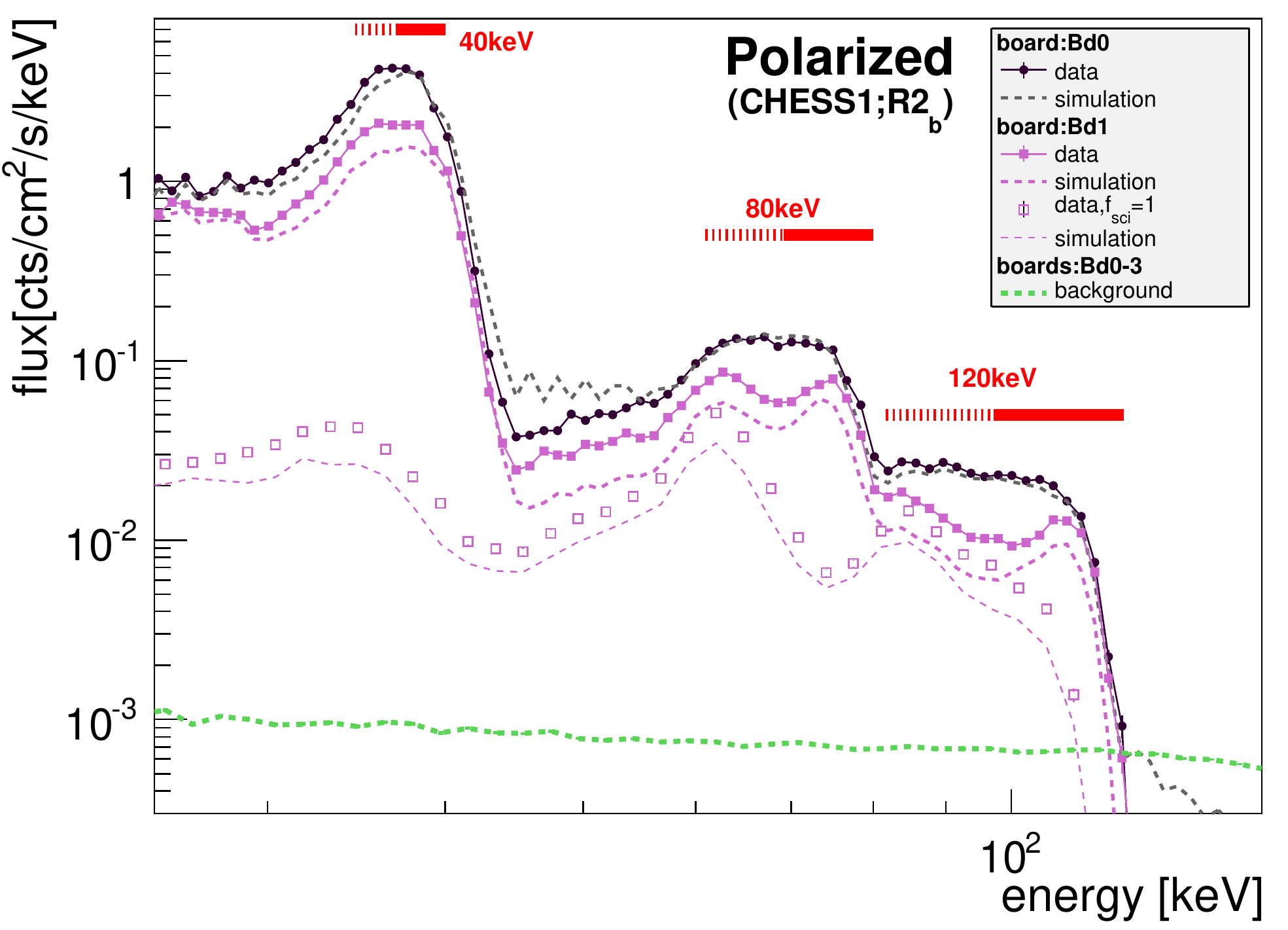}
\end{center}

\caption{Compton spectra of the polarized $40/80/120 \, \rm{keV}$ CHESS 
beam. Spectra are shown for two of the four detector sides ({\it Bd0} 
and {\it Bd1}, see Fig.~\ref{fig:X-Calibur_Configurations}, top) in ring 
{\it R2$_{\rm{b}}$}, with {\it Bd0} aligned with the plane of 
polarization ($\alpha = 0^{\circ}$). The {\it Bd1} spectrum is also 
shown with the requirement of a scintillator trigger (the discriminator 
threshold not being optimized at this point). The horizontal lines 
represent the nominal energy ranges after Compton-scattering. The 
simulated spectra are shown, as well.}

\label{fig:CHESS_ComtponSpectra_Pol}
\end{figure}

Several background runs were taken without the X-ray beam entering the 
hutch. The over-all background spectrum is shown in the left panel of 
Fig.~\ref{fig:BG_Spectra} (red data points). The {\it C1} hutch is 
shielded by (partly) lead-enforced walls which likely contribute to the 
lower background level as compared to the laboratory at Washington 
University. In fact, a signature at the lead $\rm{KL}_{1-3}$ transition 
energies can be identified in the CHESS background spectrum. Although 
the background is negligible compared to the strong X-ray signal (see 
Figs.~\ref{fig:CHESS_ComtponSpectra_UnPol} and 
\ref{fig:CHESS_ComtponSpectra_Pol} for reference), it was subtracted 
from the spectra shown in this section. It should be noted that the beam 
in the hutch may introduce an additional implicit/diffuse background by 
scattering off several components (slits, absorber foils, etc.). This 
kind of background was not determined in a dedicated measurement, but 
can potentially be higher than the CR background.

{\it Synchronization between simulations and data.} The Bragg 
monochromator only allows the $40/80/120 \, \rm{keV}$ harmonics of the 
CHESS white beam to enter the C1 hutch. The spectral intensities of the 
white beam, as well as the Bragg-reflected intensities of the 
monochromator, were calculated using the {\it X-ray Oriented Programs} 
(XOP) software 
package\footnote{http://www.esrf.eu/computing/scientific/xop2.1/}. The 
absorber foils further change the relative intensities. Given the 
density of platinum of $\rho_{\rm{Pt}} = 21.45 \, \rm{g}/\rm{cm}^{3}$, 
its mass attenuation coefficient at $40 \, \rm{keV}$ of 
$(\mu/\rho)_{\rm{Pt}} = 12.45 \rm{cm}^{2}/\rm{g}$, and the thickness of 
the Pt foils $d$, one can calculate the beam intensity $I$ entering the 
polarimeter by $I/I_{0} = \exp(-\mu/\rho \cdot \rho \cdot d)$. The mass 
attenuation coefficients $\mu/\rho$ were taken from the NIST X-ray 
database\footnote{http://www.nist.gov/pml/data/xraycoef/index.cfm}. 
However, an accurate prediction of the relative flux intensities 
entering the polarimeter requires a detailed modeling of all 
energy-dependent X-ray absorption/transmissions on the beam path 
(including the entrance windows, the Bragg monochromator, and the 
absorber foils).

Therefore, we instead used the overall spectrum measured by the whole 
polarimeter to synchronize the relative $40/80/120 \, \rm{keV}$ flux 
intensities with our Monte Carlo simulations. 
Figure~\ref{fig:CHESS_ComtponSpectra_AllDetector_UnPol} shows the 
overall X-Calibur responses (all detector rings except for R1, see 
below) to the non-polarized CHESS beam, measured with configurations 
$C_{\rm{ch1}}$ and $C_{\rm{ch3}}$. The corresponding relative 
intensities of the three mono-energetic energies in the simulations were 
matched accordingly and are in turn applied for all data versus 
simulation comparisons that follow. Note, however, that the CHESS white 
beam intensity varies. Therefore, the simulations were scaled according 
to the relative difference in the integral X-Calibur trigger rate with 
respect to the data run used to normalize the simulations. We estimate 
the systematic error on the absolute flux normalization of the 
simulations to be around $20 \%$. 
Figure~\ref{fig:CHESS_ComtponSpectra_AllDetector_UnPol} also shows the 
energy spectra with a coincident scintillator trigger ($f_{\rm{sci}} = 
1$), although obtained with the non-optimal scintillator trigger 
threshold.

{\it Compton spectra of a non-polarized beam.} The energy spectra of the 
non-polarized Compton scattered X-ray beam are shown for individual 
detector rings in Fig.~\ref{fig:CHESS_ComtponSpectra_UnPol} (left). The 
spectra are averaged over all azimuthal scattering angles $\Phi$ within 
each ring. Ring {\it R1$_{\rm{t}}$} mostly detects back-scattered events 
(the scintillator geometrically covering rings {\it R2}-{\it R8}, see 
Fig.~\ref{fig:Design}, left) which deposit the maximal amount of energy 
in the scintillator: $\Delta E_{\rm{sci}}$ up to $5.4 \, \rm{keV}$ for 
the $40 \, \rm{keV}$ beam, depositing $E_{\rm{czt}} = 34.6 \, \rm{keV}$ 
in the CZT detector. The rings further downstream detect a superposition 
of X-rays that underwent Compton-scattering under different angles. The 
rear-side rings ({\it R7} and {\it R8}), to a larger extent, detect 
X-rays which Compton scattered nearly in the forward direction~-- 
depositing only a small amount of energy in the scintillator, with an 
energy deposition in the CZT ($E_{\rm{czt}}$) close to the energy of the 
primary X-ray ($E_{\rm{line}}$). This explains the shift of the peak 
positions in the spectra towards higher energies with increasing ring 
number. The theoretical range expectations of $E_{\rm{czt}}$ are 
indicated by the horizontal lines in 
Fig.~\ref{fig:CHESS_ComtponSpectra_UnPol} (assuming an idealized energy 
resolution).

The right panel of Fig.~\ref{fig:CHESS_ComtponSpectra_UnPol} shows the 
energy spectra of individual pixel rows of ring {\it R1$_{\rm{t}}$}. It 
can be seen that the first pixel row shows a strong anomaly at energies 
$E_{\rm{czt}} > 50 \, \rm{keV}$ as compared to the pixel rows 2-4. The 
comparison with simulations further underlines the difference. A likely 
explanation of the additional continuum would be the following. The 
first row of pixels has a rather high effective area (pixel sides) in 
the plane perpendicular to the X-ray beam, as compared to other pixel 
rows. These `side' pixels will detect X-rays that underwent diffuse 
scattering in the Pt/Pb absorber foils or the polarimeter entrance plate 
(see Fig.~\ref{fig:X-Calibur_Configurations}), and succeeded in entering 
the polarimeter along a path not along the optical axis. Since this 
involves X-ray transmission through some of the material of the 
polarimeter fixture, it is more likely to happen for higher energies, 
which is where the anomaly in the spectrum becomes more prominent. The 
first pixel row was therefore excluded from the analysis of the data 
presented in this section. It should be noted, that the contamination is 
also seen in the pixel rows further downstream~-- however, with strongly 
decreasing strength.

The right panel of Fig.~\ref{fig:CHESS_ComtponSpectra_UnPol} also shows 
the spectra obtained from the same data with the additional requirement 
of a scintillator trigger ($f_{\rm{sci}} = 1$). This favors high-energy 
and/or back-scattered events which deposit a high amount of energy 
$\Delta E_{\rm{sci}}$ in the scintillator. With the $f_{\rm{sci}} = 1$ 
cut, the first pixel rows no longer show the anomaly discussed above~-- 
underlining the hypothesis of an external contamination not interacting 
with the scintillator. A more detailed study of the trigger efficiency 
of the scintillator is presented in 
Sec.~\ref{subsec:ScintillatorEfficiency}. Despite the contamination 
issue seen in the first front-side pixel rows, the Compton-spectra 
measured with X-Calibur are in good agreement with the simulations 
(although not explicitly shown for all rings in 
Fig.~\ref{fig:CHESS_ComtponSpectra_UnPol}, right, for reasons of 
visibility.).

{\it Compton spectra of a polarized beam.} A polarized X-ray beam 
introduces an azimuthal modulation with a $180^{\circ}$ periodicity to 
the measured spectra. Figure~\ref{fig:CHESS_ComtponSpectra_Pol} shows 
the Compton spectra measured on two of the four detector sides in ring 
{\it R2$_{\rm{b}}$} (with the polarimeter being oriented at an angle of 
$\alpha = 0^{\circ}$). As expected, and essential for the functionality 
of the polarimeter, the detector side located perpendicular to the 
polarization plane ({\it Bd0}) detects a higher number of 
Compton-scattered X-rays. The energy spectra obtained from simulations 
are shown in Fig.~\ref{fig:CHESS_ComtponSpectra_Pol}, as well, and are 
found to be in reasonable agreement with the data. For the data versus 
simulation comparison one has to keep in mind that the simulations were 
performed for a $100 \%$ polarized beam, corresponding to the maximal 
modulation and therefore a maximal scattering concentration in {\it 
Bd0}. The polarization fraction of the CHESS beam, on the other hand, is 
$r < 100 \%$. Therefore, the azimuthal asymmetry ({\it Bd0} versus {\it 
Bd1}) in the data is expected to be smaller compared to the simulations. 
Dividing the spectra into distinct bins in azimuth, and integrating 
counts in a given energy interval, will lead to the azimuthal scattering 
distributions that are discussed in the next paragraph.

\begin{figure*}[t!]
\begin{center}
\includegraphics[width=0.49\textwidth]{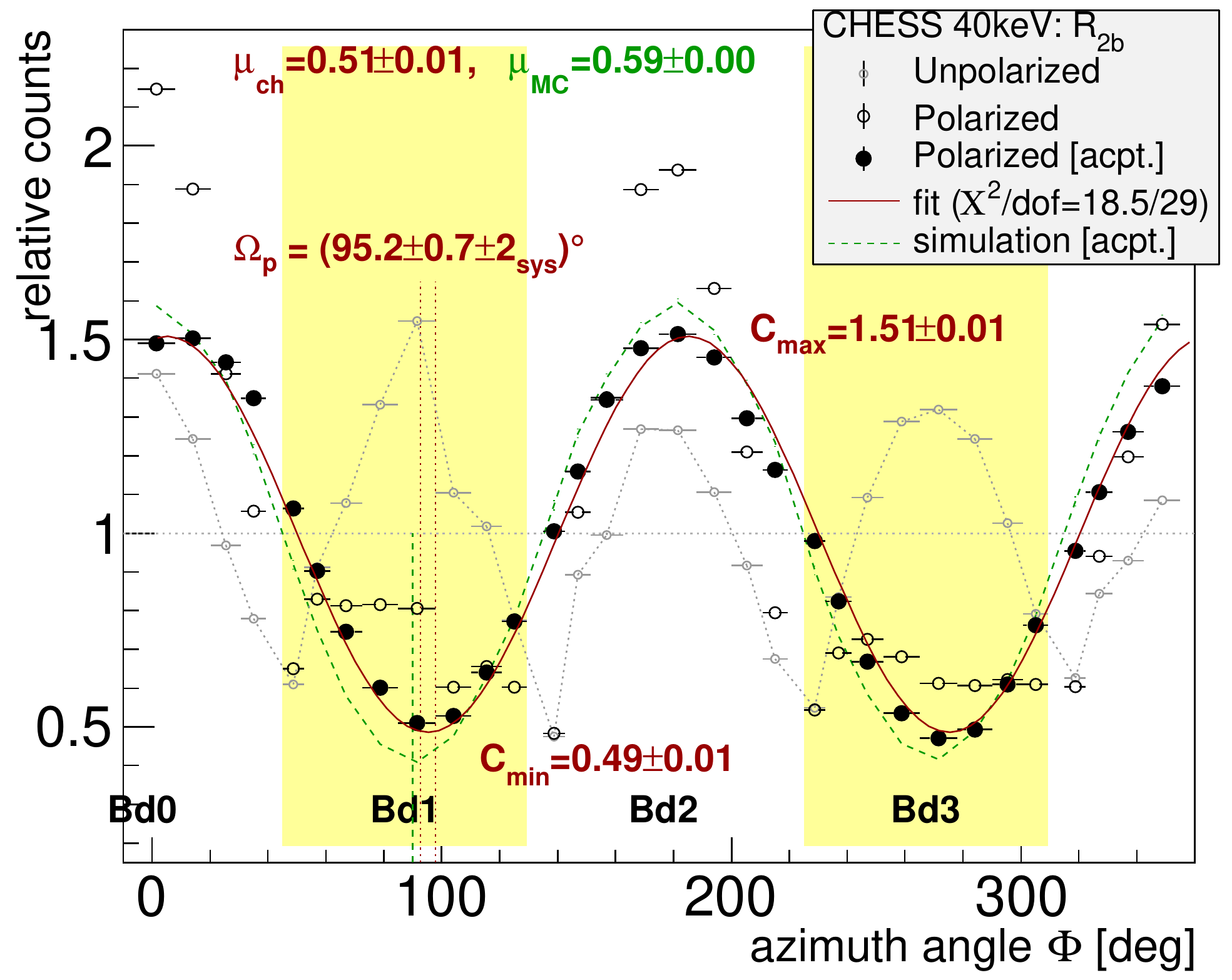}
\hfill
\includegraphics[width=0.49\textwidth]{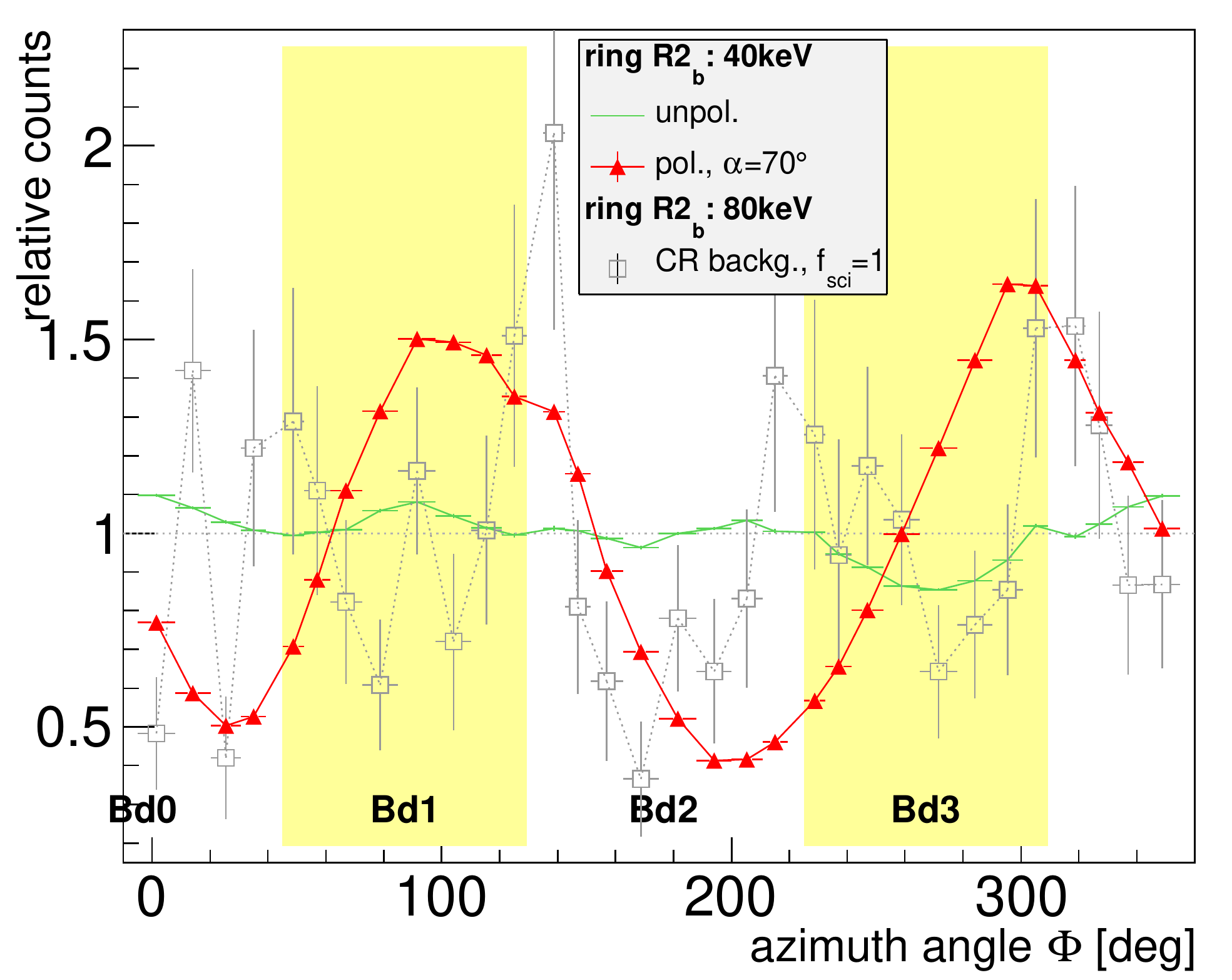}
\end{center}

\caption{{\bf Left:} Azimuthal scattering distribution (CHESS at $40 \, 
\rm{keV}$, ring R2$_{\rm{b}}$). The horizontal error bars reflect the 
azimuthal ranges $\Delta \Phi_{j}$ covered by the corresponding pixels 
$j$ (see Fig.~\ref{fig:SchematicBeamOffset_AndStokes}, left). The raw 
non-polarized beam shows the 4-fold symmetry of the four detectors boards 
{\it Bd0-Bd3} (see Fig.~\ref{fig:X-Calibur_Configurations}). The 
polarized beam has the additional $180^{\circ}$ amplification imprinted 
which is clearly extracted once corrected for the azimuthal coverage 
$\Delta \Phi_{j}$ and the pixel acceptance $a_{j}$. A sine function is 
fitted to extract $C_{\rm{min}}$, $C_{\rm{max}}$, the modulation factor 
$\mu$, and the angle of the polarization plane $\Omega_{\rm{p}}$ (dotted 
vertical lines, indicating the error range). The dashed vertical line 
indicates the nominal polarization plane. For reference, the 
distribution of a simulated $40 \, \rm{keV}$ beam is shown ($100 \%$ 
polarized, reconstructed in the same way). {\bf Right:} Normalized 
azimuthal scattering distributions derived from different data sets: a
non-polarized beam, a polarized beam measured at $\alpha = 70^{\circ}$, as 
well as the response to a cosmic-ray background run.}

\label{fig:AzimuthDistribution}
\end{figure*}

{\it Azimuthal scattering distribution.} The azimuthal scattering 
distributions were generated as outlined in 
Sec.~\ref{subsec:AnalysisPolarization} by integrating the energy range 
of Compton-scattered photons for the three harmonics of the X-ray beam 
($34.6 - 40 \, \rm{keV}$, $60.9 - 80 \, \rm{keV}$, and $81.7 - 120 \, 
\rm{keV}$, respectively). The energy intervals include an additional $2 
\, \rm{keV}$ cushion to account for the energy resolution of the CZT 
detector pixels. All measured event rates were in turn normalized on a 
pixel-by-pixel basis using the azimuthal coverage $\Delta \Phi_{j}$ and 
acceptances $a_{j}$. Following Eq.~(\ref{eqn:Acceptance}) described in 
Sec.~\ref{subsec:AnalysisPolarization}, the pixel acceptances $a_{j}$ 
were determined from the non-polarized beam for each of the above energy 
intervals.

An example of an azimuthal scattering distribution measured in the 
$34.6-40 \, \rm{keV}$ band is shown in 
Fig.~\ref{fig:AzimuthDistribution} (left). The normalized distribution 
is used to derive the plane of polarization $\Omega_{\rm{p}}$ (nominal 
value of $\Omega_{\rm{nom}} = 90^{\circ}$), as well as the relative 
scattering amplitude $\mu_{\rm{ch}}$. The distribution is compared to 
the one obtained from the simulation of a $100 \%$ polarized beam at $40 
\, \rm{keV}$. Both are found to be in good agreement~-- except for the 
slightly lower amplitude of the data which is a result of the $r < 100 
\%$ polarization fraction of the CHESS beam (see below).

Figure~\ref{fig:CHESS_2DAzimuthData} illustrates the reconstruction of 
the azimuthal scattering distributions for the complete polarimeter 
(subrings {\it R1$_{\rm{t}}$}-{\it R8$_{\rm{b}}$}), based on simulations 
(top panel) and based on measured CHESS data (bottom panel). The 
response to a non-polarized $40 \, \rm{keV}$ beam is shown on the left 
side of the figure. The second panel shows the polarimeter response to a 
polarized beam at $40 \, \rm{keV}$, not yet corrected for pixel 
acceptance $a_{j}$ and azimuthal coverage $\Delta \Phi_{j}$. Applying 
the corrections leads to the smooth distribution shown in the third 
panel, which is in good agreement if comparing the simulations to the 
data. The pixel acceptances $a_{j}$ depend on (i) the energy threshold, 
(ii) the energy resolution, and (iii) the trigger efficiency of the 
pixel $j$. The simulations take into account (i) and (ii), derived from 
the calibration measurements presented in 
Sec.~\ref{subsec:CZT_Calibration}, but not (iii). The projected count 
distributions lead to the azimuthal scattering distributions shown in 
the right panel of Fig.~\ref{fig:CHESS_2DAzimuthData}. In contrast to 
the 2D distributions shown in planar pixel coordinates, the $\Phi$ 
positions and error bars now represent the proper angular coverage of 
each pixel (compare with Fig.~\ref{fig:SchematicBeamOffset_AndStokes}, 
left). The expected $180^{\circ}$ modulation is clearly revealed, and 
the reconstructed orientation of the polarization plane agrees with the 
direction of the CHESS beam setup~-- confirming the functionality of the 
X-Calibur polarimeter.

\begin{figure*}[t!]
\begin{center}
\includegraphics[width=0.99\textwidth]{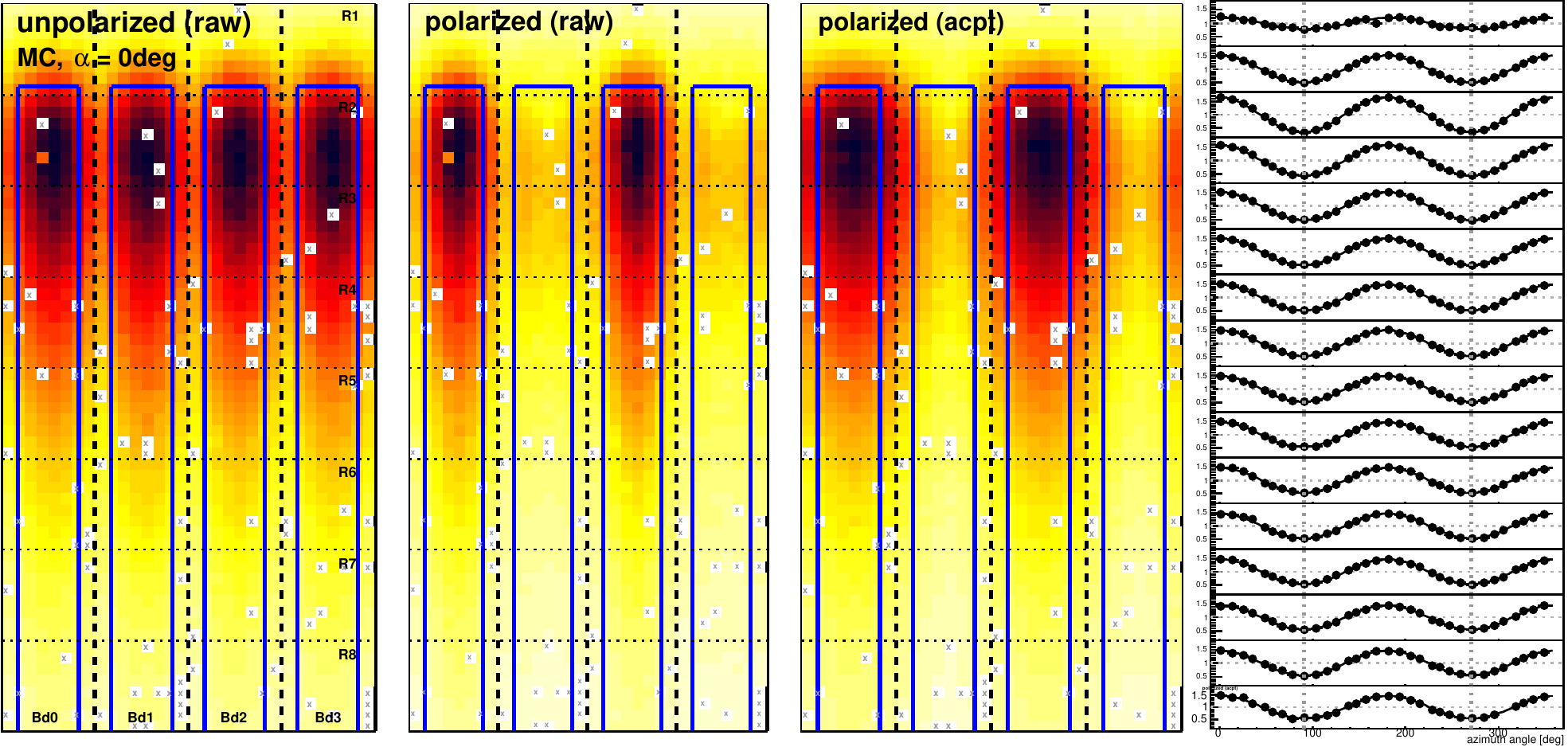} \\
\includegraphics[width=0.99\textwidth]{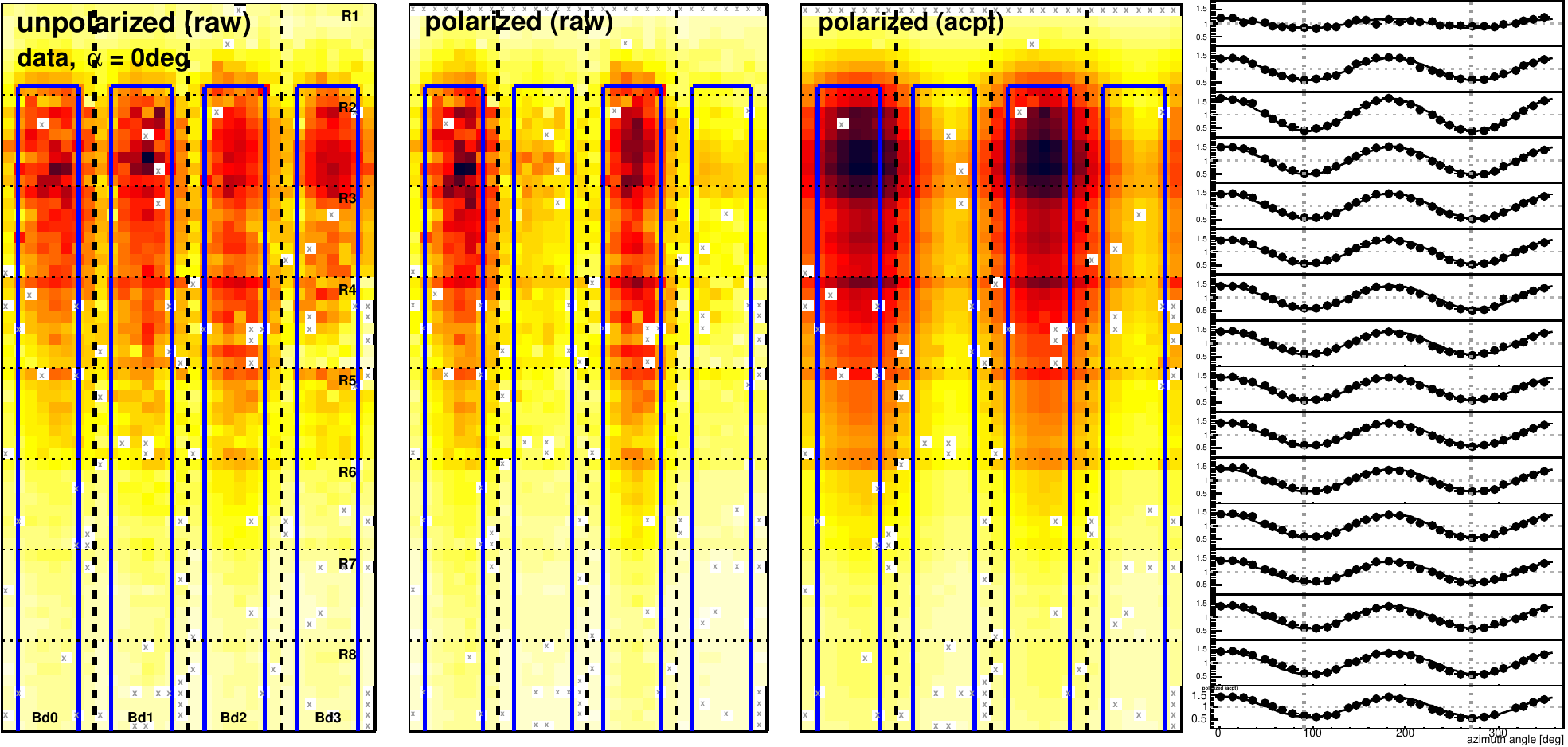} \\
\end{center}

\caption{X-Calibur 2D scattering distributions of the $40 \, \rm{keV}$ 
beam at CHESS ($C_{\rm{ch}}^{3}$, see 
Fig.~\ref{fig:X-Calibur_Configurations}, the beam enters from the top). 
Events are shown for reconstructed energies of $36-40 \, \rm{keV}$. The 
top row shows the results of simulations of a $100\%$ polarized beam. 
The bottom row shows the results of the CHESS measurement. The (blue) 
boxes in the 2D maps indicate the projected outline of the scintillator 
(Compton-scatterer). {\bf Left:} Raw count map (pixel-by-pixel) of a
non-polarized beam (neither corrected for $\Delta \Phi_{j}$ nor for pixel 
acceptance $a_{j}$). All four detector sides, {\it Bd0}-{\it Bd3}, are 
unfolded into a plane. The detector rings {\it R1} to {\it R8} are 
indicated. {\bf Second:} Count map of a raw measurement of a polarized 
beam. {\bf Third:} Count map of the polarized beam, corrected for 
$\Delta \Phi_{j}$ and $a_{j}$. {\bf Right:} Normalized azimuthal 
scattering distribution (corrected for $\Delta \Phi_{j}$ and $a_{j}$) 
for different detector rings. The vertical lines indicate the nominal 
plane $\Omega_{\rm{n}}$ of the electric field vector of the polarized 
beam.}

\label{fig:CHESS_2DAzimuthData}

\end{figure*}

\begin{figure}[t!]
\begin{center}
\includegraphics[width=0.49\textwidth]{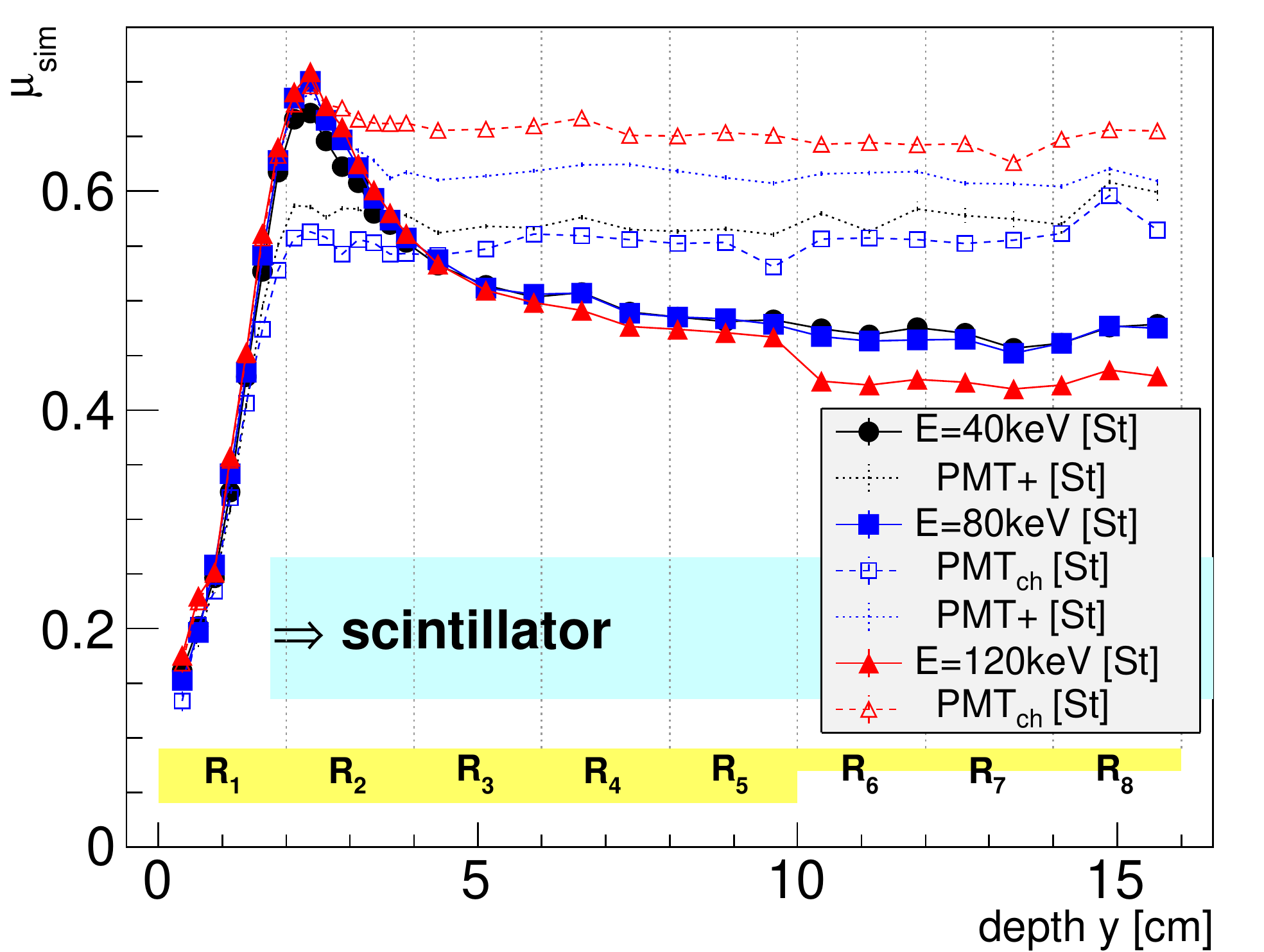}
\end{center}

\caption{Simulated modulation factors $\mu_{\rm{sim}}$ as a function of 
detector ring {\it R}, shown for different energies. The geometrical 
positions (along the optical axis $y$) are indicated for the 
scintillator and the CZT ring assembly. The simulated data were analyzed 
with the Stokes method. Results are also shown for the scintillator 
trigger requirement $f_{\rm{sci}}=1$, assuming the PMT trigger 
efficiency during the CHESS measurements (PMT$_{\rm{ch}}$) and the 
efficiency of the optimized PMT threshold (PMT+), see 
Fig.~\ref{fig:Scintillator_TriggerEfficiency}.}

\label{fig:MC_ModulationFactor}
\end{figure}

{\it Simulated modulation factors.} Each detector ring (see 
Fig.~\ref{fig:CHESS_2DAzimuthData}), or a sub ring thereof, can be seen 
as an independent detector that allows to reconstruct energy-dependent 
polarization properties of the incoming X-ray beam. 
Figure~\ref{fig:MC_ModulationFactor} shows the simulated modulation 
factors $\mu_{\rm{sim}}$ for the different energies and X-Calibur 
configurations as a function of detector ring/depth, measured along the 
optical axis. The values for $\mu_{\rm{sim}}$ were obtained using the 
Stokes analysis, and are in agreement with the corresponding results 
obtained from the analysis of the azimuthal scattering distribution (not 
shown). The simulations shown reflect the detector status of the CHESS 
measurements (dead pixels, energy resolutions, etc.), not an idealized 
detector. Ring {\it R1} detects only back-scattered X-rays with a 
correspondingly lower $\mu_{\rm{sim}}$~-- but with a better energy 
resolution since the scattering kinematics are better defined. The 
step-like structure in $\mu_{\rm{sim}}^{120\,\rm{keV}}$ between rings 
{\it R5} and {\it R6} can be explained by the strong degradation in 
energy resolution at high energies in the $2 \, \rm{mm}$ detectors 
located at rings {\it R6}-{\it R8}. The highest modulation is achieved 
in ring {\it R2}.

The simulations of the modulation factor were also done involving the 
scintillator flag $f_{\rm{sci}}$, based on the two scintillator trigger 
efficiencies as measured in 
Fig.~\ref{fig:Scintillator_TriggerEfficiency}. This reflects the 
efficiency during the CHESS measurements (not optimized), as well as the 
efficiency of the optimized scintillator trigger. The scintillator 
improves the modulation factor for most rings~-- by requiring a minimum 
energy deposition $\Delta E_{\rm{sci}}$ which effectively limits the 
allowed range of polar scattering angles. This, however, also reduces 
the event statistics (not imprinted in $\mu_{\rm{sim}}$). A scintillator 
trigger efficiency of $100 \%$ would again lead to the same distribution 
of modulation factors as no cut on $f_{\rm{sci}}$ does. The main benefit 
of the scintillator trigger lies in its ability to suppress background 
during a measurement (see for example Fig.~\ref{fig:BG_Spectra}, left). 
The modulation factor $\mu$, together with the detection rate after 
background subtraction, are the crucial characteristics that determine 
the MDP detection sensitivity of the polarimeter following 
Eq.~(\ref{eqn:MDP}).

\begin{figure}[t!]
\begin{center}
\includegraphics[width=0.49\textwidth]{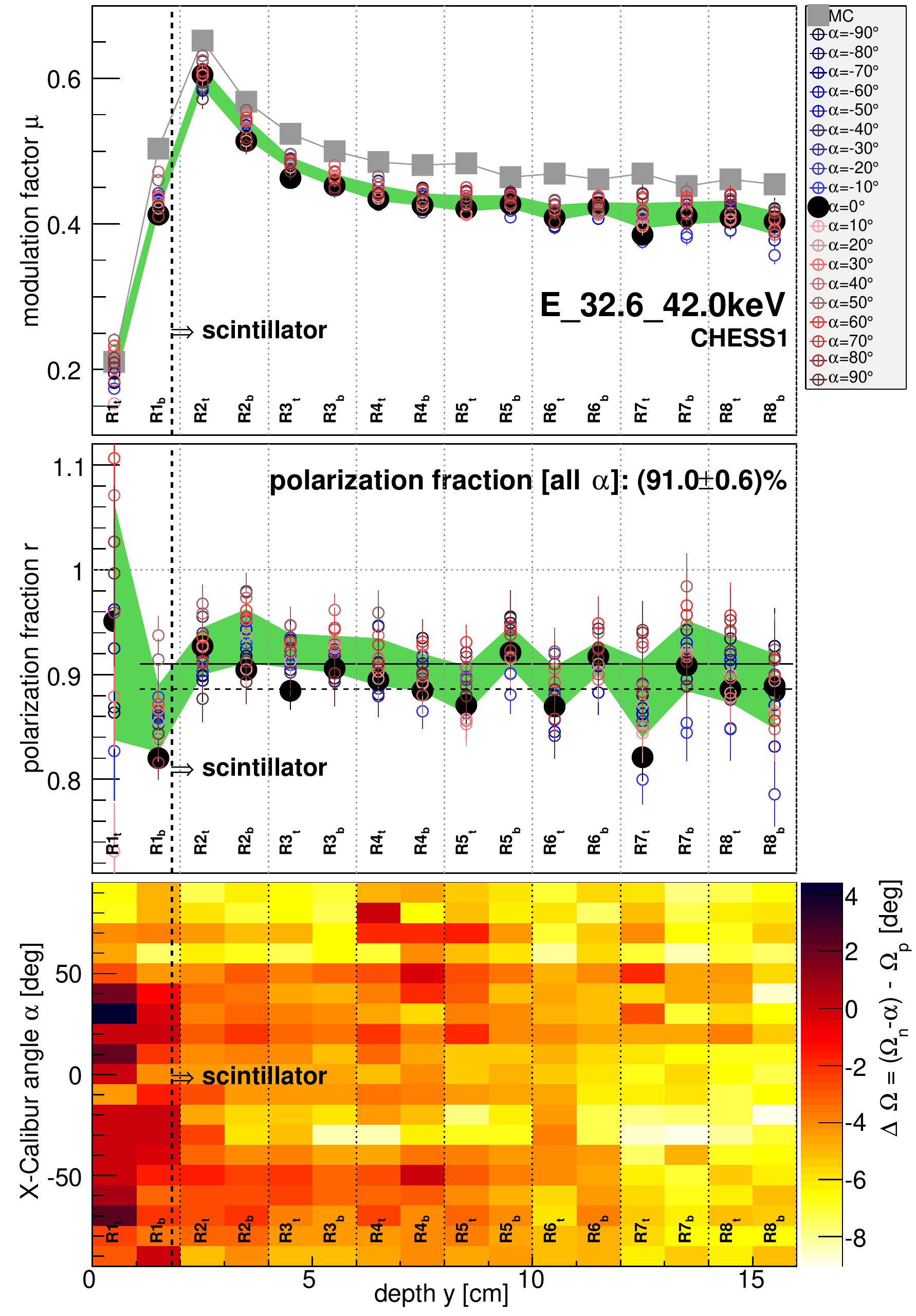}
\end{center}

\caption{Reconstructed polarization properties of the CHESS beam at $40 
\, \rm{keV}$ (configuration $C_{\rm{ch1}}$). {\bf Top:} Modulation 
factor $\mu$ versus polarimeter sub rings {\it Ri$_{\rm{t,b}}$}. The 
results were derived from simulations (MC), as well as from data taken 
under different X-Calibur orientations $\alpha$. For each $\alpha$ and 
each ring, $\mu_{\rm{ch}}$ is derived from fits to acceptance corrected 
$\Phi$-distributions (see Fig.~\ref{fig:AzimuthDistribution}, left). The 
scintillator starts covering detector rings at $y \geq 1.8 \, \rm{cm}$. 
The green error band reflects the $1 \, \rm{std.dev}$ range of measured 
points in the particular $y$ slice. {\bf Middle:} Reconstructed 
polarization fraction following Eq.~(\ref{eqn:PolFracPhiDistri}). The 
green band reflects the $1 \, \rm{std.dev}$ range of all reconstructed 
fractions in the corresponding $y$ slice. {\bf Bottom:} Residual between 
reconstructed polarization plane $\Omega_{\rm{p}}$ and true polarization 
plane $\Omega_{\rm{n}} = 90^{\circ}$, corrected for the X-Calibur 
orientation $\alpha$: $\Delta \Omega = (\Omega_{\rm{n}} - \alpha) 
-\Omega_{\rm{p}}$.}

\label{fig:AzDistri_RecoMu}
\end{figure}

{\it Polarization fraction of the CHESS beam.} The azimuthal scattering 
distributions for each sub ring and each data set were fitted to 
determine $\mu_{\rm{ch}}$. The simulated modulation factors 
$\mu_{\rm{sim}}$ (Fig.~\ref{fig:MC_ModulationFactor}) were in turn used 
to determine the polarization fraction $r$ of the CHESS beam using 
Eq.~(\ref{eqn:PolFracPhiDistri}). This is shown in 
Fig.~\ref{fig:AzDistri_RecoMu} for the $40 \, \rm{keV}$ beam for 
different orientations $\alpha$ of the polarimeter relative to the 
polarization plane. Also shown are the residuals between the 
reconstructed polarization plane and the nominal polarization plane. 
Note, that for large values of $\alpha$ the azimuthal scattering 
distribution shows an additional global slope, possibly caused by an 
$\alpha$-dependent tilt of the rotation fixture (see 
Fig.~\ref{fig:AzimuthDistribution}, right, for an example orientation of 
$\alpha = 70^{\circ}$). This will affect the reconstructed polarization 
properties and is further discussed in Sec.~\ref{subsec:Systematics}. To 
reduce this systematic effect, the CHESS azimuthal distributions were 
folded back into the $[0;180]^{\circ}$ interval before the sinusoidal 
function was fitted to the data points.

\begin{table}[t!]

\begin{tabular}{lrr}

Setup & $\Delta \Omega \, [\rm{deg}]$ & $r_{\rm{rec}} \, [\%]$ \\
\hline \hline

\noalign{\smallskip}
\multicolumn{3}{l}{{beam energy: $40 \, \rm{keV}$}} \\
\hline

$C_{\rm{ch1}}$ & $-4.7 \pm 1.9$ & $91.0 \pm 0.6$ \\
      & $-4.6 \pm 1.7$ & $88.6 \pm 0.4$ \\

\noalign{\smallskip}
\multicolumn{3}{l}{{beam energy: $80 \, \rm{keV}$}} \\
\hline

$C_{\rm{ch1}}$ & $-2.6 \pm 1.7$ & $83.5 \pm 0.8$ \\
      & $-2.0 \pm 2.1$ & $78.3 \pm 0.8$ \\
$C_{\rm{ch1}}$, $f_{\rm{sci}} = 1$ & $-2.1 \pm 2.5$ & $86.5 \pm 1.1$ \\
      & $-2.1 \pm 2.7$ & $82.8 \pm 1.0$ \\
$C_{\rm{ch4}}$ & $-2.5 \pm 1.8$ & $90.4 \pm 0.5$ \\
      & $-2.0 \pm 2.2$ & $85.5 \pm 0.6$ \\

\noalign{\smallskip}
\multicolumn{3}{l}{{beam energy: $120 \, \rm{keV}$}} \\
\hline

$C_{\rm{ch1}}$ & $-0.5 \pm 2.9$ & $89.2 \pm 1.5$ \\
      & $-0.2 \pm 3.3$ & $71.5 \pm 1.3$ \\
$C_{\rm{ch1}}$, $f_{\rm{sci}} = 1$ & $-0.7 \pm 4.6$ & $86.3 \pm 1.2$ \\
      & $-0.8 \pm 3.7$ & $80.5 \pm 1.5$ \\

\end{tabular}

\caption{Residuals between the reconstructed polarization plane 
$\Omega_{\rm{p}}$ and the nominal plane at $\Omega_{\rm{n}} = 
90^{\circ}$: $\Delta \Omega = (\Omega_{\rm{n}}-\alpha) - 
\Omega_{\rm{p}}$. The second column shows the reconstructed polarization 
fractions $r_{\rm{rec}}$ of the CHESS beam at different energies and 
different configurations/setups (partly with the $f_{\rm{sci}} = 1$ 
requirement). See Figs.~\ref{fig:AzDistri_RecoMu} and 
\ref{fig:CHESS_MorePolFractions} for an illustration. The first row per 
data set shows the results obtained from the analysis of the azimuthal 
$\Phi$-scattering distribution following 
Eq.~(\ref{eqn:PolFracPhiDistri}). The second row represents the results 
obtained from the Stokes analysis following 
Eq.~(\ref{eqn:PolarizationFromStokes}). The results are averaged over 
all orientations $\alpha$, each.}

\label{tab:CHESS_PolFrac}

\end{table}

\begin{figure}[t!]
\begin{center}
\includegraphics[width=0.49\textwidth]{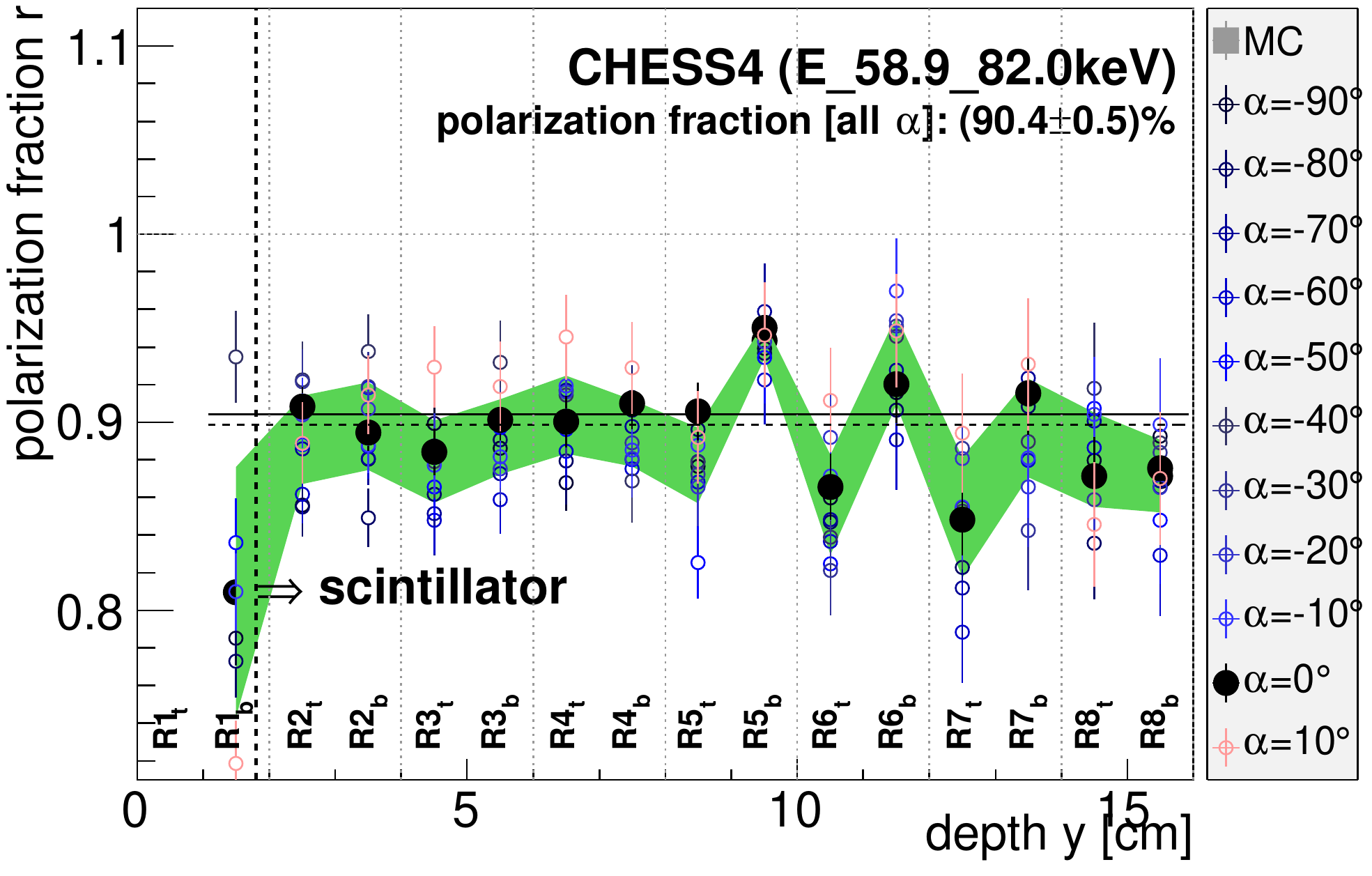} \\
\end{center}

\caption{Reconstructed polarization fractions of the CHESS beam (compare 
with Fig.~\ref{fig:AzDistri_RecoMu}) derived from the $80 \, \rm{keV}$ 
harmonic (configuration $C_{\rm{ch3}}$).}

\label{fig:CHESS_MorePolFractions}
\end{figure}

Figure~\ref{fig:CHESS_MorePolFractions} shows the corresponding 
polarization fractions reconstructed from the $80 \, \rm{keV}$ harmonic. 
The reconstructed polarization fractions obtained from all measurements 
are summarized in Tab.~\ref{tab:CHESS_PolFrac}~-- in all cases averaged 
over all measured orientations $\alpha$. Table~\ref{tab:CHESS_PolFrac} 
also shows the results derived from the Stokes analysis following 
Eq.~(\ref{eqn:PolarizationFromStokes}). Both methods are in reasonable 
agreement, whereas the Stokes analysis seems to slightly underestimate 
$r$ compared to the results obtained with the azimuthal scattering 
distribution. The systematic error on the reconstructed polarization 
fraction was estimated as described in Sec.~\ref{subsec:Systematics}. 
The polarization of the CHESS beam as reconstructed using the 
$\Phi$-distribution method is measured to be $r_{\rm{ch}}^{\Phi} = (87.8 
\pm 0.4_{\rm{stat}} \pm 4.6_{\rm{sys}}) \, \%$ with no indication of an 
energy dependence in the $40-120 \, \rm{keV}$ range. The corresponding 
value reconstructed from the Stokes analysis is $r_{\rm{ch}}^{\rm{st}} = 
(81.2 \pm 0.4_{\rm{stat}} \pm 10_{\rm{sys}}) \, \%$. The difference in 
the systematic error is explained by the fact that the Stokes analysis 
is more sensitive to the detector configuration (which was not perfectly 
constrained during the CHESS measurements). A better defined geometry 
and the rotation of the polarimeter will reduce the systematic error for 
measurements performed during the balloon flight.

{\it Polarization plane of the CHESS beam.} 
Table~\ref{tab:CHESS_PolFrac} also summarizes the average reconstructed 
polarization planes $\Omega_{\rm{p}}$ (corrected for the X-Calibur 
orientation $\alpha$). Note, that the uncertainty in setting the 
X-Calibur orientation was estimated to be $\Delta \alpha = \pm 
2^{\circ}$. Within this systematic error, as well as the statistical 
errors, the average of reconstructed polarization planes (all rings and 
all X-Calibur orientations $\alpha$) is compatible with the nominal 
plane $\Omega_{\rm{n}}$ of the CHESS beam. However, the measurements 
consistently reconstruct $\Delta \Omega < 0^{\circ}$, which may indicate 
a slight offset in the way the rotation mechanism was 
installed/calibrated. The reconstructed polarization planes shown in 
Fig.~\ref{fig:AzDistri_RecoMu} indicate a slight dependence on $\alpha$ 
and $y$. The left panel in Fig.~\ref{fig:AzimuthDistribution} shows a 
corresponding azimuthal distribution measured with the X-Calibur 
orientation of $\alpha = 70^{\circ}$, revealing a slight asymmetry. This 
result can be seen as another indication of a slight mis-alignment 
(offset and/or tilt) between the rotation axis of the scintillator and 
the optical axis of the X-ray beam. A more detailed discussion of this 
kind of systematic effect can be found in Sec.~\ref{subsec:Systematics}.

\subsection{Measurements with the X-Calibur/\INFOCUS Assembly} 
\label{subsec:FtSumnerData}

The full X-Calibur/\INFOCUS experiment was assembled and tested during a 
flight preparation campaign at the NASA {\it Columbia Scientific Balloon 
Facility} (CSBF) site in Ft.~Sumner, NM, in the fall of 2014. The 
polarimeter was installed in the rotating CsI shield assembly, which in 
turn was installed in the pressure vessel as part of the \INFOCUS X-ray 
telescope (see Fig.~\ref{fig:ActiveShieldAndTelescope}). This setup, 
$C_{\rm{ft}}$, is illustrated in 
Fig.~\ref{fig:X-Calibur_Configurations}. The CZT detector configuration 
of these measurements is listed in Tab.~\ref{tab:Detectors} and differs 
slightly from the one used in the CHESS measurements (described in 
Sec.~\ref{subsec:CHESS}). The X-ray mirror was installed on the optical 
bench located at the front part of the \INFOCUS telescope. The pressure 
vessel is located at a focal distance of $8 \, \rm{m}$.

Before the balloon flight, a beryllium (Be) window is installed in the 
top dome of the pressure vessel, to assure a pressure-sealed entrance to 
the polarimeter with high transmissivity to hard X-rays. For the 
ground-based measurements, presented in this section, the Be window was 
not installed to allow to visually align the polarimeter with the 
optical axis of the X-ray mirror. Throughout the measurements presented 
in this section, the polarimeter was rotated at $4 \, \rm{rpm}$. Due to 
the rotation, the application of the pixel acceptances $a_{j}$, defined 
in Eq.~(\ref{eqn:Acceptance}), is no longer required, since each pixel 
tests the complete, time-averaged azimuthal scattering range with 
respect to the polarization plane. As another consequence, the azimuthal 
binning can be chosen finer than in the non-rotating system in which it 
was limited to the number of $32$ CZT pixels per row. For each event, 
the hit pixel is de-rotated into the laboratory/horizon coordinate frame 
(see Fig.~\ref{fig:SchematicBeamOffset_AndStokes}, left) to determine 
the azimuthal scattering angle $\Phi$.

\begin{figure*}[t!]
\begin{center}
\includegraphics[height=0.32\textheight]{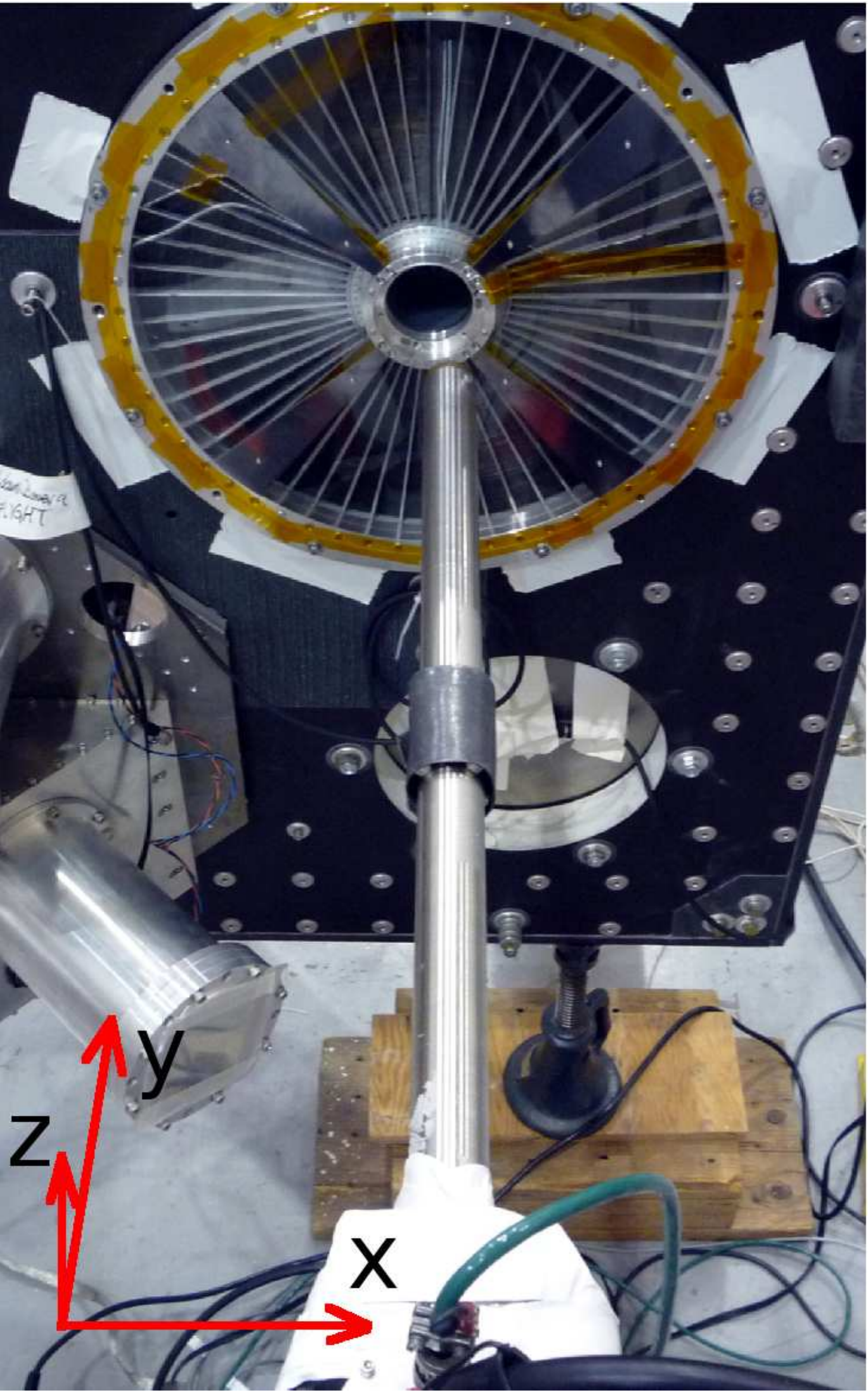}
\hfill
\includegraphics[height=0.32\textheight]{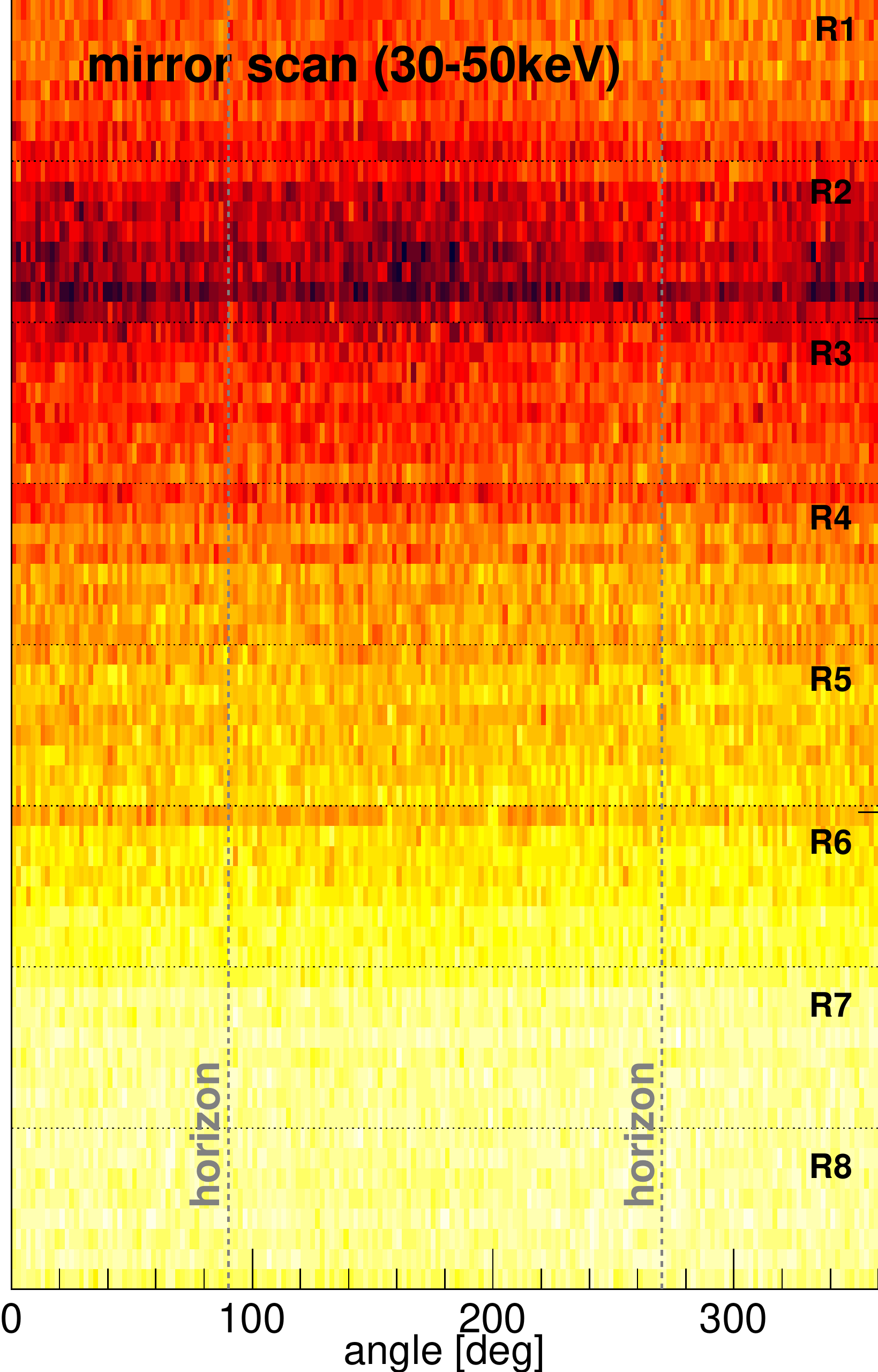}
\hfill
\includegraphics[height=0.32\textheight]{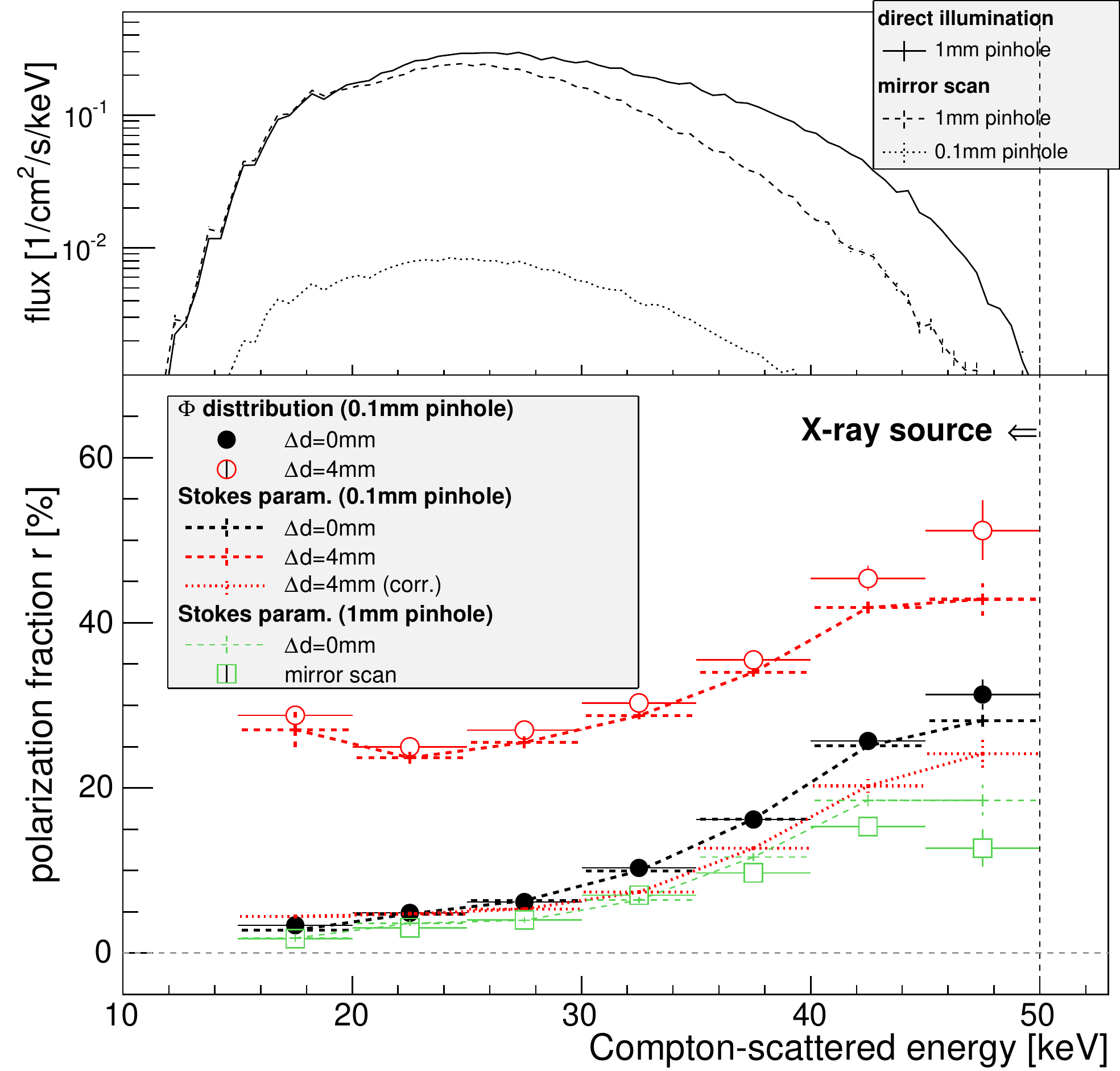}
\end{center}

\caption{\INFOCUS/X-Calibur X-ray mirror scan, configuration 
$C_{\rm{ft}}$ in Fig.~\ref{fig:X-Calibur_Configurations}. {\bf Left:} 
The collimated X-ray source is aligned with the optical axis of the 
mirror and the polarimeter which is situated in the focal plane at a 
distance of $8 \, \rm{m}$ (not visible in this photograph). The X-ray 
source can be automatically moved in the $X$/$Z$ plane to scan the 
mirror with the X-rays traveling along $Y$. {\bf Middle:} The 
Compton-scattered event distribution measured with X-Calibur (horizon 
system, de-rotated) during the scan in the energy range of $30-50 \, 
\rm{keV}$. {\bf Right:} Background subtracted measurements of the 
collimated X-ray beam. Three data runs where taken with (i) the beam 
hitting the center of the scintillator ($\Delta d = 0 \, \rm{mm}$), (ii) 
the beam hitting at an offset of $\Delta d = -4 \, \rm{mm}$, and (iii) 
the beam scanning the X-ray mirror. The top panel shows the 
Compton-scattered energy spectra (all-detector average, each). The 
bottom panel shows the energy-dependent polarization fraction 
reconstructed from the azimuthal $\Phi$-scattering distribution, as well 
as using the Stokes parameters. The given errors are statistical only.}

\label{fig:FtSumner_MirrorScan}
\end{figure*}

{\it X-ray mirror/X-Calibur alignment.} The alignment of the polarimeter 
and the mirror are crucial in order to gain the maximal sensitivity for 
polarization measurements and to reduce systematic effects. The optical 
axis and the on-axis image location of the X-ray mirror were measured by 
a CCD camera with an optical parallel beam. The X-ray mirror was placed 
in the optical beam, and its tip and tilt were adjusted such that the 
optical axis is parallel to the optical beam. Then, the CCD camera was 
placed at the center of the mirror, looking along the optical axis 
facing the polarimeter, and taking a picture of the on-axis image plane. 
The image location was recorded in pixel coordinates, which determines 
the optical axis as well as the on-axis image position. After the X-ray 
mirror was installed onto the optical bench (see 
Fig.~\ref{fig:FtSumner_MirrorScan}, left), a picture of the front 
surface of the polarimeter's scintillator was taken by the camera. The 
mirror tip and tilt were adjusted by shimming at the interface to the 
optical bench, such that the center of the scintillator is at the 
recorded location of the on-axis image of the X-ray mirror. An alignment 
between the scintillator and mirror of better than $1\, \rm{mm}$ was 
achieved which guarantees a systematic error on the reconstructed 
polarization fraction of less than $2 \%$ (see 
Sec.~\ref{subsec:Systematics}).

{\it Mirror scan.} The proper alignment of the telescope is 
tested/confirmed in a mirror scan. A movable X-ray source scans the 
surface of the mirror while the polarimeter response is measured. The 
X-ray scanning system consists of an X-ray source with a $60\, \rm{cm}$ 
long collimator, tip and tilt stages, and $X$/$Z$ travel stages to which 
the X-ray source is mounted. The X-ray source was positioned in front of 
the X-ray mirror and could be used to either directly illuminate the 
polarimeter (through the central hole in the X-ray mirror), or to scan 
the whole mirror aperture. The setup is shown in 
Fig.~\ref{fig:FtSumner_MirrorScan}, left. An Oxford 5011 electron impact 
X-ray tube is used to generate the X-rays, with an active source spot 
size of $0.05 \, \rm{mm}$ in diameter (molybdenum target, Mo). The 
produced X-ray spectrum is made of emission line (Mo-K) as well as 
bremsstrahlung. The X-ray source provides a continuum spectrum up to $50 
\, \rm{keV}$. The current was adjusted to give a reasonable event rate 
at the polarimeter located at $8 \, \rm{m}$ distance (see 
Tab.~\ref{tab:PixelRates} for reference). The collimator can be equipped 
with two interchangeable pin holes with diameters of $0.1 \, \rm{mm}$ 
and $1 \, \rm{mm}$, respectively. The beam size at $8 \, \rm{m}$ 
distance is around $2 \, \rm{mm}$ in diameter for the $0.1 \, \rm{mm}$ 
pin hole. The $1 \, \rm{mm}$ pin hole produces count rates in the 
polarimeter above one kHz at $8 \, \rm{m}$ distance ($20-50 \, 
\rm{keV}$, after air absorption). The tip and tilt stages change the 
direction of the collimated X-ray beam. The $X$/$Z$ translation stages 
allow the X-ray beam to scan over the entire aperture of the mirror 
(with the central hole in the mirror blocked by a lead absorber). The 
X-ray beam was aligned with the optical axis of the mirror. The $1 \, 
\rm{mm}$ pin hole was used to scan the mirror within $51$ horizontal 
rows along $X$. The response measured with X-Calibur is shown in 
Fig.~\ref{fig:FtSumner_MirrorScan} (middle) in the de-rotated coordinate 
system (the horizon intersects at $90^{\circ}$ and $270^{\circ}$ in this 
representation). The corresponding energy spectrum is shown in 
Fig.~\ref{fig:FtSumner_MirrorScan}, right. The results illustrate that 
the mirror was successfully aligned with the polarimeter.

{\it Direct beam illumination.} As a reference measurement to the mirror 
scan, data were taken with the polarimeter being directly illuminated by 
the collimated X-ray source ($0.1 \, \rm{mm}$ pin hole). The X-rays 
enter the scintillator along its optical axis, but do not pass the X-ray 
mirror in this measurement. The spectrum of the mirror scan shown in 
Fig.~\ref{fig:FtSumner_MirrorScan} (right, top) drops off faster 
compared to the spectrum measured from the direct beam illumination. 
This is a result of the energy-dependent effective area of the mirror. 
An additional run was taken with the X-ray beam hitting the scintillator 
off center by $\Delta d = -4 \, \rm{mm}$, which is discussed in 
Sec.~\ref{subsec:Systematics}.

The polarization parameters were reconstructed in the same way as 
described in Sec.~\ref{subsec:CHESS} (except for the acceptance 
correction $a_{j}$). The depth-dependent modulation factors (see 
Fig.~\ref{fig:MC_ModulationFactor}) obtained from the $40 \, \rm{keV}$ 
CHESS simulation were used to reconstruct the polarization fraction of 
the X-ray beam in different energy bands (assuming $\mu_{\rm{sim}}$ 
being independent of energy in the $20-50 \, \rm{keV}$ band). The 
reconstruction was done with both methods described in 
Sec.~\ref{subsec:AnalysisPolarization}, namely using the analysis of the 
azimuthal scattering distribution following 
Eq.~(\ref{eqn:PolFracPhiDistri}), as well as using the Stokes parameters 
following Eq.~(\ref{eqn:PolarizationFromStokes}). Both methods yield 
comparable results which are shown in Fig.~\ref{fig:FtSumner_MirrorScan} 
(right, bottom). An energy dependent polarization fraction is found in 
the data. However, in contrast to the mono-energetic CHESS beam 
(Sec.~\ref{subsec:CHESS}), the energy distribution of the incoming 
X-rays (before Compton-scattering in the scintillator) follows a 
continuum. To properly reconstruct the polarization fraction as a 
function of incoming X-ray energy, one would have to utilize 
reconstruction methods of forward folding or unfolding 
\cite{XCLB_LogLikelihoodAnalysis}, which is beyond the scope of this 
paper. However, a significant increase in polarization fraction can be 
observed with increasing energy. The data of the mirror scan was also 
used to reconstruct the energy-dependent polarization fraction, and is 
found to be in reasonable agreement with the direct beam measurement. 
This confirms the predictions by \citet{Katsuta2009} that an X-ray 
mirror does not substantially affect the polarization properties.

\subsection{Systematic Effects in Polarization Measurements} 
\label{subsec:Systematics}

\begin{figure*}[t!]
\begin{center}
\includegraphics[height=0.285\textheight]{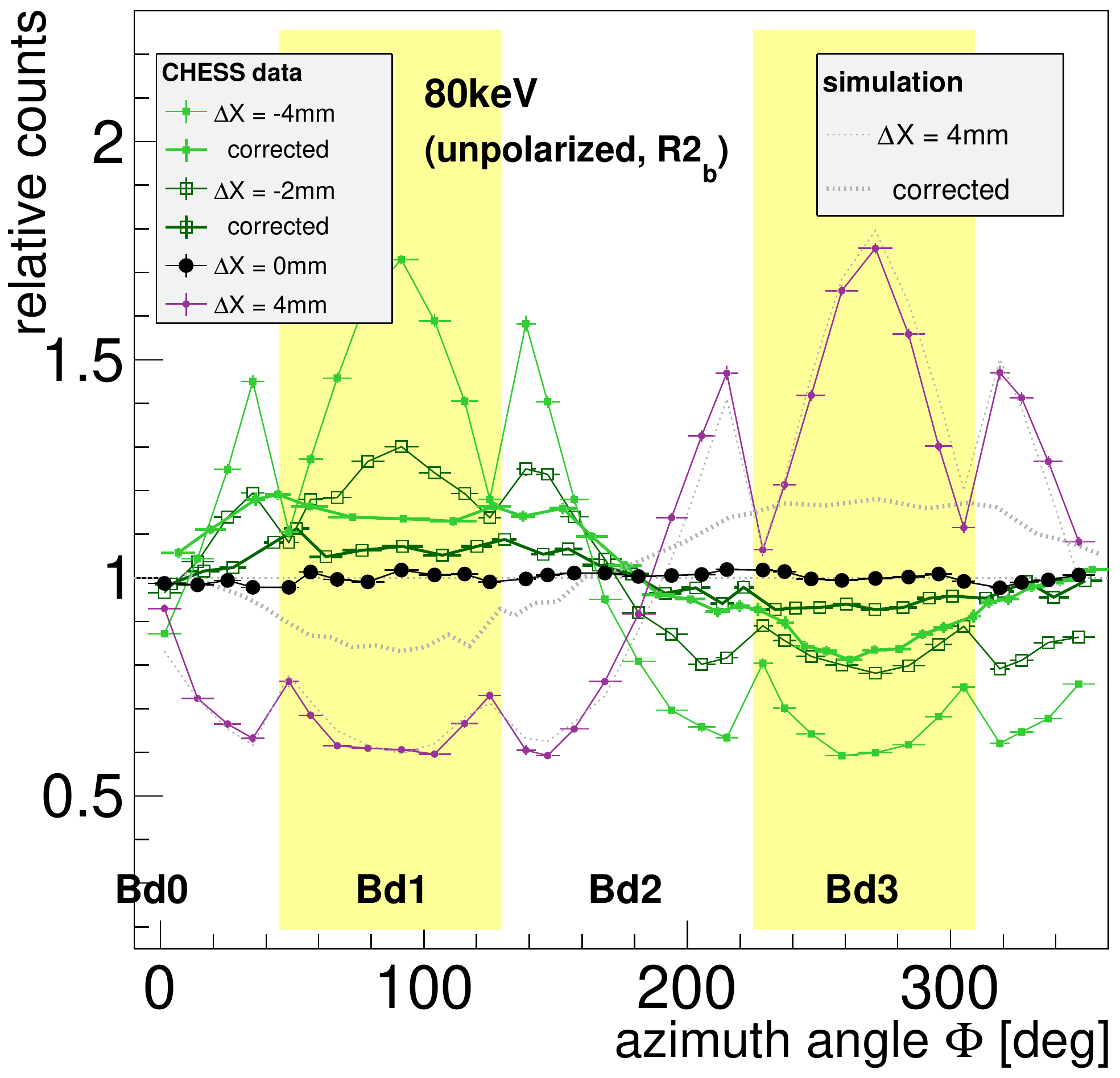}
\hfill
\includegraphics[height=0.285\textheight]{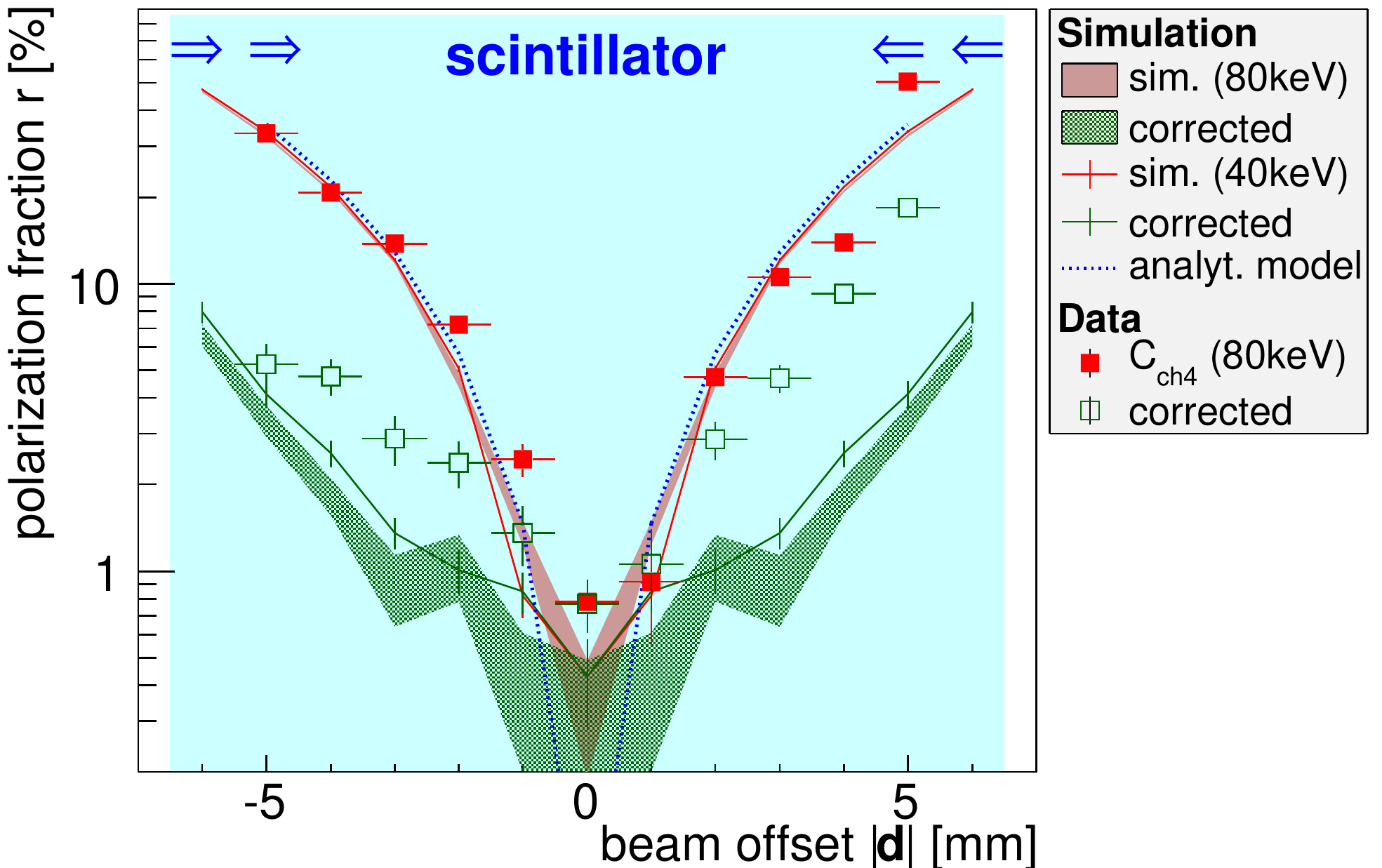}
\end{center}

\caption{{\bf Left:} Azimuthal scattering distributions of a
non-polarized X-ray beam (CHESS data and simulations, both in a 
non-rotating system) hitting the scintillator at different offsets $d$ 
relative to the optical axis. X-Calibur was oriented at $\alpha = 
0^{\circ}$, with the beam `walking' from {\it Bd1} to {\it Bd3} with 
increasing $\Delta X$ (see Fig.~\ref{fig:SchematicBeamOffset_AndStokes}, 
left, for a coordinate system). Scattering distributions are shown with 
and without the offset correction (see 
Sec.~\ref{subsec:AnalysisPolarization}). {\bf Right:} Apparent 
polarization fraction $r$ of the non-polarized beam. The light blue region 
indicates the diameter of the scintillator rod. The data were analyzed 
using the Stokes parameters as described by 
Eq.~(\ref{eqn:Stokes_Q_U_Average}). Results are shown without and with 
the corrected beam offset.}

\label{fig:PolFracSysError}
\end{figure*}

The understanding and control of systematic effects of different nature 
is crucial for the correct reconstruction of the polarization properties 
from measured data. A series of measurements was performed at CHESS (see 
Sec.~\ref{subsec:CHESS}) to study the effects of a mis-alignment 
(offset) between the X-ray beam and the optical axis of the polarimeter. 
The $X$/$Z$ stage of the table in hutch {\it C1} was used to 
systematically scan the polarimeter response for offsets ranging from 
$-5 \, \rm{mm}$ to $+5 \, \rm{mm}$ in steps of $1 \, \rm{mm}$ (the beam 
comes in along $Y$, see Figs.~\ref{fig:X-Calibur_Configurations} and 
\ref{fig:SchematicBeamOffset_AndStokes} (left) for the definition of the 
coordinate system). The scan along $X$ was performed with a polarimeter 
orientation of $\alpha = 0^{\circ}$, whereas the scan along $Z$ was 
performed with $\alpha = -90^{\circ}$. This allowed us to pairwise 
superimpose the data runs of perpendicular polarization planes that hit 
the scintillator at the same position~-- being equivalent to a 
non-polarized beam hitting at that particular offset position. The data 
runs were taken with configuration $C_{\rm{ch4}}$, testing the response 
of the $80 \, \rm{keV}$ harmonic of the CHESS X-ray beam. A set of 
simulations with an $80 \, \rm{keV}$ beam (polarized and non-polarized) 
were performed resembling the same offsets as measured at CHESS, as well 
as offset simulations at $40 \, \rm{keV}$.

{\it Non-polarized beam.} The left panel of 
Figure~\ref{fig:PolFracSysError} shows examples of azimuthal scattering 
distributions measured with different beam offsets for a non-polarized 
beam. The distributions were corrected for $\Delta \Phi_{j}$ (assuming a 
central beam) and $a_{j}$, as before. It can be seen that beam offsets 
of $\simeq 2 \, \rm{mm}$ already introduce systematic asymmetries. 
Although the offset distributions are not flat, as expected for a 
non-polarized beam, they also do not resemble the shape expected from a 
polarized beam. Given the high event statistics of the data used in the 
study, the sinusoidal fits used to determine the polarization properties 
will therefore certainly fail~-- allowing one to detect/filter the 
systematic effect. However, in the case of data with higher statistical 
uncertainty, the observed asymmetries could potentially lead to the 
reconstruction of an artificial polarization fraction of $r_{\rm{rec}} > 
0$. Note, that folding the distributions into the $[0; 180]^{\circ}$ 
interval (not shown), will substantially reduce the asymmetry. The 
Stokes analysis is per definition only considering the $[0; 
180]^{\circ}$ interval. However, it is not sensitive to deviations from 
the sinusoidal scattering distribution, and therefore does not provide a 
goodness-of-fit measure that can be used to detect imprints of a beam 
offset. Figure~\ref{fig:PolFracSysError} also shows examples of 
distributions that were offset corrected following the description in 
Sec.~\ref{subsec:AnalysisPolarization}. This, however, assumes that the 
beam offset is known.

\begin{figure*}[t!]
\begin{center}
\includegraphics[height=0.285\textheight]{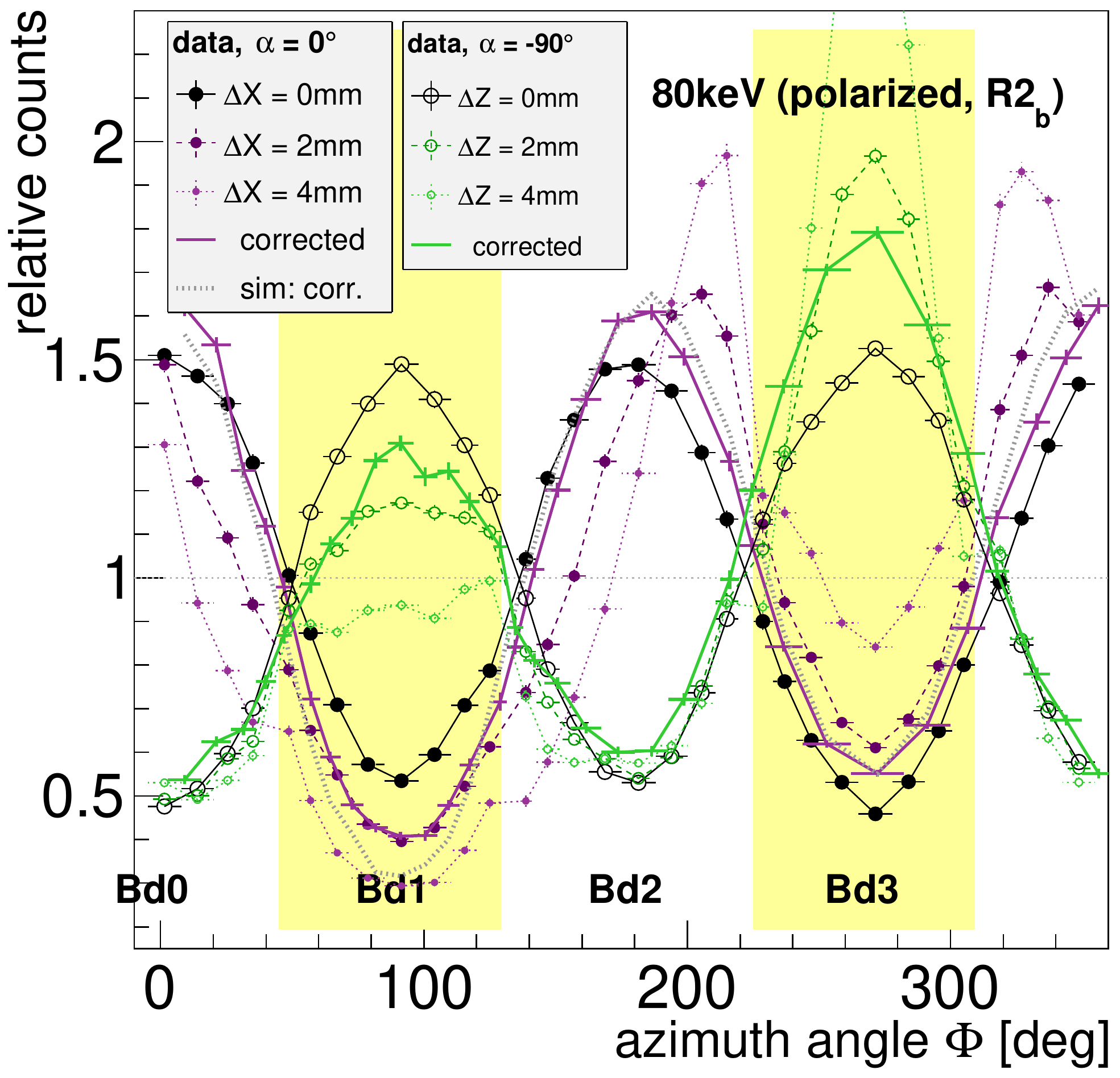}
\hfill
\includegraphics[height=0.285\textheight]{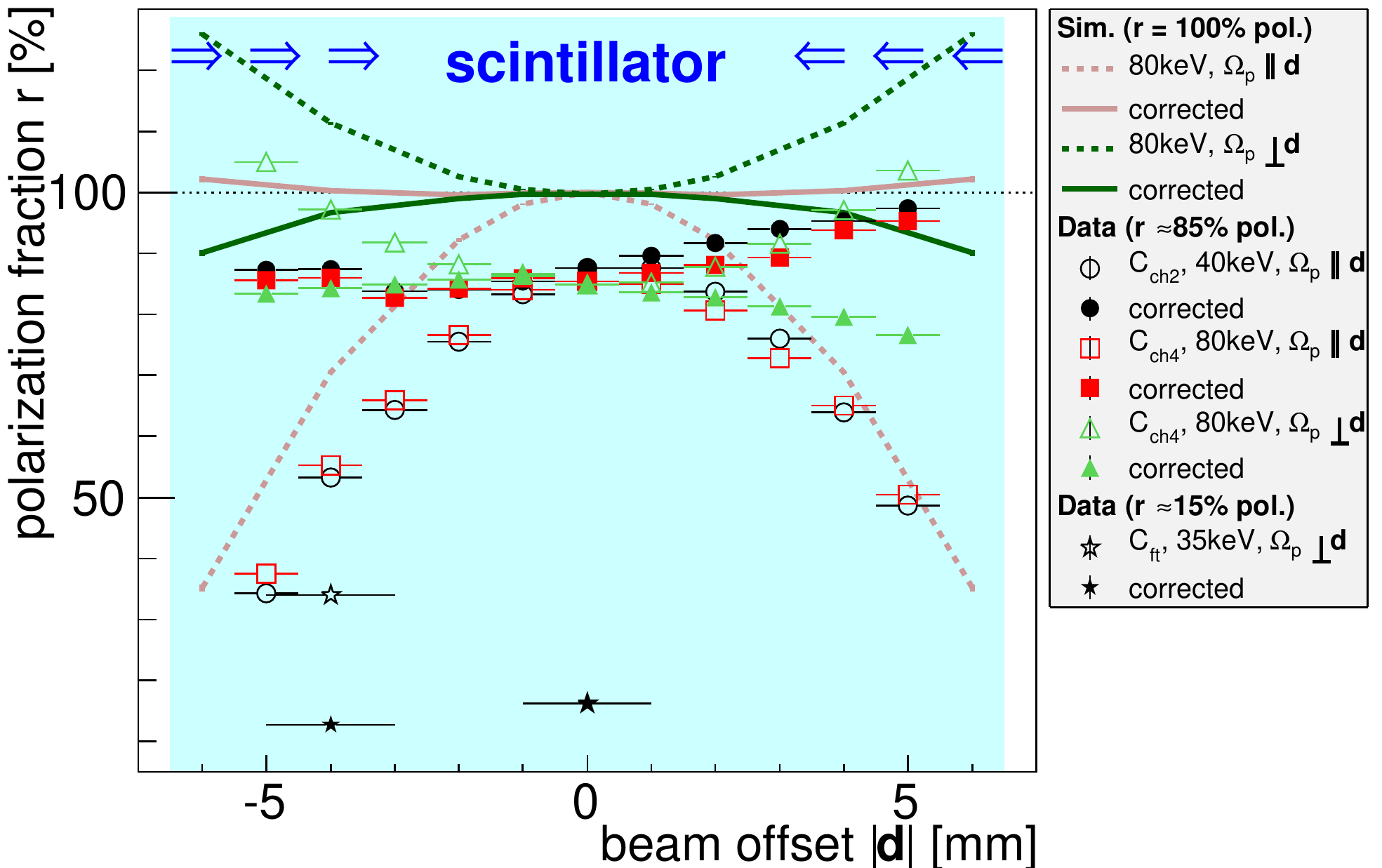}
\end{center}

\caption{{\bf Left:} Azimuthal scattering distributions of polarized 
X-ray beams hitting the scintillator at different offsets $d$. Results 
are shown for the CHESS data ($r_{\rm{ch}} \simeq 85\%$ polarized) and 
simulations ($r = 100\%$ polarized), both in a non-rotating system. 
Distributions corrected for the beam offset (see 
Sec.~\ref{subsec:AnalysisPolarization}) are shown, as well. Results are 
shown for offsets along $X$ ($\alpha = 0^{\circ}$) and $Z$ ($\alpha = 
-90^{\circ}$, corresponding to a scan along $X$ with perpendicular 
polarization). {\bf Right:} Reconstructed polarization fraction $r$ for 
the off-center beams, analyzed using the Stokes parameters as in 
Eq.~(\ref{eqn:Stokes_Q_U_Average}). The outline of the scintillator is 
indicated. Also shown are data derived from the Ft.~Sumner measurement 
shown in Fig.~\ref{fig:FtSumner_MirrorScan}, right. Results are shown 
without and with the offset correction.}

\label{fig:PolFracSysErrorPolarized}
\end{figure*}

To quantify the artificial polarization $r_{\rm rec}$ inferred when 
neglecting the beam offset, the data were analyzed using the Stokes 
parameters following Eq.~(\ref{eqn:Stokes_Q_U_Average}). 
Figure~\ref{fig:PolFracSysError} (right) shows $r_{\rm{rec}}$ as a 
function of beam offset. The results are shown for the simulations as 
well as for the CHESS data~-- both being in reasonable agreement. Also 
shown is an analytic model that simply integrates the events per 
azimuth angle interval in a plane perpendicular to the optical axis. The 
right panel of Fig.~\ref{fig:PolFracSysError} shows the same 
distribution after the beam offset correction described in 
Sec.~\ref{subsec:AnalysisPolarization} was applied. The discrepancy 
between the data and simulations after correction can be explained by an 
uncertainty in the absolute beam alignment in the data and the fact 
that, in contrast to the simulations, two separate measurements had to 
be superimposed to generate the non-polarized beam, amplifying the effect 
of unaccounted offsets or mis-alignments (any uncertainty in the offset 
not only moves the data points along the beam offset axis, but also 
along the $r$ axis).

In general, however, the first-order correction procedure greatly 
reduces the systematic effect to less than a few percent for offsets $d 
\leq 3 \, \rm{mm}$. The simulated data shown in the right panel of 
Fig.~\ref{fig:PolFracSysError} suggest that the correction works better 
for beam energies of $80 \, \rm{keV}$ as compared to $40 \, \rm{keV}$. 
This can be explained by the fact that only geometrical offsets are 
corrected for. Differences in absorption lengths in the scintillator 
material, originating from positions other than $P_{0}$ (see left panel 
in Fig.~\ref{fig:SchematicBeamOffset_AndStokes}), will cause 
second-order effects that depend on the energy of the scattered X-ray. 
However, addressing these additional correction terms is beyond the 
scope of this paper.

{\it Polarized beam.} For a polarized beam (left panel in 
Fig.~\ref{fig:PolFracSysErrorPolarized}), the offsets lead to either a 
reduction or amplification of the reconstructed polarization fraction, 
depending on the angle between the offset vector $\bf{d}$ and the plane 
of polarization $\Omega_{\rm{p}}$. The contribution of the beam offset 
to the reconstructed polarization is illustrated in the right panel of 
Fig.~\ref{fig:PolFracSysErrorPolarized} for different energies, 
including the results for the offset corrected analysis. The correction 
substantially reduces the systematic effect on $r$. A diverging trend 
can be identified in the corrected CHESS data for $d > 0 \, \rm{mm}$ 
(which is also visible in the case of the `non-polarized' beam shown in 
the right panel of Fig.~\ref{fig:PolFracSysError}). This indicates an 
inaccuracy in the experimental setup, e.g. the asymmetric/fractional 
scattering off a fixture along the beam path before entering the 
polarimeter.

An additional offset measurement was performed during the flight 
preparation in Ft.~Sumner (see Sec.~\ref{subsec:FtSumnerData}) using the 
partly polarized beam of the collimated X-ray source, hitting the 
scintillator at an offset of $d = -4 \, \rm{mm}$ (horizon system). The 
polarimeter/shield assembly was continuously rotating during the 
measurements. The corresponding fraction of apparent polarization $r$ is 
indicated in Fig.~\ref{fig:PolFracSysErrorPolarized} (right). Due to the 
rotation of the polarimeter, the offset correction (see 
Sec.~\ref{subsec:AnalysisPolarization}) was calculated on an 
event-by-event basis. The results illustrate that the correction also 
works in a rotated system, canceling the artificial fraction of 
polarization introduced by the offset. The offset-corrected polarization 
spectrum is shown in Fig.~\ref{fig:FtSumner_MirrorScan} (right), 
reproducing the measured results obtained with no beam offset.

\begin{figure}[t!]
\begin{center}
\includegraphics[width=0.49\textwidth]{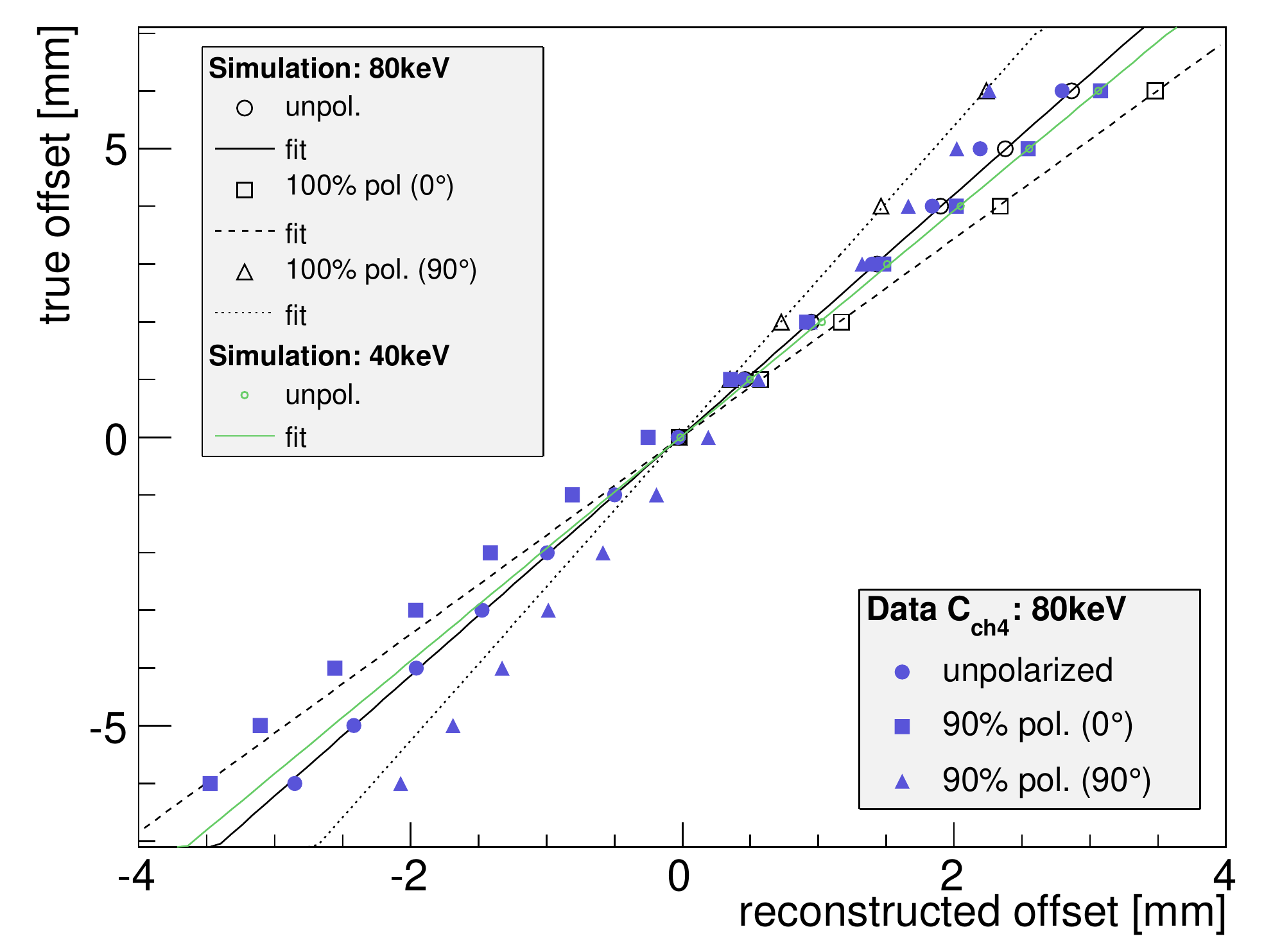}
\end{center}

\caption{Correlation between true beam offset and beam offset measured 
following Eq.~(\ref{eqn:FirstMoments}) for data and simulations.}

\label{fig:Offset_RecoMeanX}
\end{figure}

{\it Determination of the beam offset.} The geometrical offset 
correction described above assumes that the offset vector $\bf{d}$ is 
known. For sufficient event statistics and the assumption of a 
time-independent beam offset, the offset can be estimated from the data 
itself using first moments, essentially substituting $2 \Phi_{k} 
\rightarrow \Phi_{k}$ in Eq.~(\ref{eqn:StokesParams}):

\begin{eqnarray} \begin{split} \left< x \right> & = - \frac{c}{W} 
\sum_{k=1}^{N} w_{j(k)} \sin(\Phi_{k}), \\ \left< z \right> & = 
\frac{c}{W} \sum_{k=1}^{N} w_{j(k)} \cos(\Phi_{k}), \\ W & = 
\sum_{k=1}^{N} w_{j(k)}. \end{split} \label{eqn:FirstMoments} 
\end{eqnarray} Here, the constant $c=11 \, \rm{mm}$ reflects the 
distance between the scintillator center and the detector plane (see 
Fig.~\ref{fig:X-Calibur_Configurations}, top). The weights $w_{j(k)}$ 
are the same as in Eq.~(\ref{eqn:StokesSum}). As shown in 
Fig.~\ref{fig:Offset_RecoMeanX}, the measured offsets are linearly 
correlated with the true beam offsets. However, the slope of the 
correlation depends to some extent on the true polarization fraction $r$ 
of the beam and the angle between $\Omega_{\rm{p}}$ and {\bf d}. Since 
the offset of a non-polarized beam itself mimics a polarization fraction, 
there will be a residual ambiguity that prevents to completely 
disentangle $\bf{d}$ and $r$ from the data alone. Therefore, a 
time-resolved external monitoring of the beam position during the 
balloon flight is preferable.

Figure~\ref{fig:Offset_RecoMeanX} reveals a slight shift/translation 
between the slope of the reconstructed offset measured in the CHESS data 
versus the slope obtained from the simulations. This can be used to 
estimate an accuracy of the alignment achieved during the CHESS 
measurements to be $\Delta X_{\rm{sys}} \simeq 0.3 \, \rm{mm}$ for the 
X-Calibur orientation of $\alpha = 0^{\circ}$. Note, with a rotation 
axis possibly not exactly aligned with the axis of the scintillator, 
this may translate into larger offsets $\Delta X$ for different 
X-Calibur orientations $\alpha$.

{\it Systematic error on the polarization fraction.} The systematic 
error on the reconstructed polarization fraction $\Delta r_{\rm{sys}}$ 
can be estimated as follows. For the polarization measurements presented 
in this paper (Sec.~\ref{subsec:CHESS} and \ref{subsec:FtSumnerData}), 
we assume an unaccounted beam/scintillator mis-alignment of $\Delta d = 1 
\, \rm{mm}$. For a highly-polarized beam of $r = O(100\%)$ this leads to 
an error of $\Delta r_{\rm{sys,\Delta d}} \simeq 2\%$ relative to the 
reconstructed on-axis beam (see simulated curve in 
Fig.~\ref{fig:PolFracSysErrorPolarized}). For a non-polarized beam 
(simulations in Fig.~\ref{fig:PolFracSysError}) the offset leads to an 
overestimation of $\simeq 2\%$. For simplicity, we describe the 
systematic error introduced by the beam offset as independent of the 
true polarization fraction: $\Delta r_{\rm{sys,\Delta d}} = 2\%$.

The X-Calibur version used during the CHESS and Ft.~Sumner measurements 
(see Secs.~\ref{subsec:CHESS} and \ref{subsec:FtSumnerData}) allowed for 
some flexibility in adjusting the distance between the plane of the CZT 
detectors and the optical axis when assembling the polarimeter (see 
Fig.~\ref{fig:X-Calibur_Configurations}, top: $c = 11 \, \rm{mm}$). We 
measured $c_{\rm{ch}} = (11.3 \pm 0.5) \, \rm{mm}$ for the CHESS setup 
and $c_{\rm{ft}} = (10.5 \pm 0.5) \, \rm{mm}$ for the Ft.~Sumner setup. 
Conservatively, we assume $c = (11 \pm 1) \, \rm{mm}$. To study how an 
increased/decreased value of $c$ affects $r_{\rm{rec}}$, we 
correspondingly shifted pixel coordinates when analyzing a data set, but 
determined $r_{\rm{rec}}$ with the modulation factor $\mu_{\rm{sim}}$ 
obtained from simulations with the nominal value of $c = 11 \, \rm{mm}$. 
For $\Delta c = \pm 1 \, \rm{mm}$, a $100\%$ polarized beam is 
reconstructed to be $r_{\rm{rec}} = r_{-0.5\%}^{+1\%}$ (overestimation 
for a reduced distance $c$). For a non-polarized beam the effect is 
estimated to be $<0.1\%$. We therefore assume that the distance related 
systematic error\footnote{Note, that asymmetric distances of the 
different detector sides will mimic beam offsets with potentially 
stronger effects.} is proportional to $r$ with $\Delta r_{\rm{sys,c}} = 
0.01 \, r_{\rm{rec}}$. In the case of the Stokes analysis, the 
uncertainty in $c$ will introduce another systematic error: the angular 
coverage of the detector gaps, corrected for by 
Eq.~(\ref{eqn:StokesGapCorrection}), will be over/underestimated. For a 
$100\%$ polarized beam and $\Delta c = -1 \, \rm{mm}$, we find maximal 
deviation of $\Delta r_{\rm{sys,c}}^{\rm{Stk}} = \pm 6\%$, and for 
$\Delta c = +1 \, \rm{mm}$ we find $\Delta r_{\rm{sys,c}}^{\rm{Stk}} = 
\pm 4.5\%$, with the sign and strength depending on the orientation of 
the detector planes relative to the polarization vector (see 
Sec.~\ref{subsec:AnalysisPolarization}).

An error introduced by the analysis procedure can be estimated by 
comparing the results obtained with the two analysis methods presented 
in Sec.~\ref{subsec:AnalysisPolarization}. No significant differences 
were found when analyzing simulated data of a non-polarized beam. Based 
on the modulation factors of a $100\%$ polarized beam, see 
Eq.~(\ref{eq:ModulationFactor}), we estimate the systematic error to be 
on the order of $2\%$. Therefore, a relative error of $\Delta 
r_{\rm{sys,a}} = 0.02 \, r_{\rm{rec}}$ is assumed. Reasons for the 
difference can be the different treatment of dead pixels, detector gaps, 
etc. (see Sec.~\ref{subsec:AnalysisPolarization}). The total systematic 
error on the polarization fraction is:

\begin{eqnarray}
\begin{split}
\Delta r_{\rm{sys}} & = \Delta r_{\rm{sys,\Delta d}} + \Delta 
r_{\rm{sys,c}} + \left( \Delta r_{\rm{sys,c}}^{\rm{Stk}} \right) + 
\Delta r_{\rm{sys,a}} \\ & = 2\% + 0.01 \, r_{\rm{rec}} + \left(\Delta 
r_{\rm{sys,c}}^{\rm{Stk}} \right)+ 0.02 \, r_{\rm{rec}}.
\end{split}
\label{eqn:SystErrorPolFrac} 
\end{eqnarray}
This error is estimated and valid for the particular measurements 
presented in this paper. A better alignment in future measurements, a 
rotating polarimeter, and a more detailed study of the analysis methods 
will allow one to further reduce the error.

{\it Second-order systematic effects.} The tilt of the scintillator axis 
with respect to the X-ray beam will introduce another (depth-dependent) 
systematic effect. It should be mentioned that the simulations presented 
in this section assume a point-like X-ray beam, whereas the CHESS X-ray 
beam had a square-shaped footprint with side lengths of the order of $1 
\, \rm{mm}$. The systematic effects introduced by the offset/tilt might 
be weakened by properly taking into account the point spread function of 
the X-ray mirror that will be used for the astrophysical observations. 
The temperature dependence of the energy calibration of the CZT 
detectors (Sec.~\ref{subsec:TempStudies}), if not corrected for, can 
also modestly affect the reconstructed modulation factors of the data if 
the measurements are performed at temperatures other than the 
calibration temperature. However, the study of the effects discussed in 
this paragraph are beyond the scope of this paper.

\subsection{Summary of the Polarimeter Performance}

We used measurements at the CHESS X-ray beam facility to calibrate the 
polarimeter. CHESS gives a highly polarized and well collimated beam 
with well defined photon energies. We were able to demonstrate the full 
functionality of X-Calibur and measured the beam polarization to be 
$r_{\rm{ch}} = (88 \pm 5)\%$. In addition, we used a collimated X-ray 
beam from an X-ray source to make end-to-end tests of the full 
\INFOCUS/X-Calibur assembly in the field.

When uncorrected for, the $\sim$$3\%$ of defect pixels and gaps between 
detector boards produce an apparent polarization of a non-polarized beam 
of up to $5\%$ (Stokes analysis only). After correction, the apparent 
polarization goes down to $< 2 \%$. Once we rotate the polarimeter, we 
expect that the systematic effect is reduced by a factor $\sim 
\sqrt{n}$, with $n$ being the number of detected events. The systematic 
error will then be much smaller than the statistical error.

The detection principle of X-Calibur requires to focus the X-rays onto 
the center of the scatterer. If uncorrected for, an offset of $1 \, 
\rm{mm}$ leads to an apparent polarization of $\simeq 2\%$ for a
non-polarized beam. For the upcoming balloon flight, our goal is to limit 
the offset of the focal point from the center of the scatterer to $< 1 
\, \rm{mm}$, and to monitor the offset with a backward looking camera 
located close to the X-ray mirror. The camera will monitor a LED cross 
hair. We can in addition use X-ray data to constrain the location of the 
focal spot. The systematic error on the polarization fraction after 
correcting for the offset of the focal point is $< 1 \%$. Thus, for the 
upcoming balloon flight, we estimate that X-Calibur can detect $\geq 
3~\%$ polarization fractions in the $20 - 80 \, \rm{keV}$ band.

\section{Summary and Conclusions} \label{sec:Conclusion}

We designed, optimized, and built an X-ray polarimeter, X-Calibur, and 
studied its performance and sensitivity. The fully assembled X-Calibur 
polarimeter was tested (i) in the laboratory at Washington University, 
(ii) with a polarized X-ray beam at the Cornell High Energy Synchrotron 
Source (CHESS), and (iii) during a X-Calibur/\INFOCUS flight-integration 
test in Ft.~Sumner, NM. X-Calibur makes use of the fact that polarized 
photons Compton scatter preferentially perpendicular to their electric 
field orientation. It combines a detection efficiency on the order of 
$80 \%$, with a high modulation factor of $\mu \approx 0.5$ averaged 
over the whole detector assembly, and with values up to $\mu \approx 
0.7$ for select subsections of the polarimeter. Operated in a mode of 
continuous rotation, X-Calibur allows for a good control over systematic 
effects. Scattering polarimetry has the strength that it can operate 
over a wide energy range. The low energy threshold is given by the 
competition between photoelectric absorption and scattering processes, 
the mirror reflectivity limits the sensitivity at high energies.

We calibrated all 2048 CZT detector pixels of the polarimeter and 
studied their performance with respect to their energy threshold, energy 
resolution (including the contribution of electronic readout noise), and 
trigger efficiency. The CZT detectors achieve a mean energy threshold of 
$21 \, \rm{keV}$ and a mean $40 \, \rm{keV}$ energy resolution of 
$\Delta E_{\rm{czt}} \approx 4 \, \rm{keV}$ FWHM. The 
temperature-dependencies of the pixel responses were studied, as well, 
and we found that the effects can be sufficiently controlled for the 
temperature range expected during a balloon flight.

We characterized the performance of the active CsI shield and find a 
background suppression by more than one order of magnitude in the energy 
range relevant for X-Calibur. We also measured the trigger efficiency of 
the scintillator that is used as Compton scatterer and find a trigger 
threshold around $5 \, \rm{keV}$ with the potential for further 
reduction (using an improved amplification circuit). We used the CHESS 
X-ray beam to test the polarimeter. Detailed comparisons of experimental 
and simulated data allowed us to demonstrate the full functionality of 
the polarimeter and to measure the polarization of the CHESS beam. We 
studied different systematic effects that potentially affect the 
reconstructed polarization properties and estimated the systematic error 
to be smaller than $2\%$ for an upcoming balloon flight.

Our tentative observation program for an upcoming X-Calibur/\INFOCUS 
balloon flight includes galactic sources (Crab nebula, Her\,X-1, 
Cyg\,X-1, GRS\,1915, EXO\,0331) and one extragalactic source (Mrk\,421) 
for which sensitive polarization measurements will be carried through.

In principle, a similar space-borne scattering polarimeter could operate 
over the broader $3-80 \, \rm{keV}$ energy band. Here, a LiH rod would 
be used as passive scatterer. In contrast to the plastic scintillator 
used in the balloon-borne polarimeter, the LiH scatterer does not yield 
a coincidence signal. However, the lower atomic number of LiH results in 
the possibility of using the polarimeter down to energies of a few keV.

\section*{Acknowledgments}

We are grateful for NASA funding from grants NNX10AJ56G, NNX12AD51G and 
NNX14AD19G, as well as discretionary funding from the McDonnell Center 
for the Space Sciences to build the X-Calibur polarimeter. Polarization 
measurements: This work is based upon research conducted at the Cornell 
High Energy Synchrotron Source (CHESS) which is supported by the 
National Science Foundation and the National Institutes of 
Health/National Institute of General Medical Sciences under NSF award 
DMR-0936384. We would like to thank Ken Finkelstein for the excellent 
support in setting up the experiment at CHESS and for the continuous 
discussions thereafter.

\bibliography{reportBeilicke}
\bibliographystyle{ws-jai}

\end{document}